\documentclass[12pt]{article}
\usepackage{hyperref}
\usepackage{amsmath}
\usepackage{comment}
\usepackage{enumerate}
\usepackage{cite}
\numberwithin{equation}{section}
\title{Multi--charged geometries with cosmological constant}
\author{Rhucha Deshpande\footnote{rdeshpande@albany.edu}\  \   and Oleg Lunin\footnote{olunin@albany.edu}}
\date{}
\begin{document}
\def\be{\begin{equation}}
\def\bea{\begin{eqnarray}}
\def\ee{\end{equation}}
\def\eea{\end{eqnarray}}
\def\d{\partial}
\def\eps{\varepsilon}
\def\la{\lambda}
\def\b{\bigskip}
\def\nn{\nonumber \\}
\def\p{\partial}
\def\t{\tilde}
\def\h{{1\over 2}}
\def\r{{\rho}}
\def\qb{\bar q}
\def\be{\begin{equation}}
\def\bea{\begin{eqnarray}}
\def\ee{\end{equation}}
\def\eea{\end{eqnarray}}
\def\b{\bigskip}
\def\u{\uparrow}

\maketitle

\begin{center}
\ \vskip -1.2cm
{\em  Department of Physics,\\
University at Albany (SUNY),\\
1400 Washington Avenue,\\
Albany, NY 12222, USA
 }
 \end{center}

\vskip 0.08cm
\begin{abstract}
Motivated by gauged supergravities, we consider gravitational systems coupled to arbitrary numbers of fluxes and scalar fields. We show that simple ansatze for asymptotically AdS solutions in these systems fully determine the potential for the scalars, and we construct the black hole geometries which generalize the solutions known in gauged supergravities to an arbitrary number of dimensions. We also construct branes and brane intersections supported by an arbitrary number of higher--form fluxes and a correlated number of scalars.
\end{abstract}
\newpage
\tableofcontents

\newpage

\section{Introduction}

Over the last three decades, studies of black holes and black branes have led to remarkable progress in understanding string theory \cite{HorStr,PolchBraneW1,PolchBraneW2,PolchBraneW3,PolchBraneW4,PolchBraneW5}, quantum gravity \cite{BHapQGw1,BHapQGw2,BHapQGw3,BHapQGw4} and strongly coupled systems \cite{WittQCD,mald,WittAdSw1,WittAdSw2,WittAdSw3}. In particular, an important role has been played by charged objects that generalize the classic cases of the Reissner--Nordstr\"{o}m  and Kerr--Newman black holes in several directions: they contain extended sources \cite{HorStr}, multiple gauge fields \cite{SenW1,SenW2,SenW3,SenW4,SenW5,SenW6}, nontrivial scalars \cite{GibMaedaW1,GibMaedaW2,HorStrBH}, and the cosmological constant \cite{PopeAdSw1,PopeAdSw2}. While many systems without the cosmological constant arise from the ten--dimensional supergravities and their reductions, one can also extend these constructions to an arbitrary number of dimensions by starting from the Myers--Perry black holes \cite{MPerry} and dressing them with charges using techniques developed in \cite{SenW1,SenW2,SenW3,SenW4,SenW5,SenW6}. In the presence of the cosmological constant, the neutral solutions have also been constructed in arbitrary dimensions \cite{PopeAdSw1,PopeAdSw2}, but addition of charge presents significant challenges, and has only been done on the case--by--case basis. Such charged extensions were mostly constructed in the context of gauged supergravities, which are defined only in some dimensions, and whose field contents vary significantly as one looks at different dimensions. In this article we will use known solutions of several gauged supergravities as inspirations for constructing asymptotically-AdS black holes with an arbitrary number of charges in an arbitrary number of dimensions. We will also construct closely related geometries produced by $p$--branes, extending the well-known results obtained in \cite{HorStr}.

\bigskip

The great success in constructing charged black holes and black branes in string theory is partly attributed to the discovery of the U--duality group that allows one to start with a neutral seed solution of nonlinear differential equations and dress it with charges using a pure algebraic procedure\footnote{The discovery of black branes in supergravity by Horowitz and Strominger \cite{HorStr} pre-dates this algebraic method. Furthermore, the 
algebraic procedure applies only for specific couplings between gauge fields and scalars, and some solutions found in \cite{HorStr} and related work \cite{HorStrBH,GibMaedaW1,GibMaedaW2} come from more general Lagrangians.}. At the level of classical gravity for bosonic fields, such U--duality group can be formally defined in any number of dimensions, allowing one to start from a product of the neutral Myers--Perry geometry \cite{MPerry} and a torus and add several charges to the solutions. In the process of doing so, one generates gauge fields as well as several scalars. Since the reduction is performed on a torus, the gauge fields are abelian, and scalars do not have a potential. A similar reduction on spheres is more complicated, and often it leads to gauged supergravities with a nontrivial potential for the scalars. Unfortunately, such a reduction does not produce a U--duality group large enough to algebraically generate new charged solutions from the neutral ones, so the relevant configurations have been found on the case--by--case basis by solving the equations of motion. Furthermore, the actions for gauged supergravities and even their field contents strongly depend on dimensions. On the other hand, reduction on spheres also has a significant advantage: due to a nontrivial potential, the reduced theory admits solutions with anti--de--Sitter asymptotics, while the geometries obtained from the toric case are generally asymptotically--flat\footnote{One can also construct solutions with other asymptotics by taking various limits, but if all charges are turned off, then the geometries obtained via toric reduction always approach flat space at infinity.}.
Therefore, solutions constructed in gauged supergravities have played a prominent role in the exploration of the AdS/CFT correspondence \cite{mald,WittAdSw1,WittAdSw2,WittAdSw3}, and it is very interesting to find new examples of such geometries. In this article we will use solutions of gauged supergravities as inspirations for a general ansatz describing asymptotically AdS black holes with an arbitrary number of charges and find the {\it unique} action, which depends only on the numbers of fields and dimensions, that gives rise to such solutions. All known gauged supergravities are reproduced as special cases.

While many black hole geometries with AdS asymptotics are known, finding black branes with such asymptotics presents a significant challenge, even in the neutral case. This problem can be attributed to a relatively small amount of symmetry in the resulting configuration: while black holes are naturally embedded in the AdS space 
in the standard coordinates, black branes violate many symmetries of this space. Therefore, the ansatz interpolating between black branes and AdS asymptotics contains many degrees of freedom, and all attempts to obtain the solutions of equations of motion in closed analytical form have been unsuccessful\footnote{This discussion focuses on static flat branes. The geometries produced by supersymmetric branes with curved worldvolumes have been constructed \cite{LLMw1,LLMw2,LLMw3,LLMw4}, but they contain rotation, as expected from the general properties of such branes \cite{giantW1,giantW2}.}. The study of extended black objects in the presence of the cosmological constant began with exploration of black strings with AdS$_5$ asymptotics \cite{CopHor}, and it was later extended to black strings in arbitrary dimensions in \cite{MannRad,RaduChargeW1,RaduChargeW2,BlcStrLaStabW1,BlcStrLaStabW2,BlcStrLaStabW3}. 
The authors of these articles extracted many important properties of black strings with cosmological constant, but unfortunately, they had to rely on numerical simulations. More recent article \cite{DeLuBrn} extended these results to arbitrary $p$--branes and complemented the numerical analysis of \cite{CopHor,MannRad} by various analytical expansions, and in some special cases it reduced the problem to solutions of a relatively simple Abel equation. All these analyses were performed for neutral strings/branes, and in this article we extend them to the charged case. We will use the analytical expansions introduced in \cite{DeLuBrn} to prove existence and uniqueness of the charged solutions and to analyze their properties. 

In particular, we will show that under certain conditions, the extremal limits of the charged solutions may develop regular near--horizon regions, where the geometry reduces to a product of a sphere and AdS space, providing a multi--charge generalization of the well known examples of D3, M2, and M5 branes. Restricting to the near--horizon regime of charged geometries, one encounters not only black branes, but also regular wormhole--like solutions interpolating between two copies of the AdS space at two spacial infinities. Some solutions of this type were found in \cite{DeLuWrm}, and in this article we extend this construction to geometries charged under an arbitrary number of gauge fields.  Although studies of wormholes have a long history \cite{WormOldW1,WormOldW2,WormOldW3,WormOldW4,WormOldW5,WormOldW6,WormScalarW1,
WormScalarW2,WormScalarW3,WormScalarW4,WormScalarW5,WormScalarW6,WormScalarW7,
WormScalarW8}, and they intensified in the last decade
 \cite{EREPRw1,EREPRw2,EREPRw3,EREPRw4,EREPRw5}, most results have been obtained in two and four dimensions \cite{4dWHw1,4dWHw2,4dWHQw1,4dWHQw2,4dWHQw3,4dWHQw4,4dWHQw5,
 Example4dW1,Example4dW2,Example4dW3,Example4dW4}. Our analysis complements this work by studying higher--dimensional wormholes with multiple charges in the presence of a cosmological constant or a nontrivial potential for scalars\footnote{See also \cite{WHdimW1,WHdimW2,WHdimW3,WHdimW4,WHdimW5} for other approaches to wormholes in $d>4$.}. Once the cosmological parameter is switched off, the wormholes interpolating between two copies of the AdS space are no longer possible, but in this case, our geometries for multi--charged black branes become more explicit, and they extend the classic  results obtained by Horowitz and Strominger \cite{HorStr} to an arbitrary number of scalars and higher form gauge fields. 

\bigskip

This paper has the following organization. In section \ref{SecBH} we use inspiration from various supergravities to introduce a system of an arbitrary number of gauge fields and a correlated number of scalars. We show that the  harmonic ansatz for static black hole solutions leads to the unique expression for the scalar potential, which contains all known gauged supergravities as special cases. The only freedom in the potential is the overall multiplicative constant, which can be interpreted as a cosmological parameter. When all charges are switched off, the potential reduces to a cosmological constant. In section \ref{SecBrnFlat} we temporarily turn the cosmological parameter off and construct solutions for black branes with an arbitrary number of charges. In section \ref{SecSubNH} we take the near horizon limits of such solutions to obtain the AdS$_p\times$S$^n$ geometries supported by multiple fluxes, and discuss closely related multi--charged wormholes interpolating between two copies of the AdS$_d$ space in the presence of the cosmological parameter. Most results of sections \ref{SecBH} and \ref{SecBrnFlat} are obtained within specific ansatze for various components of the metric, where there is a correlation between charges under gauge fields and scalars. In section \ref{SecSubBrnGen} we relax this correlation by deriving the most general solutions consistent with symmetries of branes. The resulting explicit geometries are somewhat complicated, but we show that all cases known in the literature are recovered by taking particular limits. Finally, in section \ref{SecBrnLam} we put back the cosmological parameter and analyze properties of geometries describing charged branes in asymptotically AdS spaces. The solutions discussed in section \ref{SecBrnLam} are less explicit than those found in the rest of this article. Some technical details are presented in appendices.

\section[Multi--charged black holes with cosmological constant]{{\large{Multi--charged black holes with cosmological constant}}}
\label{SecBH}

Dimensional reduction of string theory produces gravitational systems containing multiple gauge fields and scalars, and black holes charged under these fields have been extensively studied in the past \cite{HorStrBH,SenW1,SenW2,SenW3,SenW4,SenW5,SenW6}. If the reduction is performed on a torus, then the action has only kinetic terms for the matter fields, and the multi--charged black holes can be constructed using algebraic techniques introduced in \cite{SenW1,SenW2,SenW3,SenW4,SenW5,SenW6}. Reduction on spheres leads to gauged supergravities \cite{GSR4w1,GSR4w2,GSR5w1,GSR5w2,GSR6,GSR7w1,GSR7w2}, and black hole solutions for such systems  have been found only on the case--by--case basis \cite{GSR4slnW1,GSR4slnW2,GSR4snRt,GSR5sln,GSR5snRotW1,GSR5snRotW2,GSR5snRotW3,GSR5snRotW4,
GSR5snRotW5,GSR6sln,GSR6snRot,AdCFbhW1,AdCFbhW2,CvetGub,GSR7snRot,10auth,Wu5d,Wu7d}. Some of these solutions are summarized in the Appendix \ref{AppGauSgr}, where we focused on the static case studied in this article\footnote{In that appendix we also give  references to articles containing some rotating solutions.}. In contrast to the toric reduction that generically leads to asymptotically--flat geometries, the black holes coming from gauged supergravities have AdS asymptotics, making them important for exploration of the AdS/CFT correspondence \cite{AdCFbhW1,AdCFbhW2,CvetGub}. In this section we will construct multi--charged black hole geometries which go beyond gauged supergravities. Specifically, we consider a system of an arbitrary number $k$ of gauge fields coupled to a correlated number of scalars and construct asymptotically--AdS black holes with $k$ charges. We will show that existence of simple closed--form solutions implies the unique potential for the scalar fields\footnote{A similar logic in the case of two vector fields and one scalar was used in \cite{LuGuesPot}.}, which generalizes results from all gauged supergravities summarized in the Appendix \ref{AppGauSgr}. We begin with a brief summary of the setup, then our discussion will be split into two subsections: in section \ref{SecSubBHspec} we will focus on special couplings between scalars and gauge fields inspired by string theory, and in section \ref{SecSubBHgen} the black hole solutions will be extended to general couplings.

\bigskip

Inspired by several well--known solutions of gauged supergravities reviewed in the Appendix \ref{AppGauSgr}, we consider an action describing a system of $k$ gauge fields $A_I$ and $k'$ scalars $\phi_j$ coupled to gravity:
\bea\label{ActionXphi}
S=\int d^dx\sqrt{-g}\left[R-\frac{1}{2}\sum_{j=1}^{k'}(\d{\phi}_j)^2
-\sum_{I=1}^k \frac{1}{4(X_I)^2}(F_{\mu\nu}^I)^2-{\tilde V}(\phi)
\right]\,.
\eea
Here fields $X_I$ are built from $k'$ dilatons and a constant matrix $\alpha$,
\bea\label{ActionXphiRel}
X_I=\exp\left[\sum\alpha_{Ij}\phi_j\right],
\eea
and as we will see below, there are two natural options: $k'=k$ and $k'=k-1$. 

Actions (\ref{ActionXphi}) with relations (\ref{ActionXphiRel}) commonly appear in gauged supergravities, but for our purposes it is convenient to invert (\ref{ActionXphiRel}) and use $X_J$ rather than $\phi_j$ as fundamental scalar fields. Motivated by the examples reviewed in Appendix \ref{AppGauSgr}, we require the action to be symmetric under arbitrary permutations of indices $(I,J)$. This implies that
\bea
\frac{1}{2}\sum_{j=1}^{k'}(\d{\phi}_j)^2=\beta_1\sum_{I\ne J}g^{\mu\nu}
\frac{\d_\mu X^I}{X^I}\frac{\d_\nu X^J}{X^J}+\beta_2\sum_I \frac{(\d X^I)^2}{(X^I)^2}=
w_1\left[\sum \d\psi_I\right]^2+w_2\sum (\d\psi_I)^2.\nonumber
\eea 
Here we defined a new set of scalar fields
\bea
\psi_I=-\log X^I\,.
\eea
In terms of these fields the action becomes
\bea\label{ActionPsi}
S=\int d^dx\sqrt{-g}\left[R-w_1\left[\sum \d\psi_I\right]^2-w_2\sum(\d{\psi}_I)^2
-\sum \frac{e^{2\psi_I}}{4}(F_{\mu\nu}^I)^2-{V}(\psi)
\right].
\eea
We will now use an inspiration from the known systems reviewed in Appendix 
\ref{AppGauSgr} to impose an ansatz for the black hole solutions and determine the potential ${V}(\psi)$ and parameters $(w_1,w_2)$ that lead to such geometries. 

\subsection{Special case: solutions inspired by string theory}
\label{SecSubBHspec}

Let us begin with recalling some well-known solutions of equations of motion coming from the action (\ref{ActionPsi}). Starting from the Schwarzschild--Tangherlini geometry in $d$ dimensions, adding an extra direction $y$, and performing a boost in that direction, one finds a vacuum solution describing a plane wave:
\bea\label{SchwDp1}
ds_{d+1}^2&=&-\frac{f}{H_1}dt^2+\frac{1}{f}dr^2+{r^2}d\Omega_{d-2}^2+H_1(dy+\frac{c_1s_1}{H_1}\frac{2m}{r^{d-3}}dt)^2,\\
&&f=1-\frac{2m}{r^{d-3}},\quad H_1=1+\frac{2ms_1^2}{r^{d-3}}.\nonumber
\eea
Here and below we use $s_I$ and $c_I$ as a shorthand notation for $\cosh\alpha_I$ and 
$\sinh\alpha_I$, the hyperbolic functions of the boost parameters. To add a second charge to the geometry (\ref{SchwDp1}), we perform a T duality along $y$ direction followed by another boost with parameter $\alpha_2$. The result reads
\bea\label{SchwDp2}
ds_{d+1}^2&=&-\frac{f}{H_1H_2}dt^2+\frac{1}{f}dr^2+{r^2}d\Omega_{d-2}^2+
\frac{H_2}{H_1}(dy+\frac{c_2s_2}{H_2}\frac{2m}{r^{d-3}}dt)^2,\\
B&=&\frac{c_1s_1}{H_1}\frac{2m}{r^{d-3}}dt\wedge dy,\quad e^{2\Phi}=\frac{1}{H_1},\quad
H_I=1+\frac{2ms_I^2}{r^{d-3}}.\nonumber
\eea
To view the geometry (\ref{SchwDp1}) as special case of static solutions for the action (\ref{ActionPsi}), we recall the standard rules for performing the dimensional reduction of the 
string action from $(d+1)$ to $d$ dimensions \cite{Schw,SenW1,SenW2,SenW3,SenW4,SenW5,SenW6}. For the ansatz
\bea
ds_{d+1}^2=e^{a\varphi}ds_d^2+g_{yy}(dy+A^2)^2,\quad B=A^1\wedge dy,\quad e^{-2\varphi}=e^{-4\Phi}g_{yy},\quad
a=\frac{2}{d-2}\,,
\eea
the reduced action is 
\bea
S&=&\int d^d x\sqrt{-g}\left[R-\frac{1}{d-2}(\d\varphi)^2-\frac{1}{4}(\d \ln g_{yy})^2-
\frac{1}{4}e^{-a\varphi}{F}^I_{\mu\nu}N_{IJ}{F}^{J\mu\nu}\right],\\
N&=&\mbox{diag}\left(\frac{1}{g_{yy}},g_{yy}\right)\,.\nonumber
\eea
This action indeed has the form (\ref{ActionPsi}) for $k=2$ and ${V}=0$. The reduced solution reads
\bea\label{ReducedTsT}
&&ds_d^2=H^{\frac{1}{d-2}}\left[-\frac{f}{H}dt^2+\frac{1}{f}dr^2+{r^2}d\Omega_{d-2}^2\right]\nn
&&A^I=\frac{2mc_Is_I}{r^{d-3}H_I}dt,\quad H=H_1H_2,\quad 
e^{-2\varphi}=H,\quad
e^{2\psi_I}=H^{\frac{1}{d-2}}\frac{(H_I)^2}{H}.
\eea
To determine the coefficients $(w_1,w_2)$ in (\ref{ActionPsi}), we observe that
\bea\label{ComputeW}
-\frac{1}{d-2}(\d\varphi)^2-\frac{1}{4}(\d \ln g_{yy})^2=
-\frac{d-3}{4}\left[\sum \d\psi_I\right]^2-\frac{1}{2}\sum(\d{\psi}_I)^2\,.\nonumber
\eea
This gives
\bea\label{WvaluesK2}
k=2:\quad w_1=\frac{d-3}{4},\quad w_2=\frac{1}{2}. 
\eea
Note that for $d=4,5$ the solution (\ref{SchwDp2}) and its rotating version were obtained in \cite{SenW1,SenW2,SenW3,SenW4,SenW5,SenW6} by a slightly different method of compactification to three dimensions and dualizing vector fields into scalars. The combination of boosts and dualities that led to (\ref{SchwDp2}) is an equivalent construction, but since it does not rely on dualization of vectors, it works in all dimensions.

Using the geometry (\ref{ReducedTsT}) as an inspiration, we look for solutions of the system (\ref{ActionPsi}) which have the following form:
\bea\label{AnstzHH}
&&ds_d^2=H^{\frac{1}{d-2}}\left[-\frac{f}{H}dt^2+\frac{1}{f}dr^2+{r^2}d\Omega_{d-2}^2\right],\\
&&A^I=\frac{2mc_Is_I}{r^{d-3}H_I}dt,\quad H_I=1+\frac{2ms_I^2}{r^{d-3}},
\quad H=H_1\dots H_k,\quad 
e^{2\psi_I}=\frac{(H_I)^2}{H^{(d-3)/(d-2)}}\,.\nonumber
\eea
To incorporate the cosmological constant, and more generally, a nontrivial potential 
${V}(\psi)$, we take the function $f$ to be 
\bea
f=1-\frac{2m}{r^{d-3}}+(g r)^2H.
\eea
The preceding discussion implies that the solution (\ref{ReducedTsT}) is covered by the ansatz  (\ref{AnstzHH}) with $k=2$ and $g=0$, but one can also construct embeddings for other values of $k$. Specifically, setting $s_3=s_4=\dots=s_k=0$ in (\ref{AnstzHH}), we find
\bea
e^{2\psi_{1,2}}=\frac{(H_{1,2})^2}{H^{(d-3)/(d-2)}},\quad 
e^{2\psi_{3}}=\dots=e^{2\psi_{k}}=\frac{1}{H^{(d-3)/(d-2)}},\quad
e^{2\psi}=\frac{H^2}{H^{k(d-3)/(d-2)}}.
\eea
Then the counterpart of equation (\ref{ComputeW}) is
\bea\label{ComputeWaa}
-\frac{1}{d-2}(\d\varphi)^2-\frac{1}{4}(\d \ln g_{yy})^2=
-\frac{d-3}{4(d-2)-2k(d-3)}\left[\sum\d\psi_I\right]^2-\frac{1}{2}\sum(\d{\psi}_I)^2\,.\nonumber
\eea
and the generalization of (\ref{WvaluesK2})
\bea\label{WvaluesK}
w_1=\frac{d-3}{4(d-2)-2k(d-3)},\quad w_2=\frac{1}{2}. 
\eea
Although we arrived at (\ref{WvaluesK}) by embedding the specific solution (\ref{ReducedTsT}) into the ansatz  (\ref{AnstzHH}), below we will demonstrate that these values of $(w_1,w_2)$ follow from consistency of equations of motion coming from the action (\ref{ActionPsi}), which are satisfied even in the presence of the cosmological term $g$.

Before analyzing these equations, we observe that not all $k$ dilatons $e^{2\psi_I}$ might be independent. Specifically, multiplying the dilatons, one finds
\bea
e^{2\psi_1+\dots+2\psi_k}=\frac{H^2}{H^{k(d-3)/(d-2)}}\,.
\eea
This implies that for 
\bea
k=\frac{2(d-2)}{d-3}
\eea
the dilatons are constrained. Since $k$ must be an integer, this happens only for the following values of $d$ and $k$:
\bea\label{SpecDK}
(d,k)=\Big\{(4,4);(5,3)\Big\}\,.
\eea
For all other sets $(d,k)$, the equations of motion following from (\ref{ActionPsi}) comprise the Einstein--Maxwell equations as well as equations for $k$ independent dilatons $\psi_I$. 

In the Appendix \ref{AppEOM} we demonstrate that the geometry (\ref{AnstzHH}) solves the equations of motion coming from the action (\ref{ActionPsi}) with parameters $(w_1,w_2)$ given by (\ref{WvaluesK}) and potential ${V}$ given by
\bea
V&=&g^2(d-2) \left[\frac{ z_1 P^{2}}{8 (d-2)}-\frac{(d-3)^2}{(d-2)} \sum_{I \neq J}^{k}X_I X_J 
+\frac{z_3 (d - 3)P}{(d-2)} \sum_{J=1}^{k}X_I \right],\\
P&=&\left[\prod_{I=1}^{k}X_I\right]^{-z_2}.\nonumber
\eea
The parameters $z_j$ are
\bea
z_1=2[4-2(k-1)(d-3)]z_3,\quad z_2 = \frac{3-d}{z_3},\quad z_3 =(k-2)(d-3)-2.
\eea
Here we excluded the special values of $(d,k)$ described by (\ref{SpecDK}). They will be discussed in the next subsection along with some generalizations. 

\subsection{Generalization: arbitrary scalar couplings}
\label{SecSubBHgen}

Let us generalize the solution (\ref{AnstzHH}) by starting with the action (\ref{ActionPsi}) and relaxing the condition (\ref{WvaluesK}). To have the final answer in a compact form, we introduce a fictitious parameter $\gamma$ by rescaling the scalars, i.e., we replace (\ref{ActionPsi}) by an equivalent action
\bea\label{ActionPsiGam}
S=\int d^dx\sqrt{-g}\left[R-{\tilde w}_1\left[\sum \d\psi_I\right]^2-{\tilde w}_2\sum(\d{\psi}_I)^2
-\sum \frac{e^{\gamma\psi_I}}{4}(F_{\mu\nu}^I)^2-{V}(\psi)
\right].
\eea
For $\gamma=2$, this action reduces to (\ref{ActionPsi}).

Inspired by the solution (\ref{AnstzHH}), we propose an ansatz
\bea\label{AnstzAlpha}
&&ds^2=-f H^{\nu_1}dt^2+H^{\nu_2}\frac{dr^2}{f}+H^{\nu_3}r^2d\Omega_{d-2}^2,\quad
e^{\mu\psi_I}=\frac{H^{\nu_4}}{H_I},\quad A^I=\frac{q_I}{r^{d-3}H_I}dt,\nn
&&H=\prod_I^k H_I,\quad H_I=1+\frac{2ms_I^2}{r^{d-3}},\quad 
f=1+(gr)^2H^{\nu_5}-\frac{2m}{r^{d-3}}\,,
\eea
with undetermined parameters $(\nu_1,\dots,\nu_5,\mu)$. Substitution of this ansatz into equations of motion determines the powers,
\bea\label{AlphMu1}
\mu=-\frac{\gamma}{2},\quad \nu_1=-(n-1)\nu,\quad \nu_2=\nu_3=\nu,\quad
\nu_4=-\frac{\nu_1}{2},\quad \nu_5=n\nu,
\eea
as well as parameters $(w_1,w_2)$ and the charges:
\bea\label{AlphMu2}
&&w_1=-\frac{\gamma^2n(n-1)\nu^2}{8[(n-1)k\nu-2]},\quad 
w_2=\frac{1}{8}n\gamma^2\nu,\quad q_I=2\sqrt{n\nu}ms_Ic_I.
\eea
The potential is relatively simple in terms of the harmonic functions $H_I$,
\bea\label{VgammaAsH}
V=g^2H^{n_-\nu}n(n_-)^2\left[\frac{1}{n_-}-
2{n}\left[\frac{1}{n_-}+\frac{\nu}{2}\sum\left\{\frac{1}{H_I}-1\right\}\right]^2+
\frac{\nu}{2}\sum\left[\frac{1}{H_I^2}-1\right]\right],
\eea
and in terms of $X$ it is given by
\bea\label{Vgamma}
V&=&g^2 n(n_-)^2 Y^2 \left[\frac{1}{n_-}- \frac{k \nu}{2}-{2}n\left(\frac{1}{n_-}
-\frac{k\nu}{2}+ 
\frac{\nu S_{\gamma/2}}{2Y}\right)^2 +\frac{\nu S_\gamma}{2 Y^2}\right],
\\
S_\alpha&=&\sum_{I=1}^k X_I^{\alpha},\quad Y=\prod_J^k (X_J)^z,\quad z = \frac{n_-\gamma \nu}{2 k n_-\nu-4}\,.\nonumber
\eea
Here we defined 
\bea
n_-=n-1,
\eea 
a convenient notation which will be used throughout this article. Parameter $\nu$ in (\ref{AlphMu1})--(\ref{AlphMu2}) can be rescaled by shifting the dilatons, and this freedom can be used to set $w_2=\frac{1}{2}$:
\bea\label{AlphMu3}
\nu=\frac{4}{n\gamma^2},\quad
w_1=
-\frac{(n-1)\nu}{2[n_-k\nu-2]}=\frac{n-1}{n\gamma^2-2n_-k},\quad w_2=\frac{1}{2},
\quad q_I=\frac{4ms_Ic_I}{\gamma}\,.
\eea
Our earlier result (\ref{WvaluesK}) is reproduced for $\gamma=2$. With the choice (\ref{AlphMu3}), the metric is given by
\bea\label{BHsolnFinalGam}
ds^2&=&H^\nu\left[-\frac{f}{H^{n\nu}}dt^2+\frac{dr^2}{f}+r^2d\Omega_{d-2}^2\right],\quad
e^{-\gamma\psi_I}=\frac{H^{(n-1)\nu}}{H^2_I}\,,\\
&&f=1+(gr)^2H^{n\nu}-\frac{2m}{r^{d-3}},\quad 
A^I=\frac{4mc_Is_I}{\gamma r^{d-3}H_I}dt,\quad H=\prod H_I\,.\nonumber
\eea
The geometry (\ref{BHsolnFinalGam}) along with expressions (\ref{Vgamma}) and (\ref{AlphMu3}) constitutes the main result of this section. We have outlined the logic that led to this answer, and the detailed derivation is presented in the Appendix \ref{AppEOM}. 

Some special cases of the solutions (\ref{BHsolnFinalGam}) and the potential (\ref{Vgamma}) have been encountered in gauged supergravities summarized in the Appendix \ref{AppGauSgr}. Let us briefly discuss these embeddings.
\begin{itemize}
\item The Reissner--Nordstr\"{o}m geometry with cosmological constant is recovered by going to a single gauge field in (\ref{BHsolnFinalGam}) and setting $\nu=\frac{2}{n-1}$. In this case, the scalar is turned off, and the potential (\ref{Vgamma}) reduces to a cosmological constant:
\bea
V=-g^2n(n+1).
\eea
\item A multi--charged generalization of the Reissner--Nordstr\"{o}m solution can be obtained by setting all charges equal to each other and by taking $\nu=\frac{2}{kn_-}$. This leads to vanishing scalars, and the potential again reduces to the cosmological constant.
\item Solution (\ref{4dGSsoln}) of the four--dimensional gauged supergravity (\ref{4dGSact}) is recovered by setting
\bea
n=2,\quad k=4,\quad \nu=\frac{1}{2}
\eea
in (\ref{BHsolnFinalGam}). For this value of $\nu$, the scalars are constrained by the relation 
$X_1X_2X_3X_4=1$, and the potential (\ref{Vgamma}) reproduces the SUGRA answer (\ref{4dGSpot}):
\bea
V=-{g}^2\sum_{I<J}X_IX_J.
\eea
\item Solution (\ref{5dGSsoln}) of the five--dimensional gauged supergravity (\ref{5dGSact}) is recovered by setting
\bea
n=3,\quad k=3,\quad \nu=\frac{1}{3}
\eea
in (\ref{BHsolnFinalGam}). For this value of $\nu$, the scalars are constrained by the relation 
$X_1X_2X_3=1$, and the potential (\ref{Vgamma}) reproduces the SUGRA answer (\ref{5dGSpot}):
\bea
V=-4 g^2\left[\frac{1}{X_1}+\frac{1}{X_2}+\frac{1}{X_3}\right]\,.
\eea
\item Solution (\ref{6dGSsoln}) of the six--dimensional gauged supergravity (\ref{6dGSact}) is recovered by setting
\bea
n=4,\quad k=1,\quad \nu=\frac{1}{2}
\eea
in (\ref{BHsolnFinalGam}). In this case, there is one unconstrained scalar, and the potential (\ref{Vgamma}) reproduces the SUGRA answer (\ref{6dGSpot}):
\bea
V=-g^2\left[9 X^2 +\frac{12}{X^2} - \frac{1}{X^6}\right]
\eea
\item Solution (\ref{7dGSsoln}) of the seven--dimensional gauged supergravity (\ref{7dGSact}) is recovered by setting
\bea
n=5,\quad k=2,\quad \nu=\frac{1}{5}
\eea
in (\ref{BHsolnFinalGam}). In this case, there are two unconstrained scalars, and the potential (\ref{Vgamma}) reproduces the SUGRA answer (\ref{7dGSpot}):
\bea
V=-2 g^2 \left[8X_1 X_2+\frac{4}{X_1X_2^{2}}+\frac{4}{X_1^{2}X_2}-\frac{4}{(X_1 X_2)^{4}}\right]\,.
\eea
\item An interesting set of solutions beyond gauged supergravities was found in \cite{WuAllDim}.  The static version of this solutions is given by (\ref{WUdGSsoln}), and it is reproduced by setting 
$k=1$ and $\nu=\frac{1}{d-2}$ in our geometry (\ref{BHsolnFinalGam}).
\end{itemize}
In gauged supergravities, some rotating extensions of (\ref{BHsolnFinalGam}) have been found as well \cite{GSR4snRt,GSR5snRotW1,GSR5snRotW2,GSR5snRotW3,GSR5snRotW4,GSR5snRotW5,GSR6snRot,GSR7snRot,Wu5d,Wu7d}, but unfortunately they often require actions that go beyond (\ref{ActionPsiGam}), and the additional terms have been constructed only in the case--by--case basis. For example, the action of ${\cal N}=2$ gauged supergravity in five dimensions is given by (\ref{5dGSact}),
\bea\label{Action5dSUGRA}
S&=&\int d^5 x \sqrt{-g}\left[ R - \frac{1}{2}(\d \bar{\phi})^2  - 
\sum_{I=1}^{3} \frac{1}{4X_I^2} (F_I)^2
+4 g^2 \sum_{I=1}^{3} X_I^{-1}-\frac{1}{6}\epsilon_{IJK}A^I F^J F^K\right],\nn
&&X_1=e^{-\phi_2-\phi_2},\quad X_2=e^{\phi_2-\phi_1},\quad X_3=e^{2\phi_1}\,\quad
X_1X_2X_3=1,
\eea
which reduces to a special case of (\ref{ActionPsiGam}) only in the absence of the Chern--Simons term. This term does not contribute to equations of motion for the static solutions which we are considering in this article, but it is crucial for finding the rotating solutions constructed in \cite{Wu5d}. Similar Chern--Simons couplings appear in other dimensions as well, and it would be interesting to do a reverse engineering of these terms using the logic we have implemented in this article: just as we have derived the most general potential (\ref{Vgamma}) starting from a particular ansatz (\ref{AnstzAlpha}) for the static solution, it might be possible to find the higher dimensional extensions of (\ref{Action5dSUGRA}) starting from a stationary ansatz, without appealing to gauged supergravity. An interesting example of this approach was implemented in \cite{WuAllDim}, where the most general rotating solution for one gauge field was constructed in all dimensions. In this case, the Chern--Simons term never contributes, as can be seen already in the example (\ref{Action5dSUGRA}). We hope to analyze the problem of rotating solutions with several fields in the future.

\section{Multi--charged branes with scalar couplings}
\label{SecBrn}
\label{SecBrnFlat}

In the last section we constructed geometries produced by black holes charged under multiple gauge fields and scalars, and it is natural to extend this construction to black branes charged under multiple higher form fluxes. Many solutions of this type have been constructed starting from the seminal work of Horowitz and Strominger \cite{HorStr}, and such geometries have played an important role in exploration of the gauge/gravity duality. In this section we will construct solutions describing branes with multiple charges and analyze near--horizon limits of extremal geometries and conditions under which such limits reduce to products of spheres and AdS spaces. 

Specifically, in section \ref{SecSubStack} we will use a harmonic ansatz inspired by black holes to construct black branes coupled to an arbitrary number of $p$--form fluxes. The resulting geometries have regular horizons, but in the extremal limit they may develop naked singularities. In section \ref{SecSubNH} we analyze such extremal solutions and determine the constraints on the number of fields which ensure non--singular behavior. In such a situation, the near--horizon limit produces a product of a sphere and an AdS space on the Poincare patch. 
In the same subsection we also analyze closely related wormhole--type geometries which interpolate between two copies of AdS space at two disconnected infinities. Finally, in section \ref{SecSubBrnGen} we relax the harmonic ansatz and construct the most general local solutions supported by $p$--forms and scalars consistent with symmetries of black branes. The explicit solutions are more complicated than the ones obtained within the harmonic ansatz, they have additional free parameters, and generically they lead to curvature singularities at the locations of the horizons.

\bigskip

To extend the results of the previous section to branes, we consider a generalization of the action (\ref{ActionPsiGam})
\bea\label{ActionPsiGamBrn}
S=\int d^dx\sqrt{-g}\left[R-{\tilde w}_1\left[\sum_I^k \d\psi_I\right]^2-{\tilde w}_2\sum_I^k(\d{\psi}_I)^2
-\sum_I^k \frac{e^{\gamma\psi_I}}{2(p+2)!}(F_{(p+2)}^I)^2
\right],
\eea
where $F_{(p+2)}$ are $(p+2)$--form field strengths. Note that, in contrast to (\ref{ActionPsiGam}), the potential $V(\psi)$ has been removed. As we have seen in the previous section, this potential is proportional to the cosmological constant, and in the presence of this parameter, there are no known analytical solutions even for neutral branes \cite{CopHor,MannRad,DeLuBrn}. Therefore, we do not expect to have simple solutions even in the charged case\footnote{Note, however, that one can construct charged wormholes \cite{DeLuWrm}.}, but some qualitative features of the resulting geometries will be analyzed in section \ref{SecBrnLam}.

\subsection{Solution for a single stack of multi--charged $p$--branes}
\label{SecSubStack}
Imposing the counterpart of the ansatz (\ref{AnstzAlpha}),
\bea\label{AnstzAlphaBrn}
&&ds^2=-f H^{\nu_1}dt^2+H^{\sigma}dy_{(p)}^2+H^{\nu_2}\frac{dr^2}{f}+H^{\nu_3}r^2d\Omega_{n}^2,\nn
&&e^{\mu\psi_I}=\frac{H^{\nu_4}}{H_I},\quad A_{(p+1)}^I=\frac{q_I}{r^{n-1}H_I}dt\wedge d^p y\,,\nn
&&H_I=1+\frac{2ms_I^2}{r^{n-1}},\quad 
f=1-\frac{2m}{r^{n-1}},\quad H=\prod_{I=1}^k H_I\,,
\eea
and repeating the analysis presented in the Appendix \ref{AppEOM}, we find the unique generalization of the relations (\ref{AlphMu1})--(\ref{AlphMu2}):
\bea\label{AlphMu1Brn}
&&\mu=-\frac{\gamma}{2},\quad \nu_1=\sigma=-\frac{\nu n_-}{p+1},\quad \nu_2=\nu_3=\nu,\quad
\nu_4=\frac{\nu n_-}{2},\quad \nu_5=n\nu\,,\\
&&w_1=-\frac{n_-N_p(\gamma\nu)^2}{8[n_-k\nu-2]},\quad 
w_2=\frac{N_p\gamma^2\nu}{8},\ q_I=2m\sqrt{N_p\nu}s_Ic_I,\ N_p=\frac{n+p}{p+1}\,.\nonumber
\eea
In other words, the geometry describing $k$ stacks of $p$--branes is given by 
\bea\label{BraneStackFlat}
&&ds^2=H^{\sigma}\left[-f dt^2+dy_{(p)}^2\right]+H^{\nu}\left[\frac{dr^2}{f}+r^2d\Omega_{n}^2\right],\nn
&&e^{\psi_I}=\left[\frac{(H_I)^2}{H^{\nu n_-}}\right]^{\frac{1}{\gamma}},\quad 
A_{(p+1)}^I=\frac{2m\sqrt{N_p\nu}s_Ic_I}{r^{n-1}H_I}dt\wedge d^p y\,,\quad \sigma=-\frac{\nu n_-}{p+1}\,,\\
&&H_I=1+\frac{2ms_I^2}{r^{n-1}},\quad 
f=1-\frac{2m}{r^{n-1}},\quad H=\prod_{I=1}^k H_I\,.\nonumber
\eea
The horizon is located at $r_h=(2m)^{1/n_-}$, and the entropy and temperature of the branes are given by
\bea
S=\frac{(r_h)^{n}}{4G_N}\left[\prod_I c_I\right]^{\frac{\nu n+\sigma p}{2}}\Omega_{n}V_p,\quad 
T=\frac{n-1}{r_h}\left[\prod_I c_I\right]^{N_p/2}\,.
\eea
Although generically the action (\ref{ActionPsiGamBrn}) has $k$ tensors and $k$ scalar fields, for a certain value of the parameter $\nu$,
\bea\label{NuConstr}
\nu_{constr}=\frac{2}{kn_-}\,,
\eea
the scalars in the solution (\ref{BraneStackFlat}) are subject to a constraint $\sum\psi_I=0$. This phenomenon has already been encountered for black holes in section \ref{SecBH}. 

As expected, geometries (\ref{BraneStackFlat}) reproduce well--known black branes in string theory \cite{HorStr} upon truncation to a single gauge field, i.e., by taking $k=1$ and setting parameter $\nu$ to the appropriate value. To see this, we begin with rewriting the action (\ref{ActionPsiGamBrn}) for $k=1$ using relations (\ref{AlphMu1Brn}):
\bea\label{ActionPsiGamBrnK1}
S=\int d^dx\sqrt{-g}\left[R+\frac{N_p \gamma^2\nu}{4[n_-\nu-2]}(\d\psi)^2
-\frac{e^{\gamma\psi}}{2(p+2)!}(F_{(p+2)})^2
\right]\,.
\eea
The embeddings into string theory and eleven--dimensional supergravity go as follows.
\begin{enumerate}[(a)]
\item According to (\ref{BraneStackFlat}), the scalar field $\psi$ vanishes if $n_-\nu=2$. To take this limit in (\ref{ActionPsiGamBrnK1}), one must send $\gamma$ to zero, so scalar and  tensor fields decouple. This situation is encountered for D3 branes in string theory, as well as for M2 and M5 branes in M theory. The solution (\ref{BraneStackFlat}) gives
\bea
\nu=\frac{2}{n-1},\quad \sigma=-\frac{2}{p+1}\,,
\eea
which in special cases reproduces the expected relations for D3, M2, and M5 branes.
\item Assuming that $n_-\nu\ne 2$, one can rescale the scalar field to choose a convenient normalization. There are two natural options: the canonical normalization that would give
\bea
\nu=\frac{4}{2n_-+N_p\gamma^2}\nonumber
\eea
and the choice inspired by the normalization of the dilaton in string theory and its dimensional reductions:
\bea\label{NuType2b}
\nu=\frac{2}{n_-+\frac{1}{16}N_p(n+p)\gamma^2}\,.
\eea
We will choose the latter option, which gives the kinetic term $\frac{4}{d-2}(\d\psi)^2$. 
Then comparing (\ref{ActionPsiGamBrnK1}) with the action of type IIB supergravity \cite{PolchStrW1,PolchStrW2,PolchStrW3,PolchStrW4}, one can embed fundamental strings and various branes:
\bea
\mbox{F1,\ D1}:&\gamma=\pm 1,\quad (n,p)=(7,1)&\Rightarrow\quad 
\nu=\frac{1}{4},\quad \sigma=-\frac{3}{4}\,,\nn
\mbox{NS5,\ D5}:&\gamma=\pm 1,\quad (n,p)=(3,5)&\Rightarrow\quad 
\nu=\frac{3}{4},\quad \sigma=-\frac{1}{4}\,,\\
\mbox{D7}:&\gamma=2,\quad (n,p)=(1,7)&\Rightarrow\quad 
\nu=1,\quad \sigma=0\,.\nonumber
\eea
\item
In the type IIA theory, the F1 and NS5 solutions work in the same way. For the D--branes we find
\bea
\mbox{D0}:&\gamma=\frac{3}{2},\quad (n,p)=(8,0)&\Rightarrow\quad 
\nu=\frac{1}{8},\quad \sigma=-\frac{7}{8}\,,\nn
\mbox{D2}:&\gamma=\frac{1}{2},\quad (n,p)=(6,2)&\Rightarrow\quad 
\nu=\frac{3}{8},\quad \sigma=-\frac{5}{8}\,,\\
\mbox{D4}:&\gamma=\frac{1}{2},\quad (n,p)=(4,4)&\Rightarrow\quad 
\nu=\frac{5}{8},\quad \sigma=-\frac{3}{8}\,,\nonumber\\
\mbox{D6}:&\gamma=\frac{3}{2},\quad (n,p)=(2,6)&\Rightarrow\quad 
\nu=\frac{7}{8},\quad \sigma=-\frac{1}{8}\,.\nonumber
\eea
Then the geometry (\ref{AlphMu1Brn}) with $k=1$ agrees with the well--known supergravity solutions for the non--extremal black branes.
\item As expected, in all the string/M theory cases recovered above, $\nu-\sigma=1$. It might be interesting to impose this condition and extract the consequences for $\gamma$. Using relations (\ref{NuType2b}) and (\ref{BraneStackFlat}), we find
\bea
\nu=\sigma+1\quad \Rightarrow\quad \nu=\frac{p+1}{n+p}\quad \Rightarrow \quad
\gamma^2=\frac{8-2(n-3)(p-1)}{n+p}\,.
\eea
It would be interesting to see whether these relations give physically relevant solutions beyond the cases covered in items (a)--(c). 
\item Since the geometry (\ref{BraneStackFlat}) has multiple gauge fields, it covers not only individual branes, but some smeared brane intersections as well. Specifically, let us consider a configuration D(p+4) branes stretched in the directions $(t,x^1,\dots,x^{p+4})$ and Dp branes stretched in $(t,x^1,\dots,x^{p})$. In the string frame, the metric reads
\bea
ds^2&=&\frac{1}{\sqrt{h_p h_{p+4}}}\left[-f dt^2+dx_1^2+\dots dx_p^2+h_p(dx_{p+1}^2+\dots+dx_{p+4}^2)\right]\\
&&+\sqrt{h_p h_{p+4}}\left[\frac{dr^2}{f}+r^2d\Omega^2_{4-p}\right],\qquad 
h_i=1+\frac{Q_i}{r^{3-p}}\,.\nonumber
\eea
Upon smearing the lower dimensional branes in the
$(x^{p+1},\dots,x^{p+4})$ directions and reducing the system along the four--dimensional torus spanned by these coordinates, one ends up with an effective supergravity action (\ref{ActionPsiGamBrn}) with two gauge fields: one is sourced by D(p+4) branes, and another one is sourced by the Dp branes. Generically, there are also two scalars: the original dilaton and the volume factor of the 4--torus. Dimensional
reduction from type II supergravity gives
\bea
\nu=\frac{p+1}{4},\quad \sigma=-\frac{n-1}{4},\quad n=4-p.
\eea 
In the case of the D1--D5 system, where $p=1$, there is a constraint $\psi_{2}=-\psi_1$ on the scalar due to relation $\nu n_-=1$. Therefore, there is only one independent scalar, and as we will see in the next subsection, this has important implications for the near--horizon limit of the branes. 
\item As our final example, we consider a smeared intersection of four stacks of  D3 branes: 
\bea\label{BraneScan}
\begin{array}{c|cccccccccc|}
&t&x_1&x_2&x_3&X_1&X_2&X_3&Y_1&Y_2&Y_3\\
\hline
D3_1&\bullet&&&&\bullet&\bullet&\bullet&\sim&\sim&\sim\\
D3_2&\bullet&&&&\bullet&\sim&\sim&\sim&\bullet&\bullet\\
D3_3&\bullet&&&&\sim&\bullet&\sim&\bullet&\sim&\bullet\\
D3_4&\bullet&&&&\sim&\sim&\bullet&\bullet&\bullet&\sim\\
\hline
\end{array}
\eea
Here bullets denote the directions wrapped by the branes and tildes denote the coordinates in which branes are smeared. 

The supergravity solution corresponding to the configuration (\ref{BraneScan}) in the extremal limit was constructed in \cite{KlebTseytl}, and it reads
\bea\label{D3stackGeom}
ds^2&=&-\frac{dt^2}{\sqrt{h}}+\sqrt{h}\left[dr^2+r^2 d\Omega_2^2+\frac{dX_1^2}{h_1h_2}+\frac{dX_2^2}{h_1h_3}+
\frac{dX_3^2}{h_1h_4}+\frac{dY_1^2}{h_3h_4}+\frac{dY_2^2}{h_2h_4}+\frac{dY_3^2}{h_2h_3}\right]\nn
F_5&=&\frac{1}{4}dt\wedge d\left[\frac{1}{h_1} dX_{123}+\frac{1}{h_2}dX_1dY_{23}-\frac{1}{h_3}dX_2dY_{13}+\frac{1}{h_4}dX_3 dY_{12}\right]+dual,\nn
h_i&=&1+\frac{Q_i}{r},\quad f=h_1h_2h_3h_4\,.
\eea
Reduction along six directions $(X_k,Y_k)$ produces a four--dimensional system with four gauge fields and a priori four independent scalars $\psi_I$ which is covered by the action (\ref{ActionPsiGamBrn}) with $k=4$. However, evaluating parameters $\nu$ and $\sigma$,
\bea
\nu=\frac{1}{2},\quad \sigma=-\frac{1}{2}\,,
\eea
we conclude that the scalars are subject to the constraint $\sum \psi_{i}=0$ due to the relation 
$k\nu n_-=2$. As we will show in the next subsection, this has important implications for the
near--horizon limit of the geometry (\ref{D3stackGeom}). 

\end{enumerate}

\subsection{Branes in the near--horizon limit and wormhole geometries}
\label{SecSubNH}

In the last subsection we constructed the geometries (\ref{BraneStackFlat}) which describe multi--charged black branes with regular horizons located at $r_h=2m$. As in the Schwarzschild case, one encounters a coordinate singularity at $r=r_h$, and the space can be continued using Kruskal coordinates. In this subsection we will focus on the extremal case of the geometries (\ref{BraneStackFlat}), for which the horizon is pushed to $r=0$, and explore their near--horizon limits. We will see that for a particular choice of the parameter $\nu$, such near horizon limit gives a product of the $n$--dimensional sphere and an AdS space on the Poincare patch according to the same mechanism that led to the discovery of the AdS/CFT correspondence \cite{mald}. Going to the global coordinates on AdS, one removes the coordinate singularity at the origin, but also cuts out the asymptotically--flat region \cite{WittAdSw1,WittAdSw2,WittAdSw3}. However, such a transition also allows one to introduce a cosmological constant or a nontrivial potential for scalars and extend the solutions (\ref{BraneStackFlat}) to geometries with AdS$_{p+n+2}$ rather than flat asymptotics. In this subsection we will explore such modifications, generalizing the singly-charged wormhole--type geometries which were constructed in \cite{DeLuWrm}. 

\bigskip

We begin with taking the extremal limit of the geometries (\ref{BraneStackFlat}) by sending $m$ to zero while keeping the charges $q_I\sim 2ms_I c_I$ fixed\footnote{Recall that the exact expression for $q_I$ is given by (\ref{AlphMu1Brn}).}. As usual, this implies that the boost parameters are sent to infinity. We further go to the near--horizon limit by going to small $r$, i.e., by neglecting the constant term in the harmonic functions in comparison to 
$\frac{2ms_I^2}{r^{n-1}}$. This leads to the geometries
\bea\label{BraneStackNH}
&&ds^2=H^{\sigma}\left[-dt^2+dy_{(p)}^2\right]+
H^{\nu}\left[{dr^2}+r^2d\Omega_{n}^2\right],\nn
&&e^{\gamma\psi_I}=\frac{Q_I^2}{Q^{\nu n_-}}r^{(k\nu n_--2)n_-},\quad 
 H=\frac{Q}{r^{k(n-1)}},\quad \sigma=-\frac{\nu n_-}{p+1}\,,\quad Q=\prod Q_I\,.
\eea
The fluxes are still given by (\ref{BraneStackFlat}). The solutions (\ref{BraneStackNH}) have curvature singularities at $r=0$ unless $k\nu n_-=2$: the easiest way to see this is to observe that the scalars make divergent contributions to the stress--energy tensor, but singularities can also be found by direct calculations. The regular solutions with $k\nu n_-=2$ have the following properties:
\begin{enumerate}[a)]
\item The metric describes a product of a sphere and an AdS space:
\bea\label{BraneStackNHads}
ds^2=Q^\nu\left[Q^{\sigma-\nu}r^{\frac{2n_-}{p+1}}\left[-dt^2+dy_{(p)}^2\right]+
\frac{dr^2}{r^2}\right]+Q^\nu d\Omega_{n}^2,
\eea
and the radii are given by
\bea\label{BraneStackNHradii}
R_S=Q^{\nu/2},\quad R_{AdS}=\frac{p+1}{n-1}Q^{\nu/2}\,.
\eea
The full solution (\ref{BraneStackFlat}) interpolates between (\ref{BraneStackNHads}) and asymptotically flat space at infinity.
\item Once the conditions (\ref{BraneStackNH}) is imposed, all scalars $\psi_I$ in (\ref{BraneStackNH}) become constant:
\bea\label{BraneStackNHscal}
e^{\gamma\psi_I}=\left[\frac{Q_I}{Q^{1/k}}\right]^2\,.
\eea
Although these scalars are fixed in the near--horizon limit, they have nontrivial profiles in the full solution (\ref{BraneStackFlat}), where they interpolate between (\ref{BraneStackNHscal}) and $\psi_I=0$ at infinity.
\item Although the action (\ref{AnstzAlphaBrn}) generically has $k$ gauge fields and $k$ scalars, the solution (\ref{BraneStackFlat}) with the condition $k\nu n_-=2$ implies a constraint
\bea\label{BraneStackNHconstr}
\sum\psi_I=0,
\eea
which holds everywhere, not only in the near horizon limit. Therefore, we conclude that to avoid curvature singularities in the solutions, the action (\ref{AnstzAlphaBrn}) must have $k$ gauge fields but only $k-1$ independent scalars.
\item Some string theory examples of geometries with the regularity condition $k\nu n_-=2$ were reviewed in section \ref{SecSubStack}, and not surprisingly, they contain the extremal D3, M2, M5 branes, D1--D5 system, and D3--D3--D3--D3 intersections. It would be interesting to find physical applications of other regular cases as well. Note that many string theory solutions that do not satisfy the regularity condition $k\nu n_-=2$, such as geometries produced by fundamental strings and D branes, are allowed in the full theory and lead to singularities only in the supergravity approximation.   
\end{enumerate}
Let us focus on the solutions satisfying the regularity condition $k\nu n_-=2$. The geometries (\ref{BraneStackNHads}) can be continued beyond the Poincare patch to produce a product of a sphere and a global AdS space:
\bea\label{AdStimeS}
ds^2=R_{AdS}^2\left[(\rho^2+1)d\Sigma_{p+1}^2+
\frac{d\rho^2}{\rho^2+1}\right]+R_S^2 d\Omega_{n}^2.
\eea
Here $\Sigma_{p+1}$ denotes the $(p+1)$--dimensional AdS space, and the radii are still given by (\ref{BraneStackNHradii}). In contrast to (\ref{BraneStackNHads}), the geometry (\ref{AdStimeS}) cannot be smoothly connected to the asymptotically flat space (\ref{BraneStackFlat}). Interestingly, a slight modification of (\ref{AdStimeS}) can be connected to a space with AdS$_{p+n+2}$ asymptotics, and the resulting geometry solves equations of motion coming from the action (\ref{ActionPsiGamBrn}) with an additional scalar potential, i.e., from the brane counterpart of the action 
(\ref{ActionPsiGam}). Since in the near horizon limit all scalars approach fixed values (\ref{BraneStackNHscal}), in this regime any scalar potential would look like a cosmological constant. In the presence of the cosmological term, the geometry (\ref{AdStimeS}) still solves equations of motion with sources from fluxes, but the relations (\ref{BraneStackNHradii}) are modified to \cite{LiuSabra,DeLuWrm}
\bea\label{NHradiiLam}
\frac{1}{R_S^2}=\frac{1}{Q^{\nu}}-\frac{d-1}{n-1}g^2,\quad 
\frac{1}{R^2_{AdS}}=\left[\frac{n-1}{p+1}\right]^2\frac{1}{Q^{\nu}}+\frac{d-1}{p+1}g^2,\quad d=n+p+2.
\eea
This can be viewed as a generalized Freund--Rubin ansatz \cite{FrRub}. 
Parameter $g$ is related to the cosmological constant $\Lambda$ by
\bea
\Lambda=-\frac{(d-1)(d-2)}{2}g^2\,.
\eea
Let us now go beyond the near horizon region and connect the geometry (\ref{AdStimeS}) with AdS$_{p+n+2}$ asymptotics at infinity.

\bigskip

We begin with reviewing the results for the $k=1$ case, which were obtained in \cite{DeLuWrm}. In this case, there is only one gauge field and no scalars due to the constant (\ref{BraneStackNHconstr}) so the potential is representing the genuine cosmological constant everywhere. Both the geometry (\ref{NHradiiLam}) and the asymptotic AdS$_{p+n+2}$ have the isometries of an $n$--dimensional sphere and a $p+1$--dimensional AdS space, so we will assume that the connecting solution has these isometries as well. This leads to an ansatz for the metric and the field strength:
\bea\label{WrmhlAnstz}
ds^2=Ad\Sigma_{p+1}^2+\frac{dr^2}{A}+Cd\Omega_n^2,\quad F_{(p+2)}=q \star d\Omega_n\,.
\eea 
Note that we imposed a convenient gauge by fixing the relation between AdS and radial components of the metric. Functions $A$ and $C$ in (\ref{WrmhlAnstz}) depend on the radial coordinate, and the relevant equations of motion were analyzed in \cite{DeLuWrm}. Let us briefly summarize the results of that investigation.
\begin{enumerate}[(a)]
\item Geometries (\ref{WrmhlAnstz}) solving equations of motion have curvature singularities unless there is a point $r=r_0$ where ${\dot A}={\dot C}=0$ while $A$ and $C$ remain finite. Without loss of generality, we will set $r=r_0$. 
\item For regular geometries, one must impose the boundary conditions at the origin:
\bea\label{OldBoundCond}
r=0:\quad {\dot A}={\dot C}=0,\quad A=A_0,\quad C=C_0,
\eea
and the value of $A_0$ is determined by the equations of motion:
\bea\label{A0Old}
A_0=p(p+1)\left[\frac{nn_-}{C_0}-\frac{q^2}{2C_0^n}+(d-1)(d-2)g^2\right]^{-1}\,.
\eea
Therefore, the solutions are specified by a single integration constant $C_0$. The boundary conditions can be used to integrate equations to both positive or negative values of $r$, therefore regular solutions describe wormhole--type geometries with two disconnected asymptotic regions.
\item There are three special values of $C_0$, which were labeled as $(C_\star,C_\#,C_\bullet)$ in \cite{DeLuWrm}, 
\bea
C_\star < C_\# < C_\bullet\,,
\eea
and geometries (\ref{WrmhlAnstz}) develop curvature singularities away from $r=0$ if $C_0<C_\star$ or $C_0>C_\bullet$\,.
\item For $C_0=C_\star$, one recovers the product space (\ref{AdStimeS}). 
\item For $C_0=C_\bullet$ one finds a regular interpolation between AdS$_{p+2}\times$S$^n$ and a product AdS$_{p+1}\times\Sigma^{n+1}$, where $\Sigma$ is a hyperbolic space.
\item When $C_0$ is in the range $C_\star<C_0<C_\bullet$, one find an interpolation between the AdS$_{p+2}\times$S$^n$ and an asymptotically locally AdS$_d$ space 
\cite{AsympLocAdSw1,AsympLocAdSw2}, and the standard 
AdS$_d$ space at infinity is recovered when $C_0= C_\#$.
\end{enumerate}
We refer to \cite{DeLuWrm}  for the detailed discussion of these outcomes, expressions for the special points $(C_\star,C_\#,C_\bullet)$, and the profiles of the warp factors $(A,C)$ as functions of the radial coordinate\footnote{Article \cite{DeLuWrm} also analyzed dyonic solutions, but we will not discuss them here.}. Here we will focus on extending these results to solutions with multiple charges, which have nontrivial scalar fields and the potential $V$.

\bigskip

As demonstrated above, regular geometries can be obtained only if the condition $k\nu n_-=2$ is satisfied, and this imposes the constraint (\ref{BraneStackNHconstr}) on scalar fields. For such configurations, the ansatz (\ref{WrmhlAnstz}) has a unique extension consistent with symmetries:
\bea\label{WrmhlAnstzNew}
ds^2=Ad\Sigma_{p+1}^2+\frac{dr^2}{A}+Cd\Omega_n^2,\quad F^{(I)}_{(p+2)}=q_I \star d\Omega_n,\quad \psi_I:\ \sum \psi_I=0.
\eea 
Since already in the case of one charge, it was impossible to find interpolations between different asymptotic regions in closed analytical form, here, as in \cite{DeLuWrm}, we have to rely on perturbative expansions and numerical simulations. In particular, repeating the analysis reviewed above, we conclude that the geometries (\ref{WrmhlAnstzNew}) have naked singularities unless the conditions (\ref{OldBoundCond}) are imposed. Since the equations of motion are invariant under reflection of the $r$ coordinate, the conditions (\ref{OldBoundCond}) imply the following expansions at small $r$:
\bea\label{ACPpert}
A=A_0+\sum_{j=1}^{\infty} a_j r^{2j},\quad 
C=C_0+\sum_{j=1}^{\infty} c_j r^{2j},\quad 
\psi_I=e^{(I)}_0+\sum_{j=1}^{\infty} e^{(I)}_j r^{2j}\,.
\eea
Index $I$ runs from $1$ to $(k-1)$, and the values of $e^{(I)}_0$ cab be extracted from (\ref{BraneStackNHscal}). These $(k+1)$ expansions can be substituted into three Einstein's equations and $(k-1)$ equations for the scalars, and as in the case of one charge (\ref{WrmhlAnstz}), one of Einstein's equations turns out to be redundant. Therefore, at each order in $r$, one finds $(k+1)$ linear equations for $(k+1)$ variables, leading to the unique solution. The only exception from this rule is zeroth order, where one encounters a nonlinear constraint on $(A_0,C_0)$, which is solved by a straightforward counterpart of (\ref{A0Old}): 
\bea\label{A0New}
A_0=p(p+1)\left[\frac{nn_-}{C_0}-\sum_I \frac{q_I^2}{2C_0^n}+(d-1)(d-2)g^2\right]^{-1}\,.
\eea
In particular, one concludes that solutions (\ref{WrmhlAnstzNew}) regular at $r=0$ are uniquely specified by one parameter $C_0$. Beyond the perturbative expansions (\ref{ACPpert}) one can perform numerical integration of the system and analyze the behavior of solutions at infinity. 
One finds a phase structure which is qualitatively similar to items (c)--(f) above with one major difference: now the scalar fields $\psi_I$ have nontrivial profiles. We conclude that an asymptotic behavior from each of the options (d)-(f) above leads to a {\it unique} regular solution (\ref{WrmhlAnstzNew}). The numerical values $(C_\star,C_\bullet)$ depend on the number of scalar fields and values of charges. 
\bigskip

So far we have focused on $p$--branes with $p\ge 1$, let us now make some comments about the near--horizon limit of black holes. In this case, the solutions with cosmological constant are known explicitly, so we can take the near--horizon limit in (\ref{BHsolnFinalGam}). To avoid unnecessary clutter, we will focus on the $k=1$ case and make some comments about qualitatively similar situations of arbitrary $k$ in the end. 

At finite mass, solution (\ref{BHsolnFinalGam}) describes black holes with regular horizon at $r=r_h$, where $f(r_h)=0$. The extremal limit is obtained by sending $m$ to zero while keeping $Q=2ms_1^2$ fixed. In this case one can go all the way to $r=0$, where the 
scalar field $\psi_1$ and the metric become singular unless $\nu=2/n_-$. Picking this value of 
$\nu$, we can write the extremal limit of (\ref{BHsolnFinalGam}) as
\bea\label{BhNHxi}
ds^2&=&-\frac{\rho^2+(gQ)^2\Xi^{n\nu}}{\Xi^{(n-1)\nu}}dt^2+
\frac{Q^2}{(n_-)^2}\frac{\Xi^{\nu}d\rho^2}{\rho^2+(gQ)^2\Xi^{n\nu}}+
Q^2 \Xi^\nu d\Omega_{n}^2\,.
\eea
Here we defined a convenient function 
\bea
\Xi=1+\rho\,,
\eea
which approaches one as $\rho=(r/Q)^{n-1}$ goes to zero. In the absence of the cosmological constant, the geometry (\ref{BhNHxi}) terminates at $\rho=0$, where it reduces to a product of $S^n$ and the Poincare patch of the AdS$_2$ space. For a nontrivial $g$, the leading contributions to the Ricci tensor are also reminiscent of the corresponding expressions for the product AdS$_2\times$S$^n$ with radii (\ref{NHradiiLam}),
\bea
{R^\rho}_\rho&=&{R^t}_t=-(n+1)g^2-\frac{(n-1)^2}{Q^2}+\mathcal{O}(\rho^2),\nn 
{R^a}_{b}&=&\left[-(n+1)g^2+\frac{n-1}{Q^2}\right]\delta^a_b+\mathcal{O}(\rho^2),
\eea
but the coordinate $\rho$ continues to negative values until the geometry (\ref{BhNHxi}) develops a curvature singularity at $\rho=-1$. Therefore, the extremal limit of the black hole solution (\ref{BHsolnFinalGam})  is regular if $g=0$ and singular otherwise. 

Going beyond the explicit geometry (\ref{BhNHxi}), one may hope that the wormhole ansatz (\ref{WrmhlAnstz}) can be used to connect two copies of AdS$_d$ with a 
AdS$_{p+2}\times$S$^n$ funnel in the middle, even when $p=0$. Let us now show that this is not possible due to a critical difference between the construction described earlier and the $p=1$ case. While for $p\ge 1$ equation (\ref{A0Old}) determines $A_0$ in terms of $C_0$, requirement of non--vanishing $A_0$ for $p=0$ leads to a constraint on $C_0$ itself:
\bea
\frac{nn_-}{C_0}-\frac{q^2}{2C_0^n}+(d-1)(d-2)g^2=0.
\eea
Then expanding $A$ and $C$ into Taylor series (\ref{ACPpert}) and substituting the result into equations of motion for the geometry (\ref{WrmhlAnstz}), one finds that expansion of $C$ truncates after the constant term, while $A$ has the form $A=A_0+a_1 r^2$. Therefore, the only solution with $p=0$ that satisfies the regularity conditions (\ref{OldBoundCond}) is 
AdS$_{2}\times$S$^n$. Although our discussion has focused on $k=1$, similar logic applies to arbitrary $k$, and it leads to the same conclusions: the extremal limit of black holes (\ref{BHsolnFinalGam}) is singular, and a wormhole--type geometry connecting two copies AdS$_d$ with an 
AdS$_{2}\times$S$^n$ funnel is not possible. The detailed structures of the intermediate equations, such as (\ref{BhNHxi}) for $\nu=2/(k n_-)$, are slightly more complicated.  

\bigskip

To summarize, in this subsection we have analyzed the near--horizon limits of the geometries constructed earlier, focusing on the extremal branes. We found that if the constraint $k\nu n_-=2$ is satisfied, then in such near--horizon limits one finds regular products of AdS spaces and spheres, and the coordinates of the brane solutions cover only the Poincare patch of the AdS space. We then performed continuation to the global AdS, added a cosmological constant, and demonstrated existence of regular wormhole--type geometries using the techniques developed in \cite{DeLuWrm}. Interestingly, such regular geometries are not possible for zero--branes, and the charged black hole solutions constructed in section \ref{SecBH} develop naked curvature singularities in the extremal limit.

\subsection{Smeared branes and brane intersections}
\label{SecSubInter}

In section \ref{SecSubStack} we constructed geometries (\ref{BraneStackFlat}) which solve equations of motion coming from the action (\ref{ActionPsiGamBrn}). To do so, we imposed the ansatz (\ref{AnstzAlphaBrn}) in which the number of flat directions was correlated with the rank of the field strength. Specifically, all flat directions were longitudinal to the branes. In this subsection we will extend the earlier results in two directions: we will construct the geometries describing smeared branes, in which some of the flat directions are transverse to the worldvolume, and the solutions produced by branes within branes. Some examples of the second type of solutions were briefly discussed in section \ref{SecSubStack}, where we analyzed their reductions to lower dimensions, and here we will construct the full uplifted geometries. In contrast to the string theory examples discussed in section \ref{SecSubStack}, generically we will be able to embed $s$--branes inside $p$ branes for any values of $(s,p)$, and the well known relation $p=s+4$ is recovered when the couplings of scalars are adjusted to their string theoretic values. 

Let us consider a generalization of the action (\ref{ActionPsiGamBrn}) that has two types of field strengths:
\bea\label{ActPsiGamMixed}
S&=&\int d^dx\sqrt{-g}\left[R
-\sum_I^k \frac{e^{\gamma\psi_I}}{2(p+2)!}(F_{(p+2)}^I)^2-\sum_J^{\bar k} \frac{e^{\bar\gamma{\bar\psi}_J}}{2(s+2)!}({\bar F}_{(s+2)}^J)^2\right.\\
&&\left.-{w}_1\sum_I^k(\d{\psi}_I)^2-{w}_2(\d\psi)^2
-
{\bar w}_1\sum_J^{\bar k}(\d{\bar \psi}_J)^2-{\bar w}_2(\d{\bar\psi})^2
-2w_3\d\psi\d{\bar\psi}
\right]\,.\nonumber
\eea
Here we defined two convenient combinations
\bea\label{SmearPsiDef}
\psi=\sum_I^k \psi_I,\quad {\bar\psi}=\sum_J^{\bar k} {\bar\psi}_J\,.
\eea
Without loss of generality, we will assume that $s<p$. Special cases of the action (\ref{ActPsiGamMixed}) arise in string theory in several contexts:
\begin{enumerate}[(a)]
\item In the smeared Ds--D(s+4) intersections, one often encounters two different types of the Ramond--Ramond fields sourced by different types of branes. This corresponds to $k=k'=1$ in  (\ref{ActPsiGamMixed}). Examples include D2--D6, D4--D8, and D3--D7 systems \cite{D2D6w1,D2D6w2,D2D6w3,D2D6w4,D2D6w5}. In all these cases, the fields $\psi$ and ${\bar\psi}$ are not independent since they come from a single dilaton in string theory. We will discuss implications of this constraint once the solutions are constructed below. 
\item The D1--D5 system is usually described in terms of a single three--form field strength that has both electric and magnetic components. Since in this article we are focusing on electric solutions, we will use an alternative description in terms of $F_{(3)}$ and $F_{(7)}$ with $k=k'=1$ in (\ref{ActPsiGamMixed}). As for other Ds--D(s+4) systems, the scalars are constrained. The same action, although with different coupling to scalars, describes fundamental strings dissolved in NS5 branes.
\item The supergravity action for heterotic strings contains a three--form $H_3$ and 16 abelian vector fields. This system is covered by (\ref{ActPsiGamMixed}) with $(k,{\bar k})=(1,16)$. In this case, ${\bar\gamma}=0$ and scalars ${\bar\psi}_I$ are turned off. Reduction of the action  (\ref{ActPsiGamMixed}) to lower dimensions yields a system of vector fields coupled to scalars, which has been extensively studied in the past \cite{SenW1,SenW2,SenW3,SenW4,SenW5,SenW6}. 
\end{enumerate}
Let us analyze brane configurations which solve equations of motion generated by (\ref{ActPsiGamMixed}). 

\bigskip

Solution (\ref{BraneStackFlat}) describes a single stack of $p$--branes, so it is symmetric under $SO(n+1)$ rotations with $n=d-p-2$. Similarly, one can construct a geometry describing a single stack of $s$--branes, which is invariant under $SO(n'+1)$ rotations with $n'=d-s-2$. On the other hand, one can uniformly smear the $s$--branes over a 
$(p-s)$--dimensional hyperplane. This would produce a geometry with $SO(n+1)$ invariance and $(p+1)$ flat directions. Notice that since some of these directions are transverse to the $s$--branes, and others are longitudinal, there would be no rotational invariance in the $p$ spacial directions which was present in (\ref{BraneStackFlat}). Once the smeared branes with charges from the second line of (\ref{ActPsiGamMixed}) are combined with the $p$--branes charged under the fields in the first line, one arrives at a generalization of the ansatz (\ref{AnstzAlphaBrn}):
\bea\label{AnstzBrnMixd}
&&ds^2=-f H^{\nu_1}{\bar H}^{\bar\nu_1}dt^2+H^{\sigma}{\bar H}^{\bar\sigma_1}dy_{(s)}^2+H^{\sigma}{\bar H}^{\bar\sigma_2}dz_{(p-s)}^2+
H^{\nu_2}{\bar H}^{{\bar\nu}_2}\frac{dr^2}{f}+
H^{\nu_3}{\bar H}^{{\bar\nu}_3}r^2d\Omega_{n}^2,\nn
&&e^{\mu\psi_I}=\frac{H^{\nu_4}{\bar H}^{\nu_5}}{H_I},\quad 
e^{{\bar\mu}{\bar\psi}_J}=\frac{{\bar H}^{{\bar\nu}_4}{H}^{\bar\nu_5}}{\bar H_I},\nn
&&F^I_{(p+2)}=q'_I e^{-\gamma\psi_I}\star d^n\Omega,\quad
F^J_{(s+2)}={\bar q}'_Je^{-{\bar\gamma}{\bar\psi}_J}\star\left[d^n\Omega d^{p-s}z\right],
\\
&&H_I=1+\frac{2ms_I^2}{r^{n-1}},\quad {\bar H}_J=1+\frac{2m{\bar s}_J^2}{r^{n-1}},\quad f=1-\frac{2m}{r^{n-1}}\,.\nonumber
\eea
Substitution of this ansatz into the equations of motion coming from the action (\ref{ActPsiGamMixed}) gives
\bea\label{AnswBrnMixd}
&&ds^2=H^{\sigma}{\bar H}^{\bar\sigma}\left[-f dt^2+dy_{(s)}^2\right]+H^{\sigma}{\bar H}^{\bar\nu}dz_{(p-s)}^2+
H^{\nu}{\bar H}^{{\bar\nu}}\left[\frac{dr^2}{f}+r^2d\Omega_{n}^2\right],\nn
&&e^{\gamma\psi_I}=\frac{H_I^2}{[H^{\nu}{\bar H}^{\bar\nu}]^{n_-}},\quad 
e^{{\bar\gamma}{\bar\psi}_J}=\frac{\bar H^2_I}{{H}^{\nu\la}{\bar H}^{{\bar\nu}(n_-+p-s)}},\nn
&&A_{(p+1)}^I=\frac{q_I}{r^{n-1}H_I}dtd^s y d^{p-s}z,\qquad
{\bar A}_{(s+1)}^J=\frac{{\bar q}_J}{r^{n-1}{\bar H}_J}dtd^s y,\\
&&\sigma=-\frac{n_-{\nu}}{p+1},\quad {\bar\sigma}=-\frac{(n_-+p-s){\bar\nu}}{s+1},\quad\la=\frac{n_-(s+1)}{p+1}\,,
\nn
&&q_I=2m\sqrt{N_{p,n}\nu}s_I c_I,\quad  {\bar q}_I=2m\sqrt{N_{s,n+p-s}{\bar\nu}}{\bar s}_I {\bar c}_I,\quad
N_{p,n}=\frac{n+p}{p+1}\,.\nonumber
\eea
Note that in (\ref{AnstzBrnMixd}) we did not assume specific forms of the gauge potentials, but rather started with the most general field strengths consistent with symmetries and Bianchi identities. Once the Einstein's equations fixed various powers in the metric and scalar fields in (\ref{AnswBrnMixd}), integration of the Bianchi identities yielded very simple expressions for 
$(A_{(p+1)}^I,{\bar A}_{(s+1)}^J)$. The charges $(q_I,{\bar q}_J)$ contain an additional factor of $(n-1)$ in comparison to $(q'_I,{\bar q}'_J)$. We chose to write (\ref{AnswBrnMixd}) in terms of two free parameters $(\nu,,{\bar\nu})$, but they are fixed in terms of the coefficients $(w_1,w_2,{\bar w}_1,{\bar w}_2,w_3)$ in the action (\ref{ActPsiGamMixed}). In fact, to accommodate solutions of the form (\ref{AnstzBrnMixd}), these coefficients must be constrained: in contrast to the single stack of $p$--branes discusses in section \ref{SecSubStack}, where two parameters $(w_1,w_2)$ in the action could have been traded for $\nu$ and rescaling of the scalar field, now such rescalings and changes in $(\nu,{\bar\nu})$ can absorb only four of the five constants. The parameters in the action are given by
\bea
w_2&=&\frac{(n+{\hat p})}{({\hat p}+1)}\frac{\gamma^2\nu}{8},\quad 
{\bar w}_2=(n+{\hat p})\frac{{\bar\gamma}^2{\bar\nu}}{8},\quad 
w_3=\frac{\gamma{\bar\gamma}\nu{\bar\nu}}{4E}{\hat w},\quad {\hat w}=(n+{\hat p})n_-\,,\quad
{\hat p}=p-s,\nn
w_1&=&\frac{(\gamma\nu)^2}{4E}\left[{\hat w}-\frac{{\bar k}{\bar\nu}n_-{\hat p}(n+{\hat p})^2}{2({\hat p}+1)}\right],\quad
{\bar w}_1=\frac{({\bar\gamma}{\bar\nu})^2}{4E}\left[{\hat w}-
{\hat p}(n+{\hat p})^2\left[\frac{{k}{\nu}n_-}{2}-1\right]\right],\\
E&=&2({\hat p}+1)\left[2-k\nu n_--{\bar k}{\bar\nu}(n_-+{\hat p})\right]+k{\bar k}{\hat p}n_-(n+{\hat p})\nu{\bar\nu}\,.
\nonumber
\eea
This completes our summary of the solutions (\ref{AnswBrnMixd}), let us now discuss some of their properties.

As written, the action (\ref{ActPsiGamMixed}) contains $k+{\bar k}$ gauge fields and the same number of scalars. However, solutions (\ref{AnswBrnMixd}) may impose some constraints on the scalar fields, similar to the one we have encountered in previous sections. Recalling the 
definition (\ref{SmearPsiDef}) of the fields $(\psi,{\bar\psi})$, we can compute
\bea
e^{\gamma\psi}=H^{2-k n_-\nu}{\bar H}^{-{k}n_-{\bar\nu}},\quad
e^{\bar\gamma\bar\psi}={\bar H}^{2-{\bar k} {\bar\nu}(n_-+p-s)}{H}^{-{\bar k}\la\nu}.
\eea
Assuming that all charges $(q_I,{\bar q}_J)$ are independent, we conclude that $\psi$ and 
${\bar\psi}$ are constrained if and only if 
\bea\label{ConstDilPS}
\bar\nu=\frac{2(k\nu n_--2)}{{\bar k}\left[-2(n_-+p-s)+kn_-\frac{p-s}{p+1}(n+p)\nu\right]}:\quad
{\bar\psi}=-\frac{\gamma}{\bar\gamma}\frac{{\bar k}\la\nu}{2-kn_-\nu}\psi.
\eea
In particular, these relations hold for the Ds--D(s+4) systems in string theory, where there are two gauge fields and only one independent scalar, the ten--dimensional dilaton. 

As expected, in the absence of $s$--branes, i.e., for ${\bar k}=0$, the constraints (\ref{ConstDilPS}) reduce to the relation (\ref{NuConstr}) for a single stack of $p$--branes. In section \ref{SecSubNH} we saw that the relation (\ref{NuConstr})  also implied regular near--horizon limits, but now this condition is distinct from (\ref{ConstDilPS}). Repeating the analysis presented in section \ref{SecSubNH}, we conclude that near $r=0$, the extremal version of the metric (\ref{AnswBrnMixd}) is obtained by setting 
\bea
f=1,\quad H=\frac{Q}{r^{kn_-}},\quad {\bar H}=\frac{\bar Q}{r^{{\bar k}n_-}}\,.\nonumber
\eea
The sphere decouples if $n_-(k\nu+{\bar k}{\bar\nu})=2$, while the $z$ directions decouple if 
$k\sigma+{\bar k}{\bar\nu}=0$. Combining these two relations, we find
\bea\label{BraneInterNuReg}
\nu=\frac{2(p+1)}{kn_-(n_-+p+1)},\quad {\bar\nu}=\frac{2}{{\bar k}(n_-+p+1)}\,.
\eea
With these conditions, the near--horizon geometry of (\ref{AnswBrnMixd}) reduces to $AdS_{s+2}\times S^n\times T^{p-s}$, and as expected, all scalars $(\psi_I,{\bar\psi}_J)$ saturate to constant values. As in section \ref{SecSubNH}, one can go to the global coordinates on 
$AdS_{s+2}$ but unfortunately the procedure for finding the potential term for the scalars and constructing geometries interpolating between $AdS_{s+2}\times S^n\times T^{p-s}$ and 
AdS$_d$ at infinity fails due to the presence of a torus. In the next section we will use an alternative method for constructing asymptotic AdS$_d$ with tori, but our example will use truncated versions of the action (\ref{ActPsiGamMixed}).  

For a single stack of $p$--branes discussed in section \ref{SecSubStack}, one condition (\ref{NuConstr}) ensured two properties: constrained dynamics of scalar fields and regularity of the near horizon limit of the extremal geometry. Now these two properties are ensured by (\ref{ConstDilPS}) and (\ref{BraneInterNuReg}), and the former condition is clearly weaker. Interestingly, relations (\ref{BraneInterNuReg}) imply (\ref{ConstDilPS}), so solutions which are regular in the near--horizon limit must have constrained scalars {\it everywhere}. It would be interesting to find a physical reason for this correlation. 

Let us conclude with a string theoretic example of the system satisfying the conditions (\ref{BraneInterNuReg}). Setting $k={\bar k}=1$ and $(n,p)=(3,1)$, we find that 
${\nu}={\bar\nu}=\frac{1}{2}$. This recover the well--known example of D1 branes dissolved inside D5 (or fundamental strings dissolved in NS5), and as expected, the relations (\ref{ConstDilPS}) enforcing a constraint on two scalars are satisfied since string theory has only one dilaton.

\subsection{The most general solution for static $p$--branes}
\label{SecSubBrnGen}

So far in this article we have considered various actions for gravity coupled to scalars and fluxes, and then imposed certain ansatze for the geometries. It is interesting to see whether one can find more general solutions of equations of motion, and in this section we will investigate such possibilities. To avoid unnecessary clutter, we will focus on the action (\ref{ActionPsiGamBrn}) with one vector field and one scalar, but qualitative conclusions will hold in more general cases as well. While the solution (\ref{BraneStackFlat}) for $k=1$ has only one charge parameter $q_1$, which appears in both the scalar and the field strength, we will show that the most general solution depends on two charges. A possibility of such extension was mentioned already in the original article by Horowitz and Strominger \cite{HorStr}, and explicit solutions for two--charge black holes were constructed in \cite{2301Pre,2301}. We will extend this discussion to branes and demonstrate that two--charged geometries give the most general static solutions. 

In this subsection we will focus on a simplified version of the action (\ref{ActionPsiGamBrn})\footnote{Normalization of the kinetic term for $\psi$ is inspired by the normalization of the dilaton in string theory.}:
\bea\label{ActForSingBrn}
S=\int d^dx\sqrt{-g}\left[R-\frac{4}{d-2}(\d\psi)^2
-\frac{e^{\gamma\psi}}{2(p+2)!}(F_{(p+2)})^2
\right]\,.
\eea
The most general ansatz for a single stack of black $p$--branes is given by
\bea\label{AnsatzQ}
ds^2&=&-e^{2A}dt^2+e^{2B}dy_{p}^2+e^{2R}dr^2+C d\Omega_n^2,\nn
F_{(p+2)}&=&q e^{-\gamma\psi}\star d\Omega_n\,,
\eea
where all warp factors and the scalar field $\psi$ are functions of the radial coordinate. The equations of motion for the tensor field, as well as its Bianchi identities, are automatically satisfied. Note that the ansatz (\ref{AnsatzQ}) still has a freedom in reparameterizing the radial coordinate, and a convenient gauge choice will be made below. The most general geometry (\ref{AnsatzQ}) that satisfies the equations of motion coming from (\ref{ActForSingBrn}) is derived in the Appendix \ref{AppGenBrn}, and here we just summarize the steps:
\begin{enumerate}[1.]
\item We begin with setting $q=0$. In this case, the solution (\ref{BraneStackFlat}) reduces to the Schwarzschild geometry, but one can find more general configurations with a nontrivial scalar field. As demonstrated in the Appendix \ref{AppGenBrn}, the most general solution of the system (\ref{ActForSingBrn}) that has the form  (\ref{AnsatzQ}) with $q=0$ is\footnote{In the Appendix \ref{SecAppGebSubNeutr} we construct a more general configuration (\ref{GenrSolnAnsw2}) that has five parameters $(a_1,a_2,\mu,w,c)$, but two of them can be eliminated by redefining the radial coordinate. The details are discussed at the end of section \ref{SecAppGebSubNeutr}.}
\bea\label{GenrSolnAnswMain}
ds^2&=&\left[-z^{2a_1}dt^2+z^{2a_2}dy_{p}^2\right]+
c^2e^{-2F}\left[\frac{(n-1)^2}{4Q^2}d\Omega_n^2+
\frac{dz^2}{z^2Q^{2n}}\right],\quad e^\phi=z^{c_1}\,,\nn
F&=&\frac{a_1+pa_2}{n-1}\ln z,\quad
Q=\left[z-z^{-1}\right]^{\frac{1}{n-1}}\,.
\eea
This geometry has three free parameters, $(a_1,a_2,c)$, and the constant $c_1$ is determined by solving the constraint
\bea\label{GenrSolnAnswMainCnst}
c_1^2=-\frac{d-2}{4n_-}\left[n_-(a_1^2+p a_2^2)+(a_1+pa_2)^2-n\right]\,.
\eea 

\item In the presence of charge, the equations of motion (\ref{GenrSolnSyst1}) are rather complicated, so instead of deriving the most general solution from the first principles, we looked at a perturbative expansion in powers of $q$. Then, as demonstrated in the Appendix \ref{SecAppGebSubChrg}, up to a possible $q$--dependent change of coordinates, there is a 
{\it unique solution} order--by--order in $q$. For example, when $(p,n)=(1,3)$, the first few terms in the expansion are given by (\ref{GenSolnQpert3})--(\ref{GenSolnQpert4}):
\bea\label{GenrSolnPertMain}
B&=&- (a+2 b) \ln z-\frac{2}{\gamma^2+4} \ln\left[1-\frac{(\gamma^2+4) 
z^{-(8 b+ \gamma c_1) }w q^2}{2c^2}\right]+\mathcal{O}(q^3),\nn 
R&=&-B-(a+2b)\ln z+\frac{1}{2}\ln\left[\frac{4cx^{4b+1}}{(z^2-1)^3}\right]+\mathcal{O}(q^3),\\ 
A&=&2a\ln z + B+\mathcal{O}(q^3),\quad {C}= (z^2-1)^2e^{2R}\,,\nn
\phi&=&c_1 \ln z +c_2-\gamma\left[B+(a+2 b) \ln z\right]+\mathcal{O}(q^3).\nonumber
\eea
The parameter $w$ is defined by (\ref{GenSolnQpert2}). Note that the relation between $C$ and $R$ is exact, and it is a consequence of the gauge choice which has already been made in (\ref{GenrSolnAnswMain}). Expansions similar to (\ref{GenrSolnPertMain}) can be obtained for all values of $(p,n)$, and the absence of $q$ dependence in the expressions for $(A,R,C,\phi)$ in terms of $B$ persists to all orders in $q$. Furthermore, the charge appears in the expression for $B$ only through an expansion of a simple logarithm.

\item Motivated by the expressions (\ref{GenrSolnPertMain}) and their counterparts for arbitrary values of $(p,n)$, we consider an ansatz
\bea\label{GenrSolnAnzFinMain}
&&ds^2=-e^{2A}dt^2+e^{2B}dy_{p}^2+e^{2R}dz^2+C d\Omega_n^2,\quad
F_{(p+2)}=q e^{-\gamma\psi}\star d\Omega_n\,,
\nn
&&A=a_1\ln z+\nu_A \ln H,\quad B=a_2\ln z+\nu_B \ln H,\quad C=c^2H^{\nu_C}\frac{(n_-)^2}{4Q^2}z^{a_3}\,,\nn
&& 
e^{2R}=C\frac{4}{(n_-)^2z^2Q^{2(n-1)}}\,,\quad \phi=c_1\ln z+c_2+\nu_\phi \ln H,\\
&&
Q=\left[z-z^{-1}\right]^{\frac{1}{n-1}},\quad H=1+\frac{u\,q^2}{z^\beta}\,,\quad a_3=-\frac{2(a_1+pa_2)}{n-1}\,.
\nonumber
\eea
The perturbative analysis outlined above guarantees that the {\it most general} solution of the 
system (\ref{ActForSingBrn})--(\ref{AnsatzQ}) must have this form with some specific values of the parameters $(\beta,\nu_A,\nu_B,\nu_C,\nu_\phi,u)$. An explicit substitution of (\ref{GenrSolnAnzFinMain}) into equations of motion gives
\bea\label{GenrSolnAnzFinParMain}
&&\nu_A=\nu_B=\frac{\sigma}{2},\quad \nu_C\equiv \nu=-\frac{p+1}{n-1}\sigma\,,\quad 
\nu_\phi=-\frac{\gamma(n+p)^2}{16n_-}\sigma\,,\nn
&&\sigma=-\frac{2(n-1)}{n_-(p+1)+\frac{1}{16}(n+p)^2\gamma^2}\,,\quad 
\beta=\gamma c_1-2(a_1+p a_2)\,,\\
&&u=\frac{c}{\beta^2}\left[\frac{4}{c(n_-)^2}\right]^n\left[\frac{n_-(p+1)}{2(n+p)}+
\frac{(n+p)\gamma^2}{32}\right]\,.\nonumber
\eea
Recall that the parameter $c_1$ is determined by solving the constraint (\ref{GenrSolnAnswMainCnst}). 
\end{enumerate}
To summarize, we have shown that the substitution of the most general ansatz (\ref{AnsatzQ}) consistent with symmetries of a single stack of $p$--branes into the equation of motion coming from the action (\ref{ActForSingBrn}) leads to the solution 
(\ref{GenrSolnAnzFinMain})--(\ref{GenrSolnAnzFinParMain}) specified by four parameters 
$(a_1,a_2,c,q)$. Let us now recover the standard brane geometry (\ref{BraneStackFlat}) as a special case of the solution (\ref{GenrSolnAnzFinMain})--(\ref{GenrSolnAnzFinParMain}).

\bigskip

\noindent
{\bf Recovery of the standard brane solution} 

To recover the metric  (\ref{BraneStackFlat}) from a more general case (\ref{GenrSolnAnzFinMain})--(\ref{GenrSolnAnzFinParMain}), we observe that geometry (\ref{BraneStackFlat}) implies certain relations between the warp factors in (\ref{AnsatzQ}):
\bea\label{GenSolnMapMain1}
Ce^{-\frac{2\nu}{\sigma}B}=\frac{c^2r^2}{\rho_0^2}\,\quad 
e^{2A-2B}=\rho^2_1\left[1-\frac{2m}{r^{n-1}}\right]\,\quad 
e^{R-\frac{\nu}{\sigma}B+(A-B)}=\frac{c\rho_1}{\rho_0}\frac{dr}{dz}\,.
\eea
Here $r$ is the radial coordinate used in (\ref{AnsatzQ}), and constants $(\rho_0,\rho_1)$ account for potential rescalings of various coordinates. The implications of equations (\ref{GenSolnMapMain1}) are analyzed in the Appendix \ref{AppGenBrnLim}, and they lead to the relations between $z$ and the radial coordinate $r$:
\bea\label{GenSolnMapMain2}
 r=\frac{(n-1)\rho_0}{2[1-z^2]^{1/n_-}}\quad\Rightarrow\quad z=\left[1-\left[\frac{n_-\rho_0}{2r}\right]^{n_-}\right]^{\frac{1}{2}}\,,\quad m=\frac{1}{2}\left[\frac{n_-\rho_0}{2}\right]^{n-1}\,.
\eea
Furthermore, matching the warp factor $B$ and the scalar $\phi$ between  (\ref{BraneStackFlat}) and (\ref{GenrSolnAnzFinMain})--(\ref{GenrSolnAnzFinParMain}), we find
\bea\label{GenSolnMapMain3}
a_1=1-\frac{2n_-}{n_-(p+1)+\frac{1}{16}(n+p)^2\gamma^2},\quad a_2=a_1-1\,.
\eea
Therefore, two of the four parameters $(a_1,a_2,c,q)$ become fixed. Interestingly, this significantly simplifies the expressions for $c_1$ and $\beta$, which are determined from (\ref{GenrSolnAnzFinParMain}) and (\ref{GenrSolnAnswMainCnst}):
\bea
c_1=-\frac{\gamma(n+p)^2}{8n_-}a_2,\quad \beta=2\,.
\eea
Imposing the relations (\ref{GenSolnMapMain2}), (\ref{GenSolnMapMain3}), one can rewrite the solution (\ref{GenrSolnAnzFinMain})--(\ref{GenrSolnAnzFinParMain}) as
\bea\label{GenSolnAnswMain1}
ds^2&=&e^{2B}\left[-\rho_1^2 fdt^2+dy_p^2\right]+
e^{\frac{2\nu}{\sigma}B}\frac{c^2}{\rho_0^2}\left[\frac{dr^2}{f}+r^2 d\Omega_n^2\right],\nn
\phi&=&-\frac{\gamma(n+p)^2}{8n_-}B,\quad f=1-\frac{2m}{r^{n-1}}\,,\quad
m=\frac{1}{2}\left[\frac{n_-\rho_0}{2}\right]^{n-1}\,,\\
&&\nu=-\frac{p+1}{n-1}\sigma\,,\quad \sigma=-\frac{2(n-1)}{n_-(p+1)+\frac{1}{16}(n+p)^2\gamma^2}\,.\nonumber
\eea
The last ingredient, function $B$, is given by (\ref{GenrSolnBfin}):
\bea\label{GenSolnAnswMain2}
B
&=&\frac{a_1-1}{2}\ln\left[1-\frac{2m}{r^{n-1}}-
\frac{2^{2n}(n+p)q^2}{8(n_-)^{2n}c^{2n-2}}\left\{\frac{p+1}{(n+p)^2}+\frac{\gamma^2}{16}\right\}\right]\nn
&\equiv&
\frac{a_1-1}{2}\ln\left[1-\frac{\la}{r^{n-1}}\right]+B_0.
\eea
Note that parameters $\la$ and $B_0$ are fully determined in terms of $(m,q,c)$. Therefore, the solution (\ref{GenSolnAnswMain1})--(\ref{GenSolnAnswMain2}) is specified by four parameters 
$(\rho_0,\rho_1,c,q)$, but as expected, only two of them are physical. Indeed, by rescaling time and $y$ coordinates and by shifting $B$, we can set $\rho_1=1$ and $c=\rho_0$. Then the solution (\ref{GenSolnAnswMain1})--(\ref{GenSolnAnswMain2}) reproduces the answer (\ref{BraneStackFlat}) with some complicated expression for $s_I$ in terms of physical parameters $(q,c)$.

\bigskip
\noindent
{\bf Geometries in terms of the standard radial coordinate.}

To get some intuition about geometries (\ref{GenrSolnAnzFinMain})--(\ref{GenrSolnAnzFinParMain}) and to extract their physical properties, it might be useful for go from variable $z$ to a more standard radial coordinate. Such transition is suggested by the special case (\ref{GenSolnMapMain2}), so in general we write
\bea\label{GenrSolnZasR}
z=\left[1-\left[\frac{r_0}{r}\right]^{n_-}\right]^{\frac{1}{2}}\,.
\eea
With this change of variables, the solution (\ref{GenrSolnAnzFinMain})--(\ref{GenrSolnAnzFinParMain}) becomes 
\bea\label{GenrSolnRadial}
&&ds^2=H^{\sigma}\left[-h^{a_1}dt^2+h^{a_2}dy_{p}^2\right]+
\frac{(cn_-)^2}{4}H^{\nu}h^{-\frac{(a_1+pa_2)}{n-1}}\left[\frac{dr^2}{r^2h^2Q^{2n}}+\frac{1}{Q^2}d\Omega_n^2\right],\nn
&&F_{(p+2)}=q e^{-\gamma\psi}\star d\Omega_n\,,\quad
\phi=\frac{c_1}{2}\ln h+c_2+\nu_\phi \ln H\,,
\\
&&h=1-\left[\frac{r_0}{r}\right]^{n_-},\quad
Q=\left[h+h^{-1}-2\right]^{\frac{1}{2(n-1)}},\quad  H=1+\frac{u\,q^2}{h^{\beta/2}}\,.
\nonumber
\eea
Parameters $(\nu,\sigma,\nu_\phi,\beta,u)$ are still given by (\ref{GenrSolnAnzFinParMain}). If the constraints (\ref{GenSolnMapMain3}) are imposed, then the geometry (\ref{GenrSolnRadial}) reduces to the well--known case of (\ref{BraneStackFlat}) with a single charge, and $r$ matches the standard radial coordinate. Metric (\ref{GenrSolnRadial}) has a horizon at $r=r_0$, and this point is singular unless the sphere and $y$ components of the metric saturate to constants. This happens when
\bea\label{GenSolnRegConstr}
-\frac{\beta\sigma}{2}+a_2=0,\quad 
-\frac{\nu\beta}{2}-\frac{(a_1+pa_2)}{n-1}+\frac{1}{n-1}=0.
\eea
This eliminates two free parameters, leading to the relations (\ref{GenSolnMapMain3}) obtained earlier and to the standard black branes with correlated tensor and scalar charges. This conclusion agrees with the general expectation from \cite{HorStr,HorStrBH}. However, in the 
$(n,p)=(2,0)$ case, even singular geometries that do not satisfy the constraint (\ref{GenSolnRegConstr}) have some interesting applications \cite{2301Pre,2301}, and our construction generalizes these solutions to all values of $n$ and $p$. Let us now recover the solutions of \cite{2301Pre,2301} from our geometry (\ref{GenrSolnAnzFinMain})--(\ref{GenrSolnAnzFinParMain}).

\bigskip

\noindent
{\bf Reduction to a black hole solution with three charges.}

Solution (\ref{GenrSolnAnzFinMain})--(\ref{GenrSolnAnzFinParMain}) describes a geometry of 
$p$--branes that depends on four parameters, $(a_1,a_2,c,q)$. A closely related solution describing black holes in four dimensions was recently constructed in \cite{2301}, and now we will recover these four--dimensional geometries as special cases of our solution.

The solution constructed in \cite{2301} is specified by three parameters 
$(r_+,r_-,{\bar q})$, and can be written as\footnote{We introduced a convenient notation to rewrite the geometry (16)--(17) from \cite{2301} in a compact form. We also used ${\bar q}$ to label the parameter $q$ introduced in \cite{2301} to avoid confusions with charge $q$ appearing in  (\ref{GenrSolnAnzFinMain})--(\ref{GenrSolnAnzFinParMain}).}:
\bea\label{GenSoln2301Main}
ds_E^2&=&-e^{F+2\la\phi}dt^2+e^{-F-2\la\phi}\left[\frac{\mu^4}{(\qb rQ)^4}e^{-2F}dr^2+
\frac{\mu^2}{(\qb Q)^2}d\Omega_2^2\right]\nn
\phi&=&-\frac{\la}{2(1+\la^2)}F\pm
\frac{\sqrt{1-\qb^2+\la^2}}{2\qb(1+\la^2)}G,\quad 
F=\ln\left[\frac{\rho_+\rho_-}{r^2}\right],\quad G=\ln\frac{\rho_+}{\rho_-}\\
Q&=&x^{\frac{1}{2\qb}}-x^{-\frac{1}{2\qb}}\,,\quad x=\frac{\rho_+}{\rho_-}\,,\quad
\rho_\pm=r-r_\pm,\quad \mu=r_+-r_-\,.\nonumber
\eea
Here $\la$ characterizes the coupling to the scalar, which is analogous to $\gamma$ in (\ref{ActForSingBrn}). This solution can be mapped into (\ref{GenrSolnAnzFinMain})--(\ref{GenrSolnAnzFinParMain}) with $(n,p)=(2,0)$ by changing variables as
\bea
z=x^{\frac{1}{2\qb}}\,\quad t\rightarrow \frac{r_-}{\mu}t\,,
\eea
and by imposing several relations between parameters:
\bea\label{GenSoln3chMapMain}
&&\beta=2\qb,\quad c=\frac{2\mu}{\qb},\quad 
q=\left[\frac{16\mu^2 r_+}{(4+\gamma^2)r_-}\right]^{\frac{1}{2}},\quad 
\la=\frac{\gamma}{2}\,,\nn
&&a_1=\frac{-4\beta\pm 2\sqrt{4\gamma^2-(\beta\gamma)^2+\gamma^4}}{\gamma^2+4}\,.
\eea
Detailed derivation of this map is presented in the Appendix \ref{AppGenBrnLim}. Note that while the general solution  (\ref{GenrSolnAnzFinMain})--(\ref{GenrSolnAnzFinParMain}) is specified by four parameters, $(a_1,a_2,c,q)$, for black holes the power $a_2$ becomes irrelevant, so the solutions are characterized by $(a_1,c,q)$, which are mapped to $(r_+,r_-,{\bar q})$ by the relations (\ref{GenSoln3chMapMain}). Furthermore, even for black holes, our solution (\ref{GenrSolnAnzFinMain})--(\ref{GenrSolnAnzFinParMain})  with $p=0$ is more general than the geometry (\ref{GenSoln2301Main}) since it works for all dimensions, not only for 
$(d,n)=(4,2)$. For such general black holes, the solution (\ref{GenrSolnAnzFinMain})--(\ref{GenrSolnAnzFinParMain})  can be rewritten as
\bea\label{GenrSolnAnzFinMainBH}
&&ds^2=-\left[z^{a}H\right]^{\sigma}dt^2+c^2\left[z^{a}H\right]^{\nu}
\left[\frac{dz^2}{z^2 Q^{2n}}+\frac{(n_-)^2}{4Q^2} d\Omega_n^2\right],\quad
F_{(p+2)}=q e^{-\gamma\psi}\star d\Omega_n\,,
\nn
&&\phi=c_1\ln z-\frac{(n+1)^2\gamma\sigma}{16n_-}\ln H\,,\quad
H=1+\frac{u\,q^2}{z^\beta}\,,\quad Q=\left[z-z^{-1}\right]^{\frac{1}{n-1}}\,,\\
&&\nu=-\frac{\sigma}{n-1}\,,\quad \sigma=-\frac{n-1}{n_-+\frac{1}{32}(n+1)^2\gamma^2}\,,
\quad
u=\frac{c}{\beta^2}\left[\frac{4}{c(n_-)^2}\right]^n\frac{(n_-)^2}{(n+1)\sigma}\,,\nn
&&c_1=\left[\frac{n^2}{4n_-}(1-\sigma^2 a^2)\right]^{\frac{1}{2}}\,,\quad
\beta=\gamma c_1-2\sigma a\,.\nonumber
\eea
As in the brane case, the standard radial coordinate can be defined by (\ref{GenrSolnZasR}).

\bigskip

To summarize, in this subsection we found the most general geometry (\ref{GenrSolnAnzFinMain})--(\ref{GenrSolnAnzFinParMain}) describing a single stack of $p$ branes with one charge and recovered several known solutions as special cases. By repeating this procedure for branes with multiple charges, one can find a generalization of the geometry (\ref{BraneStackFlat}), which is characterized by $k+1$ new parameters $(a_1,c_1,\dots c_k)$ in addition to $(m,s_1,\dots s_k)$, which are already present in (\ref{BraneStackFlat}). The new parameters will already appear at the level of the neutral solutions, the counterpart of (\ref{GenrSolnAnswMain}), and then they are {\it uniquely} dressed with charges. The final answer of this construction is rather complicated, so we will not present it here. We have also recovered several known geometries as special cases of the new general solution (\ref{GenrSolnAnzFinMain})--(\ref{GenrSolnAnzFinParMain}).

\section{Charged branes with cosmological constant}
\label{SecBrnLam}

In section \ref{SecBH} we constructed multi-charged black holes with nontrivial potentials for scalar fields, while in section \ref{SecBrn} we constructed multi--charged black branes without such potential. This brings up a natural question: can one find brane counterparts of the solution from section \ref{SecBH}? In this section we briefly review some previous work suggesting that existence of clean analytical expressions for such brane solutions is unlikely and discuss some interesting properties of the differential 
equations governing these brane configuration. 

\bigskip

Let us first recall some properties of the black hole solutions (\ref{BHsolnFinalGam}). In general, these geometries are supported by $k$ gauge fields and $k'$ scalars with a potential given by (\ref{Vgamma}). If all charges are set to zero, then the potential reduces to a cosmological term, and the solution (\ref{BHsolnFinalGam}) reproduces the well-known Schwarzschild--AdS geometry. If a brane counterpart of the solution (\ref{BHsolnFinalGam}) is found, then by setting all charges to zero, one should recover a black brane solution with AdS$_d$ asymptotics in a system that contains only the Einstein--Hilbert term and the cosmological constant. This system has been analyzed in the past, and unfortunately the solutions are known only numerically or though various asymptotic expansions \cite{CopHor,MannRad,DeLuBrn}. Clearly addition of charges will not improve this situation. However, there one can extract several important lessons from the previous work, so let us summarize some properties of the neutral black branes solving the Einstein's equations with a cosmological constant:
\begin{enumerate}[(a)]
\item Symmetries of a single stack of $p$--branes ensure that the metric must have the form
\bea\label{AbelAnstz}
ds^2=-e^{2A}dt^2+e^{2B}(dy_1^2+\dots+dy_p^2)+e^{2R}dr^2+r^2d\Omega_n^2\,,
\eea
where we imposed a gauge by defining the radial coordinate through the coefficient in front of the sphere. As elsewhere in this article, functions $(A,B,R)$ depend on the radial coordinate.
\item The requirement of AdS$_d$ asymptotics imposes conditions on the leading behavior of functions $(A,B,R)$ at large $r$:
\bea\label{AbelLargeR}
\mbox{large}\ r:\quad R\sim -\ln r,\quad A\sim B\sim \ln r
\eea
The first subleading corrections (the constant terms in $(A,B,R)$) are uniquely fixed by the equations of motion, but the perturbative expansions near infinity have one free parameter that accompanies $r^{-n-1}$ in $(A,B,R)$.
\item To ensure that the solution (\ref{AbelAnstz}) has a regular horizon, one must require function $e^{2A}$ to vanish at some $r=r_h>0$, where the sphere still has finite size:
\bea\label{AbelRhor}
\mbox{horizon}:\quad A\sim \frac{1}{2}\ln (r-r_h),\quad B\sim 1
\eea
The boundary conditions (\ref{AbelLargeR}), (\ref{AbelRhor}) lead to the {\it unique} solution for every value of $r_h$. This was shown numerically for black string ($p=1$) in \cite{CopHor,MannRad} and analytically for all values of $(p,n)$ in \cite{DeLuBrn}. Unfortunately the analytical proof of \cite{DeLuBrn} still does not yield an explicit solution beyond various matching expansions. 
\item In the absence of the cosmological constant, the condition (\ref{AbelRhor}) is satisfied by a neutral version of the black brane solution (\ref{BraneStackFlat}). In that case, as the horizon radius decreases to zero, one goes to the flat space that has $A=B=0$. In the presence of the cosmological constant, such solution is clearly inconsistent with the asymptotics (\ref{AbelLargeR}), but one can impose a weaker condition 
\bea\label{AbelRhorAB}
A=B
\eea 
instead. As demonstrated in \cite{DeLuBrn}, this condition indeed leads to the {\it unique} regular horizon--free solution with AdS$_d$ asymptotics, and existence of such regular solution for $(p,n)=(1,2)$ was advocated already in \cite{CopHor} using numerical integration.  While the system describing interpolation between (\ref{AbelLargeR}) and (\ref{AbelRhor}) is rather complicated, article \cite{DeLuBrn} showed that interpolation between (\ref{AbelLargeR}) and (\ref{AbelRhorAB}) for the regular solution is governed by a relatively simple Abel equation. 
\end{enumerate}
Let us now discuss modifications of these four items for the charged case. 

\bigskip

To describe charged black branes, we again impose the ansatz (\ref{AbelAnstz}) and add $k$ gauge fields and $k'$ scalars:
\bea
F^{(I)}_{(p+2)}&=&q_I e^{-\gamma\psi_I}\star d\Omega_n,\quad \psi_I\,.
\eea
At infinity, the boundary conditions are still given by (\ref{AbelLargeR}) with an additional requirement of vanishing scalars: $\psi_I(\infty)=0$. Regularity of the geometry at the horizon still implies (\ref{AbelRhor}), but now there is an additional requirement for all $\psi_I$ to remain finite. With these modifications of (\ref{AbelLargeR}) and (\ref{AbelRhor}), one can repeat the arguments of \cite{DeLuBrn} based on various expansions to prove existence and uniqueness of the solution specified by $r_h$ and a set of charges $q_I$. We will not present the arguments here, and we refer the reader to \cite{DeLuBrn} for details\footnote{Note that now we have $k'$ additional functions $\psi_I$ and $k'$ equations for them, so the construction of \cite{DeLuBrn} guarantees that there are no free parameters beyond $(r_h,q_I)$.}. Instead we will look for the charged counterparts of the regular solutions, which are amenable to analytical treatment. 

To describe the charged counterpart of the regular solution, we impose the restriction (\ref{AbelRhorAB}) in the ansatz (\ref{AbelAnstz}). While the analysis can be made for any number of charges, to avoid unnecessary clutter, we will focus on one field strength without scalars:
\bea\label{AbelAnzFF}
F_{(p+2)}&=&q\star d\Omega_n
\eea
Substituting the metric (\ref{AbelAnstz}) with $A=B$ in the Einstein's equations, we observe that function $A$ does not appear without derivatives. This suggests a change of variables:
\bea
A=\frac{1}{2}\int \frac{h}{r}dr\,.
\eea
Then combining the Einstein's equations, one finds a simple algebraic expression for $e^{2R}$:
\bea
e^{-2R}=\frac{1}{r{\dot h}}\left[(g r)^2(n+p+1)(2-h)-n_-h\right]+
\frac{q^2}{2r^{2n-1}(n+p){\dot h}}[2n_-+(1+p)h]
\eea
The remaining equations are equivalent to one first--order Abel equation \cite{Abel} for function $h$:
\bea\label{AbelEqn}
f {\dot h}+f_1 h+f_2 h^2+f_3 h^3+f_0=0.
\eea
Here functions $(f,f_0,f_1,f_2,f_3)$ are given by
\bea\label{AbelCoeff}
f&=&4(n+p)r[2nn_- -\rho^2+2Ng^2 r^{2}],\quad 
f_0=-8nn_-[n_-\rho^2+2Ng^2r^{2}],\nn
f_1&=&4nn_-\left[2n_-(n+p)-3(p+1)\rho^2\right]+P_1g^2r^{2}\,, \nn
f_2&=&8nn_-(p+1)(n+p)-2(p+1)(2n-p+3pn)\rho^2+P_2g^2r^{2}\,, \\
f_3&=&p(p+1)[2n_-(n+p)-(p+1)\rho^2+2N(gr)^2]\,.\nonumber
\eea
Here $\rho=q r^{1-n}$, $N=(n+p)(n+p+1)$, and $(P_1,P_2)$ are some complicated constants which are not important for the following discussion. The solution describing the extremal brane with $g=0$ corresponds to a simple solution of the Abel's equation,
\bea\label{AbelNHsoln}
h=-\frac{q(n-1)}{(p+1)(q+r^{n-1})}\,,
\eea
and this function approaches a finite, charge--independent limit when $r$ goes to zero: 
\bea\label{AbelNHsolnBC}
h(0)= -\frac{n-1}{p+1}.
\eea 
At infinity, function $h$ should approach $2$ to recover the AdS asymptotics, and existence and uniqueness of the flow between two fixed points of the Abel equation was proven in \cite{DeLuBrn}. We refer to that paper for details.

So far we have looked at equations with arbitrary parameters $(g,q)$, but for special values of the charge and $n=2,3$, some explicit solutions of the system (\ref{AbelAnstz}), (\ref{AbelAnzFF}) were constructed in \cite{LiuSabra}, extending some earlier work \cite{LiuSabraPre}. Here we quote only the expression for two warp factors which look especially simple,
\bea\label{LiuSabra}
n=2,3:\quad e^{2A}=e^{2B}=(gr)^2\left[1+\frac{2(p+1)}{(p+2)(p+2n-3)(gr)^2}\right]^{\frac{p+n}{p+1}}\,,
\eea
and we refer to \cite{LiuSabra} for details, such as the results for $e^{2R}$ and relations between $q$ and $g$. It would be interesting to generalize the solutions (\ref{LiuSabra}) to larger values of $n$. 

\bigskip

To summarize, in this section we have analyzed solutions describing charged 	$p$--branes in the presence of the cosmological constant or, more generally, nontrivial potentials for scalar fields, such as ones discussed in section \ref{SecBH}. Although unfortunately explicit analytical solutions seem to be beyond reach, we argued that the various expansions used in \cite{DeLuBrn} to demonstrate existence and uniqueness of neutral geometries with regular horizons extend to the charged case as well. In the latter case, such solutions are fully specified by the mass and the set of charges. In the neutral case, there was also a unique regular solution that had the symmetries of the brane ansatz and AdS$_d$ asymptotics, and properties of this geometry were governed by a single first order ODE of the Abel type. In this section we found the extension of the Abel equation to the charged case, (\ref{AbelEqn})--(\ref{AbelCoeff}), whose solutions interpolate between the near horizon limit of the extremal branes (\ref{AbelNHsoln}) and the AdS$_d$ space at infinity. The boundary condition (\ref{AbelNHsolnBC}) uniquely specifies the solution.

\section{Discussion}

In this article we explored geometries produced by multi--charged black holes and black branes with and without the cosmological constant. Getting inspiration from known solutions of gauged supergravities, we began with an ansatz for black holes charged under vector and scalar fields, and we found the unique potential for the scalars that can lead to solutions of this type. Our potential generalizes the results from all known gauged supergravities to an arbitrary number of scalar fields in an arbitrary number of dimensions. We also constructed geometries sourced by branes with an arbitrary number of charges and analyzed their properties, including their relations with higher dimensional wormholes.

\bigskip

Let us briefly summarize our results. Although asymptotically--AdS black holes have been extensively studied in the framework of gauged supergravity, it is interesting to see whether such geometries can be constructed beyond specific models. We addressed this question in section \ref{SecBH}, where we proposed ansatze for static black holes with an arbitrary number of gauge fields and derived the {\it unique} scalar potential that may support such solutions. The result represented by equation (\ref{Vgamma}) reproduces the answers for all known gauged supergravities reviewed in the Appendix \ref{AppGauSgr}. Our potential is fully specified by the number of vector fields, their couplings to scalars, and one cosmological parameter. Once the potential was determined, we also found the full gravitational solutions (\ref{BHsolnFinalGam}), which have regular horizons and AdS asymptotics.

In section \ref{SecBrn} we extended the multi--charged solutions from black holes to branes and their intersections, and in the absence of the cosmological parameter, the final results are given by equations (\ref{BraneStackFlat}) and (\ref{AnswBrnMixd}). These answers generalize all known brane intersections from string and M theory that follow the harmonic rules, but in contrast to those cases, where the relative co--dimensions of branes are always divisible by four, expressions (\ref{AnswBrnMixd}) do not have this restriction. As we showed in section \ref{SecSubInter}, the restriction on co--dimension originates from very specific string theoretic couplings of the dilaton with fluxes, and once these couplings are relaxed, one can have more general intersections. Similarly, all string theoretic branes are fully specified by their charges under gauge fields, and scalars are fully determined by these parameters. In section \ref{SecSubBrnGen} we extended this construction to uncorrelated gauge and scalar charges and constructed the most general geometry (\ref{GenrSolnAnzFinMain})--(\ref{GenrSolnAnzFinParMain}) consistent with symmetries of the branes. As expected from the general analysis of \cite{HorStrBH,HorStr}, additional uncorrelated charges lead to singular horizons, but in the case of four dimensional dilatonic black holes such singular solutions have found some interesting applications \cite{2301Pre,2301}, and we extended these most general local solutions to $p$ branes in arbitrary dimensions. 

In the presence of the cosmological parameter, the analytical solutions for branes seem to be beyond reach, but
in sections \ref{SecSubNH} and \ref{SecBrnLam} we analyzed some properties of the relevant geometries, including their existence and uniqueness, using various expansions and arguments inspired by renormalization group flows developed in \cite{DeLuBrn,DeLuWrm}. In particular, in section \ref{SecBrnLam} we reduced the problem of constructing the extremal solution to one first order Abel differential equation (\ref{AbelEqn}), and in section \ref{SecSubNH} we performed some expansions of brane solutions in the near horizon limit and closely related wormhole geometries. 

\bigskip

In this article we focused on static black holes and branes, and it would be very interesting to extend these results to the stationary case. Apart from the usual technical problems associated with rotation, our construction encounters a conceptual one. In the static case, the only relevant parts of the action of gauged supergravity are the kinetic term and the scalar potential, and we managed to generalize them to arbitrary numbers of fields and dimensions. In the rotating case, the Chern--Simons couplings start playing important roles, and they are highly model--specific. Furthermore, some of these couplings depend on the cosmological parameter. This complication can be avoided if only one gauge field is excited, and the relevant rotating solutions in all dimensions were obtained in \cite{WuAllDim}. It would be interesting to extend this work to multiple fields by generalizing at least some classes of gauged supergravity actions to arbitrary dimensions. In this article we already constructed the potential relevant for such cases, but finding the Chern--Simons couplings and explicit solutions is an interesting open problem. 

\section*{Acknowledgements}

This work was supported in part by the DOE grant DE-SC0015535.

\appendix

\section{Static black holes in gauged supergravities}
\label{AppGauSgr}

In this appendix we briefly summarize some well--known solutions of gauged supergravities, which motivate our discussion of static black holes with a cosmological parameter presented in section \ref{SecBH}. In every case we begin with the full action of gauged supergravity, and then discuss its truncation for known static configurations and quote the black hole solution. We also quote the scalar potentials which are reproduced by our general answer (\ref{Vgamma}). 

\bigskip
\noindent
{\bf Four--dimensional gauged supergravity}

We begin with the four--dimensional gauged supergravity introduced in \cite{GSR4w1,GSR4w2}. The full action is given by\footnote{In this appendix we denote the square root of the determinant of the metric by $e$ since this is common in the literature on gauged supergravity. Symbol $g$ is reserved for the cosmological parameter.}
\bea
S&=&\int ed^4 x\left[R - \frac{1}{2}(\d {\vec\phi})^2 - \frac{1}{2}
e^{-{\vec a}\cdot{\vec\phi}}(\d \chi)^2 - 
\frac{1}{2}e^{{{\vec a}_{12}}\cdot{\vec\phi}}(F_{(1)12})^2 
-  \frac{1}{2}e^{{{\vec b}_{12}}\cdot{\vec\phi}}({\cal F}^1_{(1)2})^2-g^2V\right. \nn
&&\left.- \frac{1}{4}\sum_{i=1}^{2}\left\{e^{{\vec a}_i\cdot{\vec\phi}}(F_{(2)i})^2+
e^{{\vec b}_i\cdot{\vec\phi}}({\cal F}^i_{(2)})^2\right\} -\frac{1}{2}\chi \epsilon^{\mu \nu \rho \sigma}F_{\mu \nu i}{\cal F}^i_{\rho \sigma}\right].
\eea
We use notation of \cite{10auth}, and we refer to this article for details, such as expressions for vectors $({\vec a}_i,{\vec b}_i)$. Static black holes have been constructed for a truncated version of this action,
\bea\label{4dGSact}
S=\int ed^4 x\left[R-\frac{1}{2}(\d{\vec\phi})^2+8{\bar g}^2\sum_{k=1}^3 \cosh\phi_k-
\frac{1}{4}\sum_{i=1}^4 e^{{\vec a}_I\cdot{\vec \phi}}(F_{(I)})^2\right]
\eea
where
\bea
{\vec a}_1=(1,1,1),\quad {\vec a}_2=(1,-1,-1),\quad {\vec a}_3=(-1,1,-1),\quad 
{\vec a}_4=(-1,-1,1).\quad 
\eea
The spherically-symmetric black hole solution for this action was found in \cite{GSR4slnW1,GSR4slnW2}, and it reads
\bea\label{4dGSsoln}
ds^2&=&-H^{-1/2}f dt^2+H^{1/2}\left[\frac{dr^2}{f}+r^2 d\Omega^2\right],\nn
X_i&=&\frac{H^{1/4}}{H_I},\quad A^{(I)}=\frac{\mu s_Ic_I}{H_I r}dt,\\
f&=&1-\frac{\mu}{r}+4({\bar g} r)^2 H,\quad H=H_1H_2H_3H_4,\quad 
H_I=1+\frac{\mu s_I^2}{r}\,.\nonumber
\eea
The scalars $(X_1,X_2,X_3,X_4)$ are constrained by the relation
\bea
X_1X_2X_3X_4=1,
\eea
and the potential for them is given by
\bea\label{4dGSpot}
V=-8{\bar g}^2\sum_{j=1}^3 \cosh\phi_j=-4{\bar g}^2\sum_{I<J}X_IX_J.
\eea
This solution is reproduced as a special case of our general result (\ref{BHsolnFinalGam}), (\ref{Vgamma}) with ${\bar g}=g/2$. Rotating extensions of the solution (\ref{4dGSsoln}) have been constructed in \cite{GSR4snRt}.

\bigskip
\noindent
{\bf Five--dimensional gauged supergravity}

The action of the five--dimensional gauged supergravity is given by \cite{GSR5w1,GSR5w2}
\bea\label{5dGSact}
S=\int ed^5 x\left[ R - \frac{1}{2}(\d \bar{\phi})^2  - 
\sum_{I=1}^{3} \frac{1}{4X_I^2} (F_I)^2
+4 g^2 \sum_{I=1}^{3} X_I^{-1}-\frac{1}{6}\epsilon_{IJK}A^I F^J F^K\right]\,.
\eea
Truncation to static ansatz eliminates the contribution from the Chern-Simons term, and the black hole solution was found in \cite{GSR5sln}:
\bea\label{5dGSsoln}
ds^2&=&-f H^{-2/3}+H^{1/3}\left[\frac{dr^2}{f}+r^2d\Omega_3^2\right]\,,\nn
X_I&=&\frac{H^{1/3}}{H_I},\quad A^{(I)}=\frac{\mu s_Ic_I}{H_I r^2}dt,\\
f&=&1-\frac{\mu}{r^2}+({g} r)^2 H,\quad H=H_1H_2H_3,\quad 
H_I=1+\frac{\mu s_I^2}{r^2}\,.\nonumber
\eea
Once again, the scalars are constrained by the relation
\bea
X_1X_2X_3=1,
\eea
and the potential for them is given by
\bea\label{5dGSpot}
g^2V=-4 g^2\left[\frac{1}{X_1}+\frac{1}{X_2}+\frac{1}{X_3}\right]\,.
\eea
This solution is reproduced as a special case of our general result (\ref{BHsolnFinalGam}), (\ref{Vgamma}). Rotating extensions of the solution (\ref{5dGSsoln}) were constructed in \cite{GSR5snRotW1,GSR5snRotW2,GSR5snRotW3,GSR5snRotW4,GSR5snRotW5,Wu5d}.

\bigskip
\noindent
{\bf Six--dimensional gauged supergravity}

The action of the six--dimensional gauged supergravity is given by \cite{GSR6}
\bea\label{6dGSact}
S&=&\int \left[\star R -  \frac{1}{2}\star d\phi \wedge d \phi - \frac{1}{X^2}(\star F_{(2)} \wedge F_{(2)}+\star F_{(2)}^I \wedge F_{(2)}^I) - \frac{1}{2} X^4 \star F_{(3)}\wedge F_{(3)}\right.
\nn 
&&+ g^2\star\left(9 X^2 +\frac{12}{X^2} - \frac{1}{X^6}\right)\\ \nonumber
&&\left.-A_{(2)}\wedge(\frac{1}{2}d A_{(1)} \wedge d A_{(1)}+\frac{g}{\sqrt{2}}A_{(2)} \wedge d A_{(1)}+\frac{g^2}{3}A_{(2)}\wedge A_{(2)}+\frac{1}{2}F_{(2)}^I \wedge \frac{1}{2}F_{(2)}^I)\right]
\eea
In \cite{GSR6sln} this action is truncated to two gauge fields, but for static solutions, one of them is set to zero. This results in 
\bea
S&=&\int \left[\star R -  \frac{1}{2}\star d\phi \wedge d \phi - \frac{1}{X^2}\star F_{(2)} \wedge F_{(2)} + 
g^2\star\left(9 X^2 +\frac{12}{X^2} - \frac{1}{X^6}\right)\right].
\eea
Solution is given by:
\bea\label{6dGSsoln}
ds^2&=&-H^{-3/2} f dt^2 + H^{1/2}(f^{-1}dr^2+r^2 d\Omega_4^2)\\ 
A&=&\sqrt{2}\frac{\mu s_1c_1}{H_I r^3}dt,\quad
\phi=\frac{1}{\sqrt{2}} \ln H,\quad
f= 1- \frac{\mu}{r^3}+g^2 r^2 H^2 \quad H=1+\frac{\mu s_1^2}{r^3}\,,\nonumber
\eea
and the potential for the scalar is
\bea\label{6dGSpot}
V=-g^2\left[9 X^2 +\frac{12}{X^2} - \frac{1}{X^6}\right]
\eea
This solution is reproduced as a special case of our general result (\ref{BHsolnFinalGam}), (\ref{Vgamma}). The rotating version of the solution (\ref{6dGSsoln}) was constructed in \cite{GSR6snRot}.

\bigskip
\noindent
{\bf Seven--dimensional gauged supergravity}

The action of the seven--dimensional gauged supergravity is given by \cite{GSR7w1,GSR7w2}\footnote{We use notation of \cite{Wu7d}.}
\bea\label{7dGSact}
S&=&\int \left[\star R - \star d\phi_1 \wedge d \phi_1 - 5\star d\phi_2 \wedge d \phi_2 - \frac{1}{2X_1^2}F_1\wedge \star F_1- \frac{1}{2X_2^2}F_2\wedge \star F_2\right.
\nn 
&&+2 g^2 \star \left[8X_1 X_2+4 X_1^{-1}X_2^{-2}+4 X_1^{-2}X_2^{-1}-4(X_1 X_2)^{-4}\right]\\ \nonumber
&&\left. - \frac{1}{2}(X_1 X_2)^2 *\textit{F}\wedge \textit{F}+(F_1 \wedge F_1 - g \textit{F})\wedge \textit{C}\,\right].
\eea
In the static case, one can truncate the action by dropping the last line, and the black hole solution for the resulting system was found in \cite{CvetGub,10auth}. It reads
\bea\label{7dGSsoln}
ds^2&=&-f H^{-4/5}+H^{1/5}\left[\frac{dr^2}{f}+r^2d\Omega_3^2\right]\nn
X_I&=&\frac{H^{2/5}}{H_I},\quad A^{(I)}=\frac{\mu s_Ic_I}{H_I r^4}dt,\\
f&=&1-\frac{\mu}{r^4}+({g} r)^2 H,\quad H=H_1H_2,\quad 
H_I=1+\frac{\mu s_I^2}{r^2}\,.\nonumber
\eea
The potential for the scalars is given by 
\bea\label{7dGSpot}
g^2V=-2 g^2 \left[8X_1 X_2+\frac{4}{X_1X_2^{2}}+\frac{4}{X_1^{2}X_2}-\frac{4}{(X_1 X_2)^{4}}\right]
\eea
This solution is reproduced as a special case of our general result (\ref{BHsolnFinalGam}), (\ref{Vgamma}). Rotating versions of the solution (\ref{7dGSsoln}) was constructed in \cite{GSR7snRot,Wu7d}.

\bigskip
\noindent
{\bf One--charge solution in all dimensions}

Article \cite{WuAllDim} considered an interesting system that extends gauged supergravities to all dimensions while keeping only one gauge fields. The action was postulated to be\footnote{To fit this expression in one line, we defined $\zeta=(d-1)(d-2)$.}
\bea\label{WUdGSact}
S  = \int ed^d x \left[R - \frac{\zeta}{4}(\d \phi)^2 - \frac{1}{4}e^{-(d-1)\phi}F^2+g^2(d-1)\left[(d-3)e^{\phi}+e^{-(d-3)\phi}\right]\right],
\eea
and a charged version of the Kerr-AdS solution in arbitrary dimensions \cite{PopeAdSw1,PopeAdSw2} was constructed. The static version of the solution is given by \cite{WuAllDim}:
\bea\label{WUdGSsoln}
ds^2&=&H^{\frac{1}{d-2}}\left[-(1+g^2r^2)dt^2+\frac{dr^2}{1+g^2r^2-h}+r^2d\Omega_{d-2}^2+\frac{2mc^2}{r^{d-3}H}dt^2\right]\\
A&=&\frac{2mcs}{Hr^{d-3}}dt,\quad 
h=\frac{2m}{r^{d-3}}\left[1-s^2g^2 r^2\right] \quad H=1+\frac{2ms^2}{r^{d-3}}\nonumber
\eea
and the potential is\footnote{In this case we write the potential in terms of $H$ rather than $X$ since it is easier to verify (\ref{VgammaAsH}) than (\ref{Vgamma}).} 
\bea\label{WUdGSpot}
V=-g^2(d-1)H^{-\frac{1}{d-2}}\left[d-3+H\right]
\eea
This solution is reproduced as a special case of our general result (\ref{BHsolnFinalGam}), (\ref{Vgamma}) with $k=1$ and $\nu=\frac{1}{d-2}$. Article \cite{WuAllDim} also constructed the rotating version of the solution (\ref{WUdGSsoln}) with $\left[\frac{d-1}{2}\right]$ angular momenta.

\section{Equations of motion for multi--charged black holes}
\label{AppEOM}

In this appendix we verify that the geometry (\ref{AnstzAlpha}) satisfies equations of motion coming from the action (\ref{ActionPsiGam}) with the potential given by (\ref{Vgamma}). We begin with Einstein's equations, then go to equations for the gauge potentials $A^I$,  and conclude with equations for the scalars.

\bigskip
\noindent
{\bf Einstein's equations}

Let us impose the ansatz (\ref{AnstzAlpha}) and use the Einstein's equations coming from the action (\ref{ActionPsiGam}) to determine the values of $\mu$ and five parameters $\nu_k$. To simplify the formulas below, we first assume that  $k-1$ charges are equal ($s_2=\dots=s_{k}$), and once the parameters $(\mu,\nu_j)$ are determined, the equations will be verified for arbitrary charges. 

\bigskip

We begin with quoting Einstein's equations for the system
\bea
ds^2&=&-e^{2A}dt^2+e^{2B}dr^2+C d\Omega_n^2\,,\\
A^I&=&\frac{q_I}{r^{d-3}H_I}dt,\quad e^{\mu\psi_I}=\frac{H^{\nu_4}}{H_I},\quad 
\psi=\sum_I \psi_I\,.
\nonumber
\eea
Introducing a tensor
\bea
E_{\mu\nu}=R_{\mu\nu}-w_1 \d_\mu \psi\d_\nu\psi-w_2(\sum\d_\mu\psi_I\d_\nu\psi_I)+\sum_I 
\frac{e^{\gamma\psi_I}}{2}\left[F^I_{\mu\alpha}F_\nu^{I\alpha}-\frac{1}{2(d-2)}g_{\mu\nu}F^I_{\alpha\beta}
F_I^{\alpha\beta}\right],\nonumber
\eea
we find its three independent components:
\bea\label{Eexpress}
E_{rr}&=&{\dot A}{\dot B}-{\dot A}^2-{\ddot A}-\frac{n{\ddot C}}{2C}+\frac{n{\dot B}{\dot C}}{2C}+\frac{n{\dot C}^2}{4C^2}-w_1{\dot\psi}^2-w_2\sum {\dot\psi}_I^2+\frac{(n_-)^3e^{2B}}{2nC^n}\sum q_I^2 e^{-\gamma \psi_I}\,,\nn
E_{\theta\theta}&=&n_-+\frac{e^{-2B}}{2}\left[({\dot B}-{\dot A}){\dot C}-\frac{(n-2){\dot C}^2}{2C}-{\ddot C}\right]-\frac{(n_-)^2}{2nC^{n-1}}\sum q_I^2 e^{-\gamma \psi_I}\,,\\
E_{tt}&=&e^{2A-2B}\left[{\ddot A}+{\dot A}^2-{\dot A}{\dot B}+
\frac{n{\dot A}{\dot C}}{2C}\right]-\frac{(n_-)^3e^{2A}}{2nC^{n}}\sum q_I^2 e^{-\gamma \psi_I}\,.\nonumber
\eea 
The ansatz (\ref{AnstzAlpha}) with $p=(k-1)$ equal charges gives
\bea\label{eAfactApp}
e^{2A}=f H^{\nu_1},\quad e^{2B}=\frac{H^{\nu_2}}{f},\quad C=H^{\nu_3}r^2,\quad
\psi_I=\frac{1}{\mu}\ln\frac{H^{\nu_4}}{H_I},\quad H=H_1 H_2^{p}\,.
\eea
Recall that 
\bea
H_I=1+\frac{2Ms_I^2}{r^{d-3}},\quad 
f=1+(gr)^2H^{\nu_5}-\frac{2M}{r^{d-3}}\,.
\eea
We will proceed in several steps:
\begin{enumerate}[1.]
\item Since the potential $V$ is proportional to $g^2$, the Einstein's equations for $g=0$ give $E_{\mu\nu}=0$. In the items 1--5 we will focus on this $g=0$ case. Substituting the factors (\ref{eAfactApp}) into (\ref{Eexpress}) and expanding the result to the first order in $M$, we find
\bea
E_{\theta\theta}=\frac{n_-M}{r^{n-1}}\left[\nu_1+\nu_2+(n-2)\nu_3\right]
(s_1^2+p\,s_2^2)+\mathcal{O}(M^2).
\eea
This leads to our first relation between parameters $\nu_k$:
\bea
\nu_2=-\left[\nu_1+(n-2)\nu_3\right].
\eea
We will use it to eliminate $\nu_2$ in all expressions below.
\item Expanding 
$E_{\theta\theta}$ to the third order in $M$ and requiring the result to vanish, we 
arrive at two equations (one at order $M^2$ and one at order $M^3$), which can be solved for $q_1$ and $q_2$. Substituting the results into (\ref{Eexpress}), we find
\bea
E_{tt}=-\frac{4M^3s_2^4}{r^{3n-1}}(p n_-)^2[\nu_1+n_-\nu_3]\left[\nu_1+n_-\nu_3-\frac{1}{p}\right]\left[1+\mathcal{O}(s_3^2)+\mathcal{O}(s_1^2)\right]+\mathcal{O}(M^4).\nonumber
\eea
Since the powers $\nu_j$ can't depends on the parameters $s_j$ specifying the charges, we arrive at two options:
\bea\label{Nu1Options}
\nu_1=-n_-\nu_3+\frac{1}{p}\quad \mbox{or}\quad \nu_1=-n_-\nu_3\,.
\eea
\item Let us begin with the first option in (\ref{Nu1Options}). Recalling the expression for the charges $q_1$ and $q_2$ from step 2 and setting $s_1=0$ in the results, we find
\bea
\left.q_1^2\right|_{s_1=0}&=&-\frac{np}{n_-\gamma}[pn_-\nu_3-2]
\left[n_-\mu\nu_3+\gamma\nu_4-\frac{\gamma}{p}\right]s_2^2\,,\nn
\left.q_2^2\right|_{s_1=0}&=&\frac{n}{n_-\gamma}[pn_-\nu_3-2]
\left[n_-\mu\nu_3+\gamma\nu_4\right]s_2^2\,.
\eea
On the general grounds, we expect to have $q^2_2\propto s^2_2(1+s_2^2)$, which does not match the dependence above. Even ignoring this observation and proceeding with equations 
$E_{\mu\nu}=0$ beyond the leading orders in $M$, one arrives at a logical inconsistency, but we will skip unnecessary details. The outcome of this analysis is inconsistency of the first option in (\ref{Nu1Options}).

\item Let us now consider the second branch in (\ref{Nu1Options}):
\bea
\nu_1=-n_-\nu_3\,.
\eea
Recalling the expressions for $q_1$ and $q_2$ from step 2 and setting $s_1=0$ or $s_2=0$ in the result, we find
\bea\label{q12Split}
\left.q_1^2\right|_{s_1=0}&=&-\frac{np\nu_3}{\gamma}
\left[-2\mu-\gamma +pn_-\mu\nu_3+p\gamma\nu_4\right]s_2^2(1+s_2^2),\nn
\left.q_2^2\right|_{s_2=0}&=&-\frac{n\nu_3}{\gamma p}
\left[-2\mu+n_-\mu\nu_3+\gamma(\nu_4-1)\right]s_1^2(1+s_1^2).
\eea
Since $q_I$ should depend only on $s_I$, the expressions above must vanish.
Setting $\nu_3=0$ is not a good option since it leads to $q_1=q_2=0$. Setting the square brackets to zero, we find\footnote{For $k=2$, both square brackets in equations (\ref{q12Split}) are the same, so vanishing of the relations (\ref{q12Split}) determines only $\nu_4$. In this case, the value of $\mu$ is determined by the requirement for $s_2$ to be absent in $q_1^2$ beyond the limit  (\ref{q12Split}).}
\bea
\mu=-\frac{\gamma}{2},\quad \nu_4=\frac{n_-}{2}\nu_3\,.
\eea 
With these conditions, we arrive at the final expressions for the charges:
\bea
q_1^2=4n\nu_3 s_1^2(1+s_1^2)M^2,\quad q_2^2=4n\nu_3 s_2^2(1+s_2^2)M^2.
\eea
At this point, equations $E_{\theta\theta}=0$ and $E_{tt}=0$ are satisfied.
\item Expanding the equation $E_{rr}=0$ in powers of $M$, we find two constraints on parameters 
$(w_1,w_2)$, and once these constraints are imposed, all Einstein's equations are satisfied. Collecting expressions for all relevant parameters, we find
\bea\label{nuMost}
&&\nu_1=-n_-\nu_3,\quad \nu_2=\nu_3,\quad \mu=-\frac{\gamma}{2},\quad \nu_4=\frac{n_-}{2}\nu_3,\nn 
&&w_1=-\frac{nn_-(\nu_3\gamma)^2}{8[kn_-\nu_3-2]},\quad 
w_2=\frac{n\nu_3\gamma^2}{8}\,.
\eea
\item Once the cosmological constant is turned on, the equations of motion coming from the action (\ref{ActionPsiGam}) are
\bea\label{EmnV}
E_{\mu\nu}=\frac{1}{d-2}V g_{\mu\nu}\,.
\eea
In particular, this implies that 
\bea
\frac{E_{\theta\theta}}{E_{rr}}=\frac{g_{\theta\theta}}{g_{rr}}\,.\nonumber
\eea
An explicit calculation shows that this relation is satisfied if and only if
\bea\label{nu5}
\nu_5=n\nu_3\,.
\eea
\item Once the relations (\ref{nuMost}) and (\ref{nu5}) are imposed, an explicit evaluation of both sides of equation (\ref{EmnV}) gives
\bea\label{VpotApp}
V=g^2H^{n_-\nu}n(n_-)^2\left[\frac{1}{n_-}-
2{n}\left[\frac{1}{n_-}+\frac{\nu}{2}\sum\left\{\frac{1}{H_I}-1\right\}\right]^2+
\frac{\nu}{2}\sum\left[\frac{1}{H_I^2}-1\right]\right]\,.
\eea
Here $\nu=\nu_3$. Although we outlined the derivation of  (\ref{nuMost}), (\ref{nu5}), and (\ref{VpotApp}) assuming that $k-1$ charges are equal, direct calculations show that the final results work for arbitrary charges $q_I$. 
\end{enumerate}
\noindent
To summarize, we have demonstrated that the parameters in the ansatz (\ref{eAfactApp}) are given by
\bea\label{AlphMu1App}
&&\mu=-\frac{\gamma}{2},\quad \nu_1=-(n-1)\nu,\quad \nu_2=\nu_3=\nu,\quad
\nu_4=-\frac{\nu_1}{2},\quad \nu_5=n\nu\,,\nn
&&w_1=-\frac{\gamma^2n(n-1)\nu^2}{8[kn_-\nu-2]},\quad 
w_2=\frac{1}{8}n\gamma^2\nu,\quad q_I=2\sqrt{n\nu}Ms_Ic_I\,.
\eea
Although $\nu$ appears as a free parameter in these expressions, it can be removed by rescaling the scalar fields. In particular, this freedom can be used to set $w_2=\frac{1}{2}$, leading to 
\bea\label{NuFixedApp}
\nu=\frac{4}{n\gamma^2}\,.
\eea
We will use the condition (\ref{NuFixedApp}) in the main body of the paper, but in this appendix we will treat  $\nu$ as a free parameter.

The potential (\ref{VpotApp}) was written in terms of harmonic functions $H_I$, and it terms of scalar fields it becomes
\bea\label{VgammaApp}
V&=&g^2 n(n_-)^2 Y^2 \left[\frac{1}{n_-}- \frac{k \nu}{2}-{2}n\left(\frac{1}{n_-}
-\frac{k\nu}{2}+ 
\frac{\nu S_{\gamma/2}}{2Y}\right)^2 +\frac{\nu S_\gamma}{2 Y^2}\right],
\\
S_\alpha&=&\sum_{I=1}^k X_I^{\alpha},\quad Y=\prod_J^k (X_J)^z,\quad z = \frac{n_-\gamma \nu}{2 k n_-\nu-4}\,.\nonumber
\eea
We will now verify that once the relations (\ref{AlphMu1App}) and (\ref{VgammaApp}) are imposed, equations for the vector and scalar fields are satisfied.

\bigskip
\noindent
{\bf{Equations for vector fields}}

The equations for the vector fields read
\bea\label{MaxwellEqn}
d\left[\star e^{\gamma \psi_I}dA^I\right]=0,
\eea
and for the ansatz (\ref{AnstzAlpha}) the left hand side becomes
\bea
&&\frac{1}{\sqrt{-g}}\d_r\left[(r^{d-2}H^{\nu(d-n)/2})(H_I^2H^{-(n-1)\nu}) H^{\nu(n-2)}\d_r A^I_t\right]=
\frac{1}{\sqrt{-g}}\d_r\left[r^{d-2}(H_I)^2\d_r \frac{q_I}{r^{d-3}H_I}\right]\nn
&&\qquad=-\frac{q_I}{\sqrt{-g}}\d_r\left[r^{4-d}\d_r (r^{d-3}H_I)\right]=
-\frac{q_I}{\sqrt{-g}}\d_r\left[r^{4-d}\d_r (r^{d-3})\right]\,.
\eea
This expression vanishes, confirming the equation (\ref{MaxwellEqn}). 

\bigskip
\noindent
{\bf{Equations for scalar fields}}

Equation of motion for the scalar field $\psi_I$ is
\bea\label{DilEqnApp}
2 \left[w_{2 \gamma} \nabla^2 \psi_I + w_{1 \gamma} \nabla^2 \psi\right] = \frac{\gamma}{2} e^{\gamma \psi_I} F_I^2 + \frac{\d V_{\gamma}}{\d \psi_I}\,.
\eea
Explicit evaluation of the left hand side gives
\bea\label{DilEqnApp1}
&&2 \left[w_{2 \gamma} \nabla^2 \psi_I + w_{1 \gamma} \nabla^2 \psi\right]=
\frac{2}{\gamma}\left[2w_{2 \gamma} \nabla^2 \ln H_I + w_{2\gamma}[-n_-\nu+(2-kn_-\nu)\frac{w_{1 \gamma}}{w_{2\gamma}}] \nabla^2\ln H\right]\nn
&&\quad=\frac{4w_{2\gamma}}{\gamma}\nabla^2 \ln H_I=\frac{n\gamma\nu}{2}\frac{1}{\sqrt{-g_d}}\d_r\left[r^{d-2}f\frac{\d_r H_I}{H_I}\right]\nn
&&\quad=\frac{n\gamma\nu}{2\sqrt{-g_d}}\left[-\frac{(2Ms_Ic_I(d-3))^2}{H_I^2r^{d-2}}+
g^2\d_r\left[r^{d}H^{n\nu}\frac{\d_r H_I}{H_I}\right]\right]\,.
\eea
Recall that 
\bea
\sqrt{-g_d}=r^{d-2}H^{\nu}.
\eea
The $g$--independent part of (\ref{DilEqnApp1}) matches the contribution from the two--form field strength to the right hand side of (\ref{DilEqnApp}):
\bea\label{DilEqnApp2}
\frac{\gamma e^{\gamma\psi_I}}{2}(F^I)^2&=&-\frac{\gamma}{2}
H^{\nu(n-2)}\frac{H_I^2}{H^{(n-1)\nu}}
\left[\d_r\frac{q_I}{r^{d-3} H_I}\right]^2=
-\frac{\gamma(q_I)^2}{2H^{\nu} r^{4(d-3)}H_I^2}[\d_r(r^{d-3} H_I)]^2
\nn
&=&-\frac{\gamma(q_I(d-3))^2}{2H_I^2 H^\nu r^{2(d-2)}}=
-\frac{\gamma n\nu(2Ms_Ic_I(d-3))^2}{2H_I^2 H^\nu r^{2(d-2)}}\,.
\nonumber
\eea
Therefore we conclude that
\bea\label{ScalEqnInter}
&&2 \left[w_{2 \gamma} \nabla^2 \psi_I + w_{1 \gamma} \nabla^2 \psi\right]- \frac{\gamma}{2} e^{\gamma \psi_I} F_I^2=\frac{ng^2\gamma\nu}{2\sqrt{-g_d}}
\d_r\left[r^{d}H^{n\nu}\frac{\d_r H_I}{H_I}\right]\nn
&&\quad=\frac{n(n_- g)^2\gamma\nu r^{d-2}h_I H^{n\nu}}{2\sqrt{-g_d}}\left[
n\nu \sum_J h_J+\frac{d-1}{n_-}-\frac{1}{H_I}\right]\,.
\eea
Here we defined
\bea
h_J=\frac{1}{H_J}-1\,.
\eea
The derivative of the potential (\ref{Vgamma}) is
\bea
\frac{\d V}{\d X_I}= -\frac{g^2\gamma \nu}{2 X_I}n(n_-)^2 (X_I^{\frac{\gamma}{2}}-Y)
\left[-X_I^{\frac{\gamma}{2}}+\frac{d-1}{d-3}Y 
+(d-2)\nu (\sum X_J^{\frac{\gamma}{2}} - k Y)\right]\,.
\eea
Expressing $X_J$ in terms of $H_J$, we conclude that the right hand side of (\ref{DilEqnApp2}) is equal to $\d V_{\gamma}/\d X_I$. This completes verification of the scalar equation (\ref{DilEqnApp}). 

\bigskip
\noindent
{\bf Equations for scalar fields in special cases} 

Equations (\ref{DilEqnApp}) were derived from the action (\ref{ActionPsiGam}) assuming that all scalar fields $\psi_I$ are independent. However, this assumption fails for some values of $k$. Specifically, recalling the expression for $\psi_I$, we find
\bea
e^{\mu\psi_I}=\frac{H^{\nu_4}}{H_I},\quad \Rightarrow\quad \psi= \frac{k\nu_4-1}{\mu}\ln H\,.
\eea
The scalar fields are constrained if $k\nu_4=1$, then (\ref{AlphMu1App}) implies that
\bea\label{NuConstrApp}
\nu=\frac{2}{kn_-}\,.
\eea
If the parameter $\nu$ is fixed by the relation (\ref{NuFixedApp}), then the last constraint can be written as a condition on the number of vector fields:
\bea
k = \frac{ (d-2) \gamma^2}{2(d-3)}\,.
\eea
Only $k-1$ scalars are independent. Variation of the action (\ref{ActionPsiGam}) with respect to these scalars gives a counterpart of equations (\ref{ScalEqnInter})
\bea\label{ScalEqnInterSpec}
2w_{2 \gamma} \nabla^2 \psi_I - \frac{\gamma}{2} e^{\gamma \psi_I} F_I^2
=\frac{n(n_- g)^2\gamma\nu r^{d-2}h_I H^{n\nu}}{2\sqrt{-g_d}}\left[
n\nu \sum_J h_J+\frac{d-1}{n_-}-\frac{1}{H_I}\right]+\la\,.
\eea
Here $\la$ is a Lagrange multiplier enforcing the constraint $\sum \psi_I=0$. 

With the condition (\ref{NuConstrApp}), the potential (\ref{VpotApp}) simplifies to
\bea\label{VpotSpecApp}
V&=&g^2H^{2/k}n(n_-)^2\left[-
\frac{2n}{(kn_-)^2}\left[\sum\frac{1}{H_I}\right]^2+
\frac{1}{kn_-}\sum\frac{1}{H_I^2}\right]\\
&=&-\frac{g^2}{k^2}n\left[4n\sum_{I\ne J}(X_IX_J)^{\gamma/2}+(2n-kn_-)\sum_I X_I^\gamma\right].\nonumber
\eea
Recall that
\bea
X_I=e^{-\psi_I}=\left[\frac{H^{\nu_4}}{H_I}\right]^{2/\gamma}=\left[\frac{H^{1/k}}{H_I}\right]^{2/\gamma}
\eea
once the constraint (\ref{NuConstrApp}) is taken into account. Derivative of the potential (\ref{VpotSpecApp}) is
\bea
\frac{\d V}{\d \psi_I}&=&\frac{\gamma g^2}{k^2} n \left[2n\sum_{J\ne I}(X_IX_J)^{\gamma/2}+(2n-kn_-) X_I^\gamma\right]+{\tilde \la}\nn
&=&\frac{\gamma g^2}{k^2}\frac{H^{2/k}}{H_I}\left[\sum_{J}\frac{2n}{H_J}-\frac{kn_-}{H_I}\right]+{\tilde \la}\,,\nonumber
\eea
where ${\tilde\la}$ is a Lagrange multiplier enforcing the constraint $\sum \psi_I=0$. Direct calculations show that equations of motion for all scalar fields are satisfied.

\section{Derivation of the most general brane solution}
\label{AppGenBrn}

In this appendix we will analyze equations of motion coming from the action (\ref{ActForSingBrn}) and construct the most general solution for the ansatz (\ref{AnsatzQ}) consistent with symmetries of branes. The construction will go in two stages: in section \ref{SecAppGebSubNeutr} we will focus on solutions with vanishing field strength, and in section \ref{SecAppGebSubChrg} we will add the charge. Once the most general solution is constructed, in section \ref{AppGenBrnLim} we will recover some special cases which have appeared in the literature.

\bigskip

We begin with imposing the ansatz (\ref{AnsatzQ}) 
\bea\label{AnsatzQApp}
ds^2&=&-e^{2A}dt^2+e^{2B}dy_{p}^2+e^{2R}dr^2+C d\Omega_n^2,\nn
F_{(p+2)}&=&q e^{-\gamma\psi}\star d\Omega_n\,,
\eea
and writing the equations of motion coming from the action (\ref{ActForSingBrn})
\bea\label{GenrSolnSyst1}
&&\frac{n{\dot C}^2}{4C^2}+\frac{n}{2C}\left[{\dot R}{\dot C}-{\ddot C}\right]+
{\dot R}{\dot E}-\ddot E-({\dot A}^2+p{\dot B}^2)-\frac{4{\dot\phi}^2}{d-2}+\frac{n_-q^2}{2(n+p)}\frac{e^{2R-\gamma\phi}}{C^n}=0,\nn
&&{\ddot A}+{\dot A}\left[\frac{n{\dot C}}{2C}-{\dot R}+p{\dot B}+{\dot A}\right]-
\frac{n_-q^2}{2(n+p)}\frac{e^{2R-\gamma\phi}}{C^n}=0,\nn
&&{\ddot B}+{\dot B}\left[\frac{n{\dot C}}{2C}-{\dot R}+p{\dot B}+{\dot A}\right]-
\frac{n_-q^2}{2(n+p)}\frac{e^{2R-\gamma\phi}}{C^n}=0,\\
&&-\frac{(n-2){\dot C}^2}{4C}+\frac{1}{2}
\left[{\dot R}{\dot C}-{\ddot C}-{\dot C}{\dot E}\right]+(n-1)e^{2R}-
\frac{(p+1)q^2}{2(n+p)}\frac{e^{2R-\gamma\phi}}{C^{n-1}}=0,\nn
&&{\ddot\phi}+{\dot\phi}\left[{\dot E}-{\dot R}\right]+\frac{n{\dot C}{\dot\phi}}{2C}+
\frac{\gamma q^2}{16}\frac{n+p}{C^n}e^{2R-\gamma\phi}=0.\nonumber
\eea
Here we defined a convenient combination $E=A+pB$. 

\subsection{The neutral case}
\label{SecAppGebSubNeutr}

Let us first set $q=0$ in the equations above. In the absence of charges, the system (\ref{GenrSolnSyst1}) is invariant under the transformations
\bea\label{GenSolnSymm}
R\rightarrow R+\lambda,\quad C\rightarrow C e^{2\la},
\eea
so it is convenient to go from $(C,R)$ to new functions $(C_p,R_p)$ defined by
\bea
C=e^{2C_p+2R},\quad R=\int R_p dr.
\eea
The symmetry (\ref{GenSolnSymm}) ensures that once the system (\ref{GenrSolnSyst1}) is rewritten in terms $(C_p,R_p)$, there are no remaining integrals. Furthermore, functions $(A,B)$ appear in the system (\ref{GenrSolnSyst1}) only through their derivatives\footnote{This is expected from the ansatz (\ref{AnsatzQ})): constant shifts in $(A,B)$ can be absorbed into rescaling of tine and $y$ coordinates.}, so one can lower the order of the system by writing
\bea
A=\int A_p dr, \quad B=\int B_p dr. 
\eea
Now we recall that the ansatz (\ref{AnsatzQApp}) preserves its form under reparameterizations of the radial direction, and this symmetry can be used to choose a convenient gauge. There is a generic situation, where $A-B$ is not constant and two special cases:
\begin{enumerate}[(a)]
\item If $A-B$ is a constant, but $A$ is not, then one can fix the gauge in (\ref{AnsatzQ}) by setting 
\bea
A=B=\ln r.
\eea
The first two equations in (\ref{GenrSolnSyst1}) become equivalent, and they give
\bea
R_p=-\frac{nr{\dot C}_p-p}{(n-1)r}\,.
\eea
Then the last two equations in  (\ref{GenrSolnSyst1})  decouple, and one arrives at the solution
\bea
&&\phi=\phi_0+\phi_1 \ln r,\quad 
C_p=\ln r+\ln\left[c_1 r^{\zeta}+\frac{c_2}{c_1}r^{-\zeta}\right],\\
&&c_2=-\frac{(n-1)^2}{4\zeta}\quad
\phi_1=\left[\frac{n\zeta^2-(n+p)(p+1)}{n-1}\right]^{\frac{1}{2}}\,.\nonumber
\eea
This solution characterized by one physical parameter $\zeta$ is a degenerate ``extremal'' case of a more general geometry constructed below, so we will not discuss it further. 
\item If both $A$ and $B$ are constants, then one can impose the gauge
\bea
A=B=0,\quad C_p=\ln r\,.
\eea
Then the fourth equation in (\ref{GenrSolnSyst1}) can be integrated to give
\bea
R_p=-\frac{2r}{r^2-c^2 r^{2n}}\,.
\eea
The remaining equations determine the dilaton:
\bea
\phi=\left[\frac{(d-2)n}{n-1}\right]^{\frac{1}{2}}\mbox{arctanh}\left[cr^{n-1}\right]+\phi_0\,.
\eea
We will not discuss this degenerate option further.
\end{enumerate}
Going back to the general case, when $A-B$ is a nontrivial function of the radial coordinate, we 
impose the gauge
\bea
B=A+\ln x\,.
\eea
We denoted the new radial coordinate by $x$ to distinguish it from a more standard gauge choice $C=r^2$. The first four equations in (\ref{GenrSolnSyst1}) can be combined to give a remarkably simple relation
\bea
A_p=-\frac{n-1}{p+1}R_p-\frac{n}{p+1}{\dot C}_p-\frac{p-1}{p+1}\,\frac{1}{r}
\eea
This expression ensures that the last equation in  (\ref{GenrSolnSyst1})  decouples leading to the solution for the scalar field,
\bea
\phi=c_0+c_1 \ln x\,.
\eea
Then two equations corresponding to the time and brane directions become equivalent and lead to another simple relation
\bea
R_p=-\frac{n{\dot C}_p}{n-1}-\frac{c_2}{x}
\eea
This leaves two non--linear second order differential equations for $C_p$, one of which looks deceivingly simple:
\bea\label{SecOrdDecv}
r{\ddot C}_p+{\dot C}_p+r(n-1)^2 e^{-2C_p}=0
\eea
The second equation is more complicated, but it can be combined with (\ref{SecOrdDecv}) to yield
\bea\label{SecOrdDecvFO}
r^2{\dot C}_p^2-2r{\dot C}_p-\frac{m-1}{n(p+1)}\left[(n+p)(n_-c_2-2)c_2+\frac{c_1^2}{2}(p+1)\right]-(n_-)^2r^2 e^{-2C_p}=0.
\eea
Using several changes of variables, one can show that the solution of the last equation has the form
\bea
C_p = c_{p0}+c_{p1}\ln x + \ln \left[x^{\nu}+c_{p2}\right]
\eea
Then substitution into (\ref{SecOrdDecv}) and (\ref{SecOrdDecvFO}) determines the coefficients,
\bea
C_p = a_{0}+\left[1-\frac{\nu}{2}\right]\ln x + \ln \left[x^{\nu}-\frac{(n_-)^2}{\nu^2}e^{-2a_0}\right]
\eea
as well as parameter $c_1$
\bea\label{GenrSolnAnsw1Cons}
c_1^2=\frac{d-2}{4}\left[-c_2[c_2n_-+2]\frac{n+p}{p+1}+\frac{n}{4n_-}(\nu^2-4)\right].
\eea
Collecting all information above, we arrive at the final answer:
\bea\label{GenrSolnAnsw1}
&&A=\frac{1-p-n_-c_2}{p+1}\ln x,\quad B=A+\ln x,\quad \phi=c_0+c_1 \ln x\,,\nn
&&R=\frac{2c_2n_-+n(\nu-2)}{2n_-}\ln x-\frac{n}{n-1}\ln {\tilde Q},\\
&&\ln C=2a_0+\frac{2c_2n_-+(\nu-2)}{n_-}\ln x-\frac{2}{n-1}\ln {\tilde Q},\quad
Q=x^\nu-\frac{(n_-)^2}{\nu^2}e^{-2a_0}\nonumber
\eea
Note that we have chosen convenient integration constants while going from $(A_p,R_p)$ to  $(A,R)$: such constants can be adjusted by coordinate transformations, such as rescaling of the time direction. The solution (\ref{GenrSolnAnsw1}) has four parameters $(c_0,c_1,c_2,a_0)$ which are subject to the constraint (\ref{GenrSolnAnsw1Cons}). The parameter $c_0$ can be eliminated by shifting the scalar field. 

While we have shown that the expressions (\ref{GenrSolnAnsw1}) with the constraint (\ref{GenrSolnAnsw1Cons}) represent the most general neutral solution of the system (\ref{GenrSolnSyst1}), the final answer becomes cleaner in a somewhat different notation. Specifically, we change the radial variable $x\rightarrow a x^b$ to make the expressions for $(A,B)$ more symmetric\footnote{Note that this changes that gauge $B=A+\ln x$, which was used to derive (\ref{GenrSolnAnsw1}).}. Then the geometry described by (\ref{GenrSolnAnsw1}) can be written as
\bea\label{GenrSolnAnsw2}
ds^2&=&\left[-x^{2a_1}dt^2+x^{2a_2}dy_{p}^2\right]+c^2e^{-2F}\left[\frac{(n-1)^2}{4w\mu^2Q^2}d\Omega_n^2+\frac{dx^2}{x^2Q^{2n}}\right],\quad e^\phi=x^{c_1}\nn
F&=&\frac{a_1+pa_2}{n-1}\ln x,\quad
Q=\left[w x^\mu-x^{-\mu}\right]^{\frac{1}{n-1}}
\eea
The parameter $c_1$ is determined by solving the constraint
\bea\label{GenrSolnAnswC1cns}
c_1^2=-\frac{d-2}{4n_-}\left[n_-(a_1^2+p a_2^2)+(a_1+pa_2)^2-n\mu^2\right]
\eea 
The solution (\ref{GenrSolnAnsw2}) has five free parameters, $(a_1,a_2,\mu,w,c)$. By scaling $x$ and appropriate rescaling of time and brane coordinates, we can set $w=1$. Furthermore, if one defines a new coordinate $z=x^\mu$, then the parameter $\mu$ disappears from the solution. This leaves three parameters $(a_1,a_2,c)$.

\subsection{Addition of charge}
\label{SecAppGebSubChrg}

Let us now add the charge to the geometry (\ref{GenrSolnAnsw2}). Unfortunately, the equations are more complicated than in the neutral case, so to find the {\it most general} solution we will proceed in two steps:
\begin{enumerate}
\item Treating the charge as a perturbation, we find the first few corrections to (\ref{GenrSolnAnsw2}) and prove that the most general solution arranges in a certain pattern order--by--order in $q$.
\item Once the pattern is proven, it can be used to impose the most general ansatz consistent with the perturbative expansion, and we will solve the resulting equations of motion.
\end{enumerate}
Let us begin with the first step.

To avoid unnecessary clutter, we will focus on black strings in six dimensions $(p,n)=(1,3)$, although similar expansions with $(p,n)$ dependent coefficients will hold in the general case. Furthermore, we will set $\mu=1$. In this special case, the functions involved in the neutral solution (\ref{GenrSolnAnsw2}) can be rewritten as
\bea\label{GenBrnSpecNeut}
&&A = - (a+2 b) \ln x \quad  B = (a-2 b) \ln x,\quad 
e^{2R} =\frac{4c\, x^{1+4 b}}{(x^2-1)^3} \quad C= \frac{4c\,x^{1+4 b}}{(x^2-1)} \nn
&&\phi = c_1 \ln x+ c_2 \quad a= \frac{1}{2} \sqrt{3 - 32 b^2 - 2 c_1^2} 
\eea
Looking at the metric (\ref{GenrSolnAnsw2}), we observe that its radial and sphere components contain information about 
$(a_1,a_2,p)$, while their ratio does not:
\bea
\frac{g_{\Omega\Omega}}{g_{xx}}=\frac{(n-1)^2x^2}{4w\mu^2}Q^{2(n-1)}
\eea
Therefore, even in the charged case, it might be convenient to impose the gauge
\bea\label{GenBrnSpecGauGen}
e^{-2R}C=\frac{(n-1)^2x^2}{4w\mu^2}Q^{2(n-1)}\,.
\eea
In our special case (\ref{GenBrnSpecNeut}), this reduces to
\bea\label{GenBrnSpecGau}
e^{-2R}C= (x^2-1)^2\,.
\eea
Note that this condition does not fully fix the gauge: while $C$ is a scalar under repameterizations of $x$, function $R$ is not. Constructing the scalar quantity,
\bea
\frac{e^{2R}dx^2}{C}=\frac{dx^2}{(x^2-1)^2}
\eea
we can identify the residual reparameterization symmetry by finding the most general function 
${\bar x}(x)$ that satisfies the relation
\bea
\frac{dx^2}{(x^2-1)^2}=\frac{d{\bar x}^2}{({\bar x}^2-1)^2}\nonumber
\eea
The solution is specified by one constant $u$:
\bea\label{GenBrnSpecGauExtr}
{\bar x}=\frac{x- u + u  x}{1 + u - u  x }
\eea
This freedom will play an important role below. Note that a similar freedom is also present in the general case (\ref{GenBrnSpecGauGen}). 

In the gauge (\ref{GenBrnSpecGau}), the perturbation of the solution (\ref{GenrSolnAnsw2}) by charge is given by
\bea\label{GenBrnSpecPertAnz}
&&A = - (a+2 b) \ln x + q^2 f_{1}\quad  B = (a-2 b) \ln x + q^2 f_{2}\nn
&&e^{2R}= \frac{4c\,x^{1+4 b}}{(x^2-1)^3} \left[1 + q^2 f_{3}\right]^2 \quad 
C=\frac{4c\,x^{1+4 b}}{(x^2-1)} \left[1 + q^2 f_{3}\right]^2 \\ 
&&\phi = c_1 \ln x + c_2+ q^2 f_{4} \quad F_{(p+2)} = q \star d\Omega_n\nonumber
\eea
and functions $(f_1,f_2,f_3,f_4)$ can be expanded in the power series in $q^2$. In the first order in $q^2$, one finds a solution that depends on seven integration constants, but {\it all of them} can be absorbed by diffeomorphisms and shifts of the scalar field:
\begin{enumerate}[1.]
\item Constants $(a,b,c,c_2)$ in (\ref{GenBrnSpecPertAnz}) can be shifted by $q$--dependent terms. Recall that the constant $c_1$ was constrained in the neutral case, and similar constraints persist in all powers in $q$, so $c_1$ is not an independent parameter.

\item Coordinate $t$ and $y$ can be rescaled by $q$--dependent factors

\item Reparameterization (\ref{GenBrnSpecGauExtr}) that preserves the gauge condition cab be used to remove the seventh parameter.  
\end{enumerate}
Once all integration constants are absorbed, one arrives at the {\it unique} solution in the first order in $q^2$: 
\bea\label{GenSolnQpert1}
&&A = (a-2b) \ln x +\frac{q^2w x^{-8 b - \gamma c_1}}{c^2}\,,\quad
B = - (a+2 b) \ln x +\frac{q^2w x^{-8 b - \gamma c_1}}{c^2}\,,\nn
&&R = \frac{1}{2}\ln \left[\frac{4c\,x^{1+4 b}}{(x^2-1)^3}\right] - 
\frac{q^2w x^{-8 b - \gamma c_1}}{c^2}\,, \quad
C=\frac{4c\, x^{1+4 b}}{(x^2-1)} -\frac{8q^2w x^{-8 b - \gamma c_1}}{ 
c (x^2-1)}\,, \\
&&\phi = c_1 \ln x +c_2-  \frac{q^2 \gamma w x^{-8 b - \gamma c_1}}{c^2}\,.\nonumber
\eea
Here we defined 
\bea\label{GenSolnQpert2}
w=\frac{e^{-\gamma p_2}}{64(8 b+ \gamma c_1)^2}\,.
\eea
Expanding the equations to higher powers of $q$, one again funds seven integration constants in each order, so all this freedom can be absorbed into shifting the parameters of the original solutions (\ref{GenBrnSpecPertAnz}) and residual diffeomorphisms. Therefore, one finds the unique solution order--by--order in the charge expansion, and the first few terms suggest very simple relations between various warp factors. For example, at the second order one finds 
\bea\label{GenSolnQpert3}
&&B = - (a+2 b) \ln x + \frac{q^2w x^{-8 b - \gamma c_1}}{ c^2}+
\frac{(\gamma^2+4) x^{-2(8 b+ \gamma c_1) }w^2 q^4}{4c^4}+\mathcal{O}(q^6)\\ 
&&A = 2a\ln x + B+\mathcal{O}(q^6),\quad \phi=c_1 \ln x +c_2-
\gamma\left[B+(a+2 b) \ln x\right]+\mathcal{O}(q^6)\nn 
&&R=-B-(a+2b)\ln x+\frac{1}{2}\ln\left[\frac{4cx^{4b+1}}{(x^2-1)^3}\right],
\quad {C}= (x^2-1)^2e^{2R}\,.\nonumber 
\eea
As can be seen from these expressions and subsequent terms, the charge parameter $q$ enters only through one warp factor $B$, while the expressions of $(A,R,C,\phi)$ in terms of $B$ maintain their unperturbed form. Furthermore, subsequent terms in the expansion of $B$ indicate that a certain expression truncates at the first order in $q$:
\bea\label{GenSolnQpert4}
\left[e^B\right]^{-\frac{\gamma^2+4}{2}}=\left[x^{a+2b}\right]^{\frac{\gamma^2+4}{2}}
\left[1-\frac{(\gamma^2+4) x^{-(8 b+ \gamma c_1) }w q^2}{2c^2}\right].
\eea
The unique expressions similar to (\ref{GenSolnQpert3})--(\ref{GenSolnQpert4}) emerge for all values of $(p,m,\mu)$, and they lead to an ansatz for the {\it most general} solution consistent with symmetries of the problem:
\bea\label{GenrSolnChrgAnzFin}
&&ds^2=-e^{2A}dt^2+e^{2B}dy_{p}^2+e^{2R}dz^2+C d\Omega_n^2,\quad
F_{(p+2)}=q e^{-\gamma\psi}\star d\Omega_n\,,
\nn
&&A=a_1\ln z+\nu_A \ln H,\quad B=a_2\ln z+\nu_B \ln H,\quad C=c^2H^{\nu_C}\frac{(n_-)^2}{4Q^2}z^{a_3}\,,\nn
&& 
e^{2R}=C\frac{4}{(n_-)^2z^2Q^{2(n-1)}}\,,\quad \phi=c_1\ln z+c_2+\nu_\phi \ln H,\\
&&
Q=\left[z-z^{-1}\right]^{\frac{1}{n-1}},\quad H=1+\frac{u\,q^2}{z^\beta}\,,\quad a_3=-\frac{2(a_1+pa_2)}{n-1}\,.
\nonumber
\eea
Parameters $(\beta,\nu_A,\nu_B,\nu_C,\nu_\phi,u)$ are determined by solving equations of motion\footnote{Without loss of generality, we have set $c_2=0$.}:
\bea\label{GenrSolnChrgAnzFinPar}
&&\nu_A=\nu_B=\frac{\sigma}{2},\quad \nu_C=-\frac{p+1}{n-1}\sigma\,,\quad 
\nu_\phi=-\frac{\gamma(n+p)^2}{16n_-}\sigma\,,\nn
&&\sigma=-\frac{2(n-1)}{n_-(p+1)+\frac{1}{16}(n+p)^2\gamma^2}\,,\quad 
\beta=\gamma c_1-2(a_1+p a_2)\,,\\
&&u=\frac{c}{\beta^2}\left[\frac{4}{c(n_-)^2}\right]^n\left[\frac{n_-(p+1)}{2(n+p)}+
\frac{(n+p)\gamma^2}{32}\right]\,.\nonumber
\eea
Recall that the parameter $c_1$ is determined by solving the constraint (\ref{GenrSolnAnswC1cns}) with 
$\mu=1$.

\subsection{Relations to known solutions}
\label{AppGenBrnLim}

In this subsection we will analyze some special cases of the geometry (\ref{GenrSolnChrgAnzFin})--(\ref{GenrSolnChrgAnzFinPar}) that reduce to known solutions. Specifically we will look at three examples: the neutral solution without the dilaton, the geometry (\ref{BraneStackFlat}) describing a single stack of $p$--branes, and the four--dimensional black holes with two charges, which were recently constructed in \cite{2301}. 

\bigskip 
\noindent
{\bf Neutral branes}

We begin with recovering the neutral case of the brane solution (\ref{BraneStackFlat}). This geometry (\ref{BraneStackFlat}) describes the Schwarzschild solution with trivial dilaton and constant warp factor factor $B$, so 
one arrives at a truncated version of the constraint (\ref{GenrSolnAnswC1cns}) by setting $c_1=a_2=0$ in that relation:
\bea
c_1=a_2=0\quad\Rightarrow\quad a_1=1.
\eea 
In this special case, the neutral solution (\ref{GenrSolnAnsw2}) becomes\footnote{As in (\ref{GenrSolnChrgAnzFin}), we set $w=1$ and defined a new coordinate $z=x^\mu$.}
\bea
ds^2&=&\left[-z^{2}dt^2+dy_{p}^2\right]+c^2z^{-2/n_-}\left[\frac{(n_-)^2}{4Q^2}d\Omega_n^2+\frac{dz^2}{z^2Q^{2n}}\right],\\ 
e^\phi&=&1,\ 
Q=\left[z-z^{-1}\right]^{\frac{1}{n-1}}\,.\nonumber
\eea
Defining a new coordinate
\bea\label{GenSolnDefRntr}
r=\frac{cn_-}{2Q}z^{-1/n_-}=\frac{cn_-}{2}\left[z^2-1\right]^{-\frac{1}{n-1}}\,,
\eea
we arrive at the standard Schwarzschild-Tangherlini geometry:
\bea\label{GenrSolnAnswNtrRspec}
ds^2=-h dt^2+dy_{p}^2+\frac{dr^2}{h}+r^2d\Omega_n^2,\quad h=1+\left[\frac{cn_-}{2r}\right]^{n-1}
\eea
The change of variables (\ref{GenSolnDefRntr}) can also be applied to the most general uncharged solution (\ref{GenrSolnAnsw2}), and the result reads
\bea\label{GenrSolnAnswNtrR}
ds^2&=&\left[-h^{a_1}dt^2+h^{a_2}dy_{p}^2\right]+h^{a_3}\left[\frac{dr^2}{f}+r^2d\Omega_n^2\right],\quad 
e^{2\phi}=h^{c_1}\\
h&=&1+\left[\frac{cn_-}{2r}\right]^{n-1}\equiv1-\left[\frac{r_0}{r}\right]^{n-1},\quad a_3=\frac{1-a_1-pa_2}{n-1}\,.\nonumber
\eea
Parameter $c_1$ is determined from the constraint (\ref{GenrSolnAnswC1cns}) with $\mu=1$. The geometry (\ref{GenrSolnAnswNtrR}) is asymptotically--flat, and it is specified by three parameters: $(a_1,a_2,r_0)$. To avoid a curvature singularity at $r=r_0$, one must set $c_1=a_3=0$, leading to a constraint 
\bea
a_2\left[a_2-\frac{2}{p+1}\right]=0.
\eea
The option $a_2=0$ gives back (\ref{GenrSolnAnswNtrRspec}). The second option, $a_2=\frac{2}{p+1}$ leads to a curvature singularity, unless $p=1$, where one gets a double analytic continuation of the solution (\ref{GenrSolnAnswNtrRspec}) with $(a_1,a_2)=(0,1)$. 

\bigskip

In the presence of charge, application of the change of coordinates (\ref{GenSolnDefRntr}) to the geometry (\ref{GenrSolnChrgAnzFin}) gives
\bea\label{GenrSolnAnswRt1}
ds^2&=&H^\sigma\left[-h^{a_1}dt^2+h^{a_2}dy_{p}^2\right]+H^\nu h^{a_3}\left[\frac{dr^2}{h}+r^2d\Omega_n^2\right],\nn
e^{2\phi}&=&h^{c_1}H^{2\nu_\phi}\,\quad
H=1+\frac{u\,q^2}{h^{\beta/2}},\quad h=1+\frac{\la}{r^{n-1}}\,,\\
\nu&=&-\frac{p+1}{n-1}\sigma\,,\quad 
\nu_\phi=-\frac{\gamma(n+p)^2}{16n_-}\sigma\,,\quad a_3=\frac{1-a_1-pa_2}{n-1}\,.
\nonumber
\eea
The expressions for $(\sigma,\beta,u)$ are given by (\ref{GenrSolnChrgAnzFinPar}), and $c_1$ is determined by solving the constraint (\ref{GenrSolnAnswC1cns}) with $\mu=1$. The free parameters in (\ref{GenrSolnAnswRt1}) are 
$(a_1,a_2,\la,q)$. 

\bigskip
\noindent
{\bf A single stack of $p$--branes}

Let us now recover the  geometry (\ref{BraneStackFlat}) produced by a single type of branes from the general solution (\ref{GenrSolnChrgAnzFin})--(\ref{GenrSolnChrgAnzFinPar}). To do so, we first restrict (\ref{BraneStackFlat}) to a single type of branes,
\bea\label{TempBraneStack}
ds^2={\tilde H}^{\sigma}\left[-f dt^2+dy_{(p)}^2\right]+{\tilde H}^{\nu}\left[\frac{dr^2}{f}+r^2d\Omega_{n}^2\right],\quad e^{\gamma\psi}={\tilde H}^{2-\nu n_-},\quad \sigma=-\frac{\nu n_-}{p+1}
\eea
Rewriting this metric in the general form 
\bea
ds^2=-e^{2A}dt^2+e^{2B}dy_{p}^2+e^{2R}dz^2+C d\Omega_n^2\,,
\eea
we extract several relations between the warp factors and the radial coordinate $r$:
\bea
Ce^{-\frac{2\nu}{\sigma}B}=\frac{c^2r^2}{\rho_0^2}\,\quad 
e^{2A-2B}=\rho^2_1\left[1-\frac{2m}{r^{n-1}}\right]\,\quad 
e^{R-\frac{\nu}{\sigma}B+(A-B)}=\frac{c\rho_1}{\rho_0}\frac{dr}{dz}\,.
\eea
Here constants $(\rho_0,\rho_1)$ account for potential rescalings of various coordinates. 
This overdetermined system has solutions if and only if
\bea
\beta=-2(p+1)a_1-\frac{(p+n)^2a_1\gamma^2}{8(n-1)}+c_1\gamma+2p,\quad
2m=\left[\frac{(n-1)\rho_0}{2}\right]^{n-1}\,,
\eea
which implies that
\bea
 r=\frac{(n-1)\rho_0}{2[1-z^2]^{1/n_-}}\quad\Rightarrow\quad z=\left[1-\left[\frac{n_-\rho_0}{2r}\right]^{n_-}\right]^{\frac{1}{2}}\,.
\eea
Not surprisingly, this relation looks similar to (\ref{GenSolnDefRntr}). Requiring the proper form of $B$ and $\phi$, we find
\bea
c_1=\frac{\gamma(n+p)^2}{8n_-}(1-a_1),\quad 
a_1=1-\frac{2n_-}{n_-(p+1)+\frac{1}{16}(n+p)^2\gamma^2}\quad\Rightarrow\quad \beta=2.
\eea
With these expressions, the constraint (\ref{GenrSolnAnswC1cns}) is automatically satisfied and
\bea\label{GenrSolnBfin}
B&=&\frac{a_1-1}{2}\ln\left[1-\frac{2m}{r^{n-1}}-
\frac{2^{2(n-3)}(n+p)\gamma^2q^2}{2(n_-)^{2n}c^{2n-2}}-
\frac{2^{2n-3}(p+1)q^2}{(n+p)(n_-)^{2n}c^{2n-2}}\right]\nn
&=&
\frac{a_1-1}{2}\ln\left[1-\frac{\la}{r^{n-1}}\right]+B_0.
\eea
The first term recovers the expected answer, and $B_0$ can be absorbed into rescaling of the 
$y$ coordinates. The dilaton is
\bea
\phi=-\frac{\gamma(n+p)^2}{8n_-}B\,.
\eea
To summarize, we conclude that the standard solution (\ref{TempBraneStack}) is reproduced once the specific values of the parameters $(a_1,a_2)$ in  (\ref{GenrSolnChrgAnzFin})--(\ref{GenrSolnChrgAnzFinPar})  are chosen:
\bea
a_1=1-\frac{2n_-}{n_-(p+1)+\frac{1}{16}(n+p)^2\gamma^2},\quad a_2=a_1-1\quad
\Rightarrow\quad \beta=2\,.
\eea
The two remaining parameters, $(q,c)$, fully specify the solution, and the standard normalization  of $(t,y)$ coordinates corresponds to $(\rho_0,\rho_1)=(c,1)$. 

\bigskip
\noindent
{\bf Four--dimensional black hole with dilatonic charge.}

Let us now recover the solution  which describes the four--dimensional black hole with a magnetic and dilatonic charge found in \cite{2301}. This geometry, in the coordinates defined in \cite{2301}, is given by (\ref{GenSoln2301Main}).

Let us compare (\ref{GenSoln2301Main}) to 
(\ref{GenrSolnChrgAnzFin})--(\ref{GenrSolnChrgAnzFinPar}) with $(n,p)=(2,0)$:
\bea
&&ds^2=-e^{2A}dt^2+e^{2R}dz^2+C d\Omega_n^2,
\nn
&&A=a_1\ln z+\frac{\sigma}{2} \ln H,\quad C=\frac{c^2}{4Q^2}e^{-2A}\,,\quad 
e^{2R}=\frac{4}{z^2Q^{2}}C,
\nn
&& 
\phi=c_1\ln z+c_2-\frac{\gamma\sigma}{4}\ln H,\quad Q=\left[z-z^{-1}\right],\quad H=1+\frac{u\,q^2}{z^\beta}\,,\\
&&c_1=\sqrt{1-a_1^2}\,,\quad \beta=\gamma c_1-2 a_1\,,\quad
u=\frac{1}{\beta^2 c}\left[4+{\gamma^2}\right]\,.\nonumber
\eea
Comparing three components of the metric, one arrives at the map between coordinates and parameters of two solutions:
\bea
z=x^{\frac{1}{2\qb}}\,,\quad \beta=2\qb,\quad c=\frac{2\mu}{\qb},\quad 
q=\left[\frac{16\mu^2 r_+}{(4+\gamma^2)r_-}\right]^{\frac{1}{2}},\quad 
\la=\frac{\gamma}{2}\,.
\eea
There is also a rescaling of the time coordinate:
\bea
t\rightarrow \frac{r_-}{\mu}t\,.
\eea
Various function are given by
\bea\label{GenSolnTemp1}
&&F=2\ln\left[\frac{z^{\bar q}(r_+-r_-)}{r_+-r_- z^{2\bar q}}\right]\,,\quad G=2{\bar q}\ln z,
\nn
&&\phi=-\frac{\gamma}{\gamma^2+4}F+\frac{4a_1+a_1\gamma^2+8\qb}{2\gamma\qb(\gamma^2+4)}G+2\ln\frac{r_-}{\mu}\,.
\eea
The independent parameters are $(c,a_1,q)$ or $(r_+,r_-,{\bar q})$, and constant $c_1$ is determined by solving the constraint
\bea
(\gamma^2+4)a_1^2+8\beta a_1+4(\beta^2-\gamma^2)=0.
\eea
Alternatively, one can view the constraint as an equation $a_1$, and solving it one finds
\bea
a_1=\frac{-4\beta\pm 2\sqrt{4\gamma^2-(\beta\gamma)^2+\gamma^4}}{\gamma^2+4}\quad
\Rightarrow\quad
\frac{4a_1+a_1\gamma^2+8\qb}{2\gamma\qb(\gamma^2+4)}=\pm \frac{\sqrt{1-\qb^2+\la^2}}{\qb(1+\la^2)}\,.
\eea
The last relation ensures that (\ref{GenSolnTemp1}) recovers the correct expression for the dilaton from (\ref{GenSoln2301Main}). This completes the verification of the map between the geometry found in \cite{2301} and a special case of our solution (\ref{GenrSolnAnzFinMain})--(\ref{GenrSolnAnzFinParMain}).


\begin{thebibliography}{99}
\bibitem{HorStr}
G.~T.~Horowitz and A.~Strominger,
``Black strings and P-branes,''
Nucl. Phys. B \textbf{360}, 197-209 (1991).
%
\bibitem{PolchBraneW1}
J.~Dai, R.~G.~Leigh and J.~Polchinski,
``New Connections Between String Theories,''
Mod. Phys. Lett. A \textbf{4}, 2073-2083 (1989).
\bibitem{PolchBraneW2}
R.~G.~Leigh,
``Dirac-Born-Infeld Action from Dirichlet Sigma Model,''
Mod. Phys. Lett. A \textbf{4}, 2767 (1989).
\bibitem{PolchBraneW3}
P.~Horava,
``Strings on World Sheet Orbifolds,''
Nucl. Phys. B \textbf{327}, 461-484 (1989).
\bibitem{PolchBraneW4}
P.~Horava,
``Background Duality of Open String Models,''
Phys. Lett. B \textbf{231}, 251-257 (1989).
%
\bibitem{PolchBraneW5}
%
J.~Polchinski,
``Dirichlet Branes and Ramond-Ramond charges,''
Phys. Rev. Lett. \textbf{75}, 4724-4727 (1995),
arXiv:hep-th/9510017 [hep-th].
%
\bibitem{BHapQGw1} 
A.~Strominger and C.~Vafa,
``Microscopic origin of the Bekenstein-Hawking entropy,''
Phys. Lett. B \textbf{379}, 99-104 (1996),
arXiv:hep-th/9601029.
\bibitem{BHapQGw2} 
C.~G.~Callan and J.~M.~Maldacena,
  ``D-brane Approach to Black Hole Quantum Mechanics,''
  Nucl.\ Phys.\  B {\bf 472}, 591 (1996), arXiv:hep-th/9602043.
\bibitem{BHapQGw3} 
 G.~T.~Horowitz and A.~Strominger,
  ``Counting States of Near-Extremal Black Holes,''
  Phys.\ Rev.\ Lett.\  {\bf 77}, 2368 (1996), arXiv:hep-th/9602051.
\bibitem{BHapQGw4} 
 J.~C.~Breckenridge, D.~A.~Lowe, R.~C.~Myers, A.~W.~Peet, A.~Strominger and C.~Vafa,
  ``Macroscopic and Microscopic Entropy of Near-Extremal Spinning Black Holes,''
  Phys.\ Lett.\  B {\bf 381}, 423 (1996), arXiv:hep-th/9603078.
  %
\bibitem{WittQCD}
E.~Witten,
``Branes and the dynamics of QCD,''
Nucl. Phys. B \textbf{507}, 658-690 (1997),
arXiv:hep-th/9706109.
\bibitem{mald}
J.~M.~Maldacena,
  ``The large N limit of superconformal field theories and supergravity,''
  Adv.\ Theor.\ Math.\ Phys.\  {\bf 2}, 231 (1998)
  [Int.\ J.\ Theor.\ Phys.\  {\bf 38}, 1113 (1999)], arXiv:hep-th/9711200.
\bibitem{WittAdSw1}
  S.~S.~Gubser, I.~R.~Klebanov and A.~M.~Polyakov,
  ``Gauge theory correlators from non-critical string theory,''
  Phys.\ Lett.\  B {\bf 428}, 105 (1998), arXiv:hep-th/9802109.
\bibitem{WittAdSw2}
  E.~Witten,
  ``Anti-de Sitter space and holography,''
  Adv.\ Theor.\ Math.\ Phys.\  {\bf 2}, 253 (1998), arXiv:hep-th/9802150.
\bibitem{WittAdSw3}
 O.~Aharony, S.~S.~Gubser, J.~M.~Maldacena, H.~Ooguri and Y.~Oz,
  ``Large N field theories, string theory and gravity,''
  Phys.\ Rept.\  {\bf 323}, 183 (2000), arXiv:hep-th/9905111.

\bibitem{SenW1}
A.~Sen,
  ``Strong - weak coupling duality in four-dimensional string theory,''
  Int.\ J.\ Mod.\ Phys.\ A {\bf 9}, 3707 (1994)
  hep-th/9402002.
\bibitem{SenW2}
A.~Sen,
  ``Strong - weak coupling duality in three-dimensional string theory,''
  Nucl.\ Phys.\ B {\bf 434}, 179 (1995)
  hep-th/9408083.
\bibitem{SenW3}
  A.~Sen,
  ``Black hole solutions in heterotic string theory on a torus,''
  Nucl.\ Phys.\ B {\bf 440}, 421 (1995)
  hep-th/9411187.
\bibitem{SenW4}
M.~Cvetic and D.~Youm,
  ``Dyonic BPS saturated black holes of heterotic string on a six torus,''
Phys.\ Rev.\ D {\bf 53}, 584 (1996)
  hep-th/9507090.
\bibitem{SenW5}
M.~Cvetic and A.~A.~Tseytlin,
  ``Solitonic strings and BPS saturated dyonic black holes,''
  Phys.\ Rev.\ D {\bf 53}, 5619 (1996)
  hep-th/9512031.
\bibitem{SenW6}
M.~Cvetic and D.~Youm,
  ``All the static spherically symmetric black holes of heterotic string on a six torus,''
  Nucl.\ Phys.\ B {\bf 472}, 249 (1996)
  hep-th/9512127.
%
\bibitem{GibMaedaW1}
G.~W.~Gibbons,
``Antigravitating Black Hole Solitons with Scalar Hair in N=4 Supergravity,''
Nucl. Phys. B \textbf{207}, 337-349 (1982).
%
\bibitem{GibMaedaW2}
G.~W.~Gibbons and K.~i.~Maeda,
``Black Holes and Membranes in Higher Dimensional Theories with Dilaton Fields,''
Nucl. Phys. B \textbf{298}, 741-775 (1988)
%
\bibitem{HorStrBH}
D.~Garfinkle, G.~T.~Horowitz and A.~Strominger,
``Charged black holes in string theory,''
Phys. Rev. D \textbf{43}, 3140 (1991).
%
%
\bibitem{PopeAdSw1}
G.~W.~Gibbons, H.~Lu, D.~N.~Page and C.~N.~Pope,
``The General Kerr-de Sitter metrics in all dimensions,''
J. Geom. Phys. \textbf{53}, 49-73 (2005),
arXiv:hep-th/0404008.
\bibitem{PopeAdSw2}
G.~W.~Gibbons, H.~Lu, D.~N.~Page and C.~N.~Pope,
``Rotating black holes in higher dimensions with a cosmological constant,''
Phys. Rev. Lett. \textbf{93}, 171102 (2004),
arXiv:hep-th/0409155.
%
\bibitem{MPerry}
R.~C.~Myers and M.~J.~Perry,
  ``Black Holes in Higher Dimensional Space-Times,''
  Annals Phys.\  {\bf 172}, 304 (1986).
%
\bibitem{LLMw1}
H.~Lin, O.~Lunin and J.~M.~Maldacena,
``Bubbling AdS space and 1/2 BPS geometries,''
JHEP \textbf{10}, 025 (2004),
arXiv:hep-th/0409174.
\bibitem{LLMw2}
A.~Donos,
``A Description of 1/4 BPS configurations in minimal type IIB SUGRA,''
Phys. Rev. D \textbf{75}, 025010 (2007),
arXiv:hep-th/0606199.
\bibitem{LLMw3}
B.~Chen, S.~Cremonini, A.~Donos, F.~L.~Lin, H.~Lin, J.~T.~Liu, D.~Vaman and W.~Y.~Wen,
``Bubbling AdS and droplet descriptions of BPS geometries in IIB supergravity,''
JHEP \textbf{10}, 003 (2007),
arXiv:0704.2233 [hep-th].
\bibitem{LLMw4}
O.~Lunin,
``Brane webs and 1/4-BPS geometries,''
JHEP \textbf{09}, 028 (2008),
arXiv:0802.0735 [hep-th].
\bibitem{giantW1}
J.~McGreevy, L.~Susskind and N.~Toumbas,
``Invasion of the giant gravitons from Anti-de Sitter space,''
JHEP \textbf{06}, 008 (2000),
arXiv:hep-th/0003075.
\bibitem{giantW2}
M.~T.~Grisaru, R.~C.~Myers and O.~Tafjord,
``SUSY and goliath,''
JHEP \textbf{08}, 040 (2000),
arXiv:hep-th/0008015.
%
\bibitem{CopHor}
K.~Copsey and G.~T.~Horowitz,
``Gravity dual of gauge theory on $S^2\times S^1\times R$,''
JHEP \textbf{06}, 021 (2006)
arXiv:hep-th/0602003 [hep-th].
\bibitem{MannRad}
R.~B.~Mann, E.~Radu and C.~Stelea,
``Black string solutions with negative cosmological constant,''
JHEP \textbf{09}, 073 (2006),
arXiv:hep-th/0604205 [hep-th].
%
\bibitem{RaduChargeW1}
Y.~Brihaye, E.~Radu and C.~Stelea,
``Black strings with negative cosmological constant: Inclusion of electric charge and rotation,''
Class. Quant. Grav. \textbf{24}, 4839-4870 (2007)
arXiv:hep-th/0703046 [hep-th].
\bibitem{RaduChargeW2}
Y.~Brihaye and T.~Delsate,
``Charged-rotating black holes and black strings in higher dimensional Einstein-Maxwell theory with a positive cosmological constant,''
Class. Quant. Grav. \textbf{24}, 4691-4710 (2007),
arXiv:gr-qc/0703146 [gr-qc].
%
\bibitem{BlcStrLaStabW1}
Y.~Brihaye, T.~Delsate and E.~Radu,
``On the stability of AdS black strings,''
Phys. Lett. B \textbf{662}, 264-269 (2008),
arXiv:0710.4034 [hep-th].
\bibitem{BlcStrLaStabW2}
T.~Delsate,
``Perturbative non uniform black strings in AdS(6),''
Phys. Lett. B \textbf{663}, 118-124 (2008),
arXiv:0802.1392 [hep-th].
\bibitem{BlcStrLaStabW3}
T.~Delsate,
``New stable phase of non uniform black strings in AdS(4),''
JHEP \textbf{12}, 085 (2008)
arXiv:0808.2752 [hep-th].
%
\bibitem{DeLuBrn}
R.~Deshpande and O.~Lunin,
``Black branes with cosmological constant,''
JHEP \textbf{05}, 136 (2022),
arXiv:2112.12930 [hep-th].
%
\bibitem{DeLuWrm}
R.~Deshpande and O.~Lunin,
``Charged wormholes in higher dimensions,''
Nucl. Phys. B \textbf{996}, 116355 (2023)
arXiv:2212.11962 [hep-th].
%
%
\bibitem{WormOldW1}
A.~Einstein and N.~Rosen,
``The Particle Problem in the General Theory of Relativity,''
Phys. Rev. \textbf{48}, 73-77 (1935).
\bibitem{WormOldW2}
C.~W.~Misner and J.~A.~Wheeler,
``Classical physics as geometry: Gravitation, electromagnetism, unquantized charge, and mass as properties of curved empty space,''
Annals Phys. \textbf{2}, 525-603 (1957).
%
\bibitem{WormOldW3}
M.~S.~Morris, K.~S.~Thorne and U.~Yurtsever,
``Wormholes, Time Machines, and the Weak Energy Condition,''
Phys. Rev. Lett. \textbf{61}, 1446-1449 (1988).
%
\bibitem{WormOldW4}
I.~R.~Klebanov, L.~Susskind and T.~Banks,
``Wormholes and the Cosmological Constant,''
Nucl. Phys. B \textbf{317}, 665-692 (1989).
%
\bibitem{WormOldW5}
M.~Visser,
``Traversable wormholes: Some simple examples,''
Phys. Rev. D \textbf{39}, 3182-3184 (1989).
%
\bibitem{WormOldW6}
M.~Visser,
``Quantum wormholes,''
Phys. Rev. D \textbf{43}, 402-409 (1991).
%
\bibitem{WormScalarW1}
C.~Barcelo and M.~Visser,
``Traversable wormholes from massless conformally coupled scalar fields,''
Phys. Lett. B \textbf{466}, 127-134 (1999),
arXiv:gr-qc/9908029.
\bibitem{WormScalarW2}
C.~Barcelo and M.~Visser,
``Scalar fields, energy conditions, and traversable wormholes,''
Class. Quant. Grav. \textbf{17}, 3843-3864 (2000),
arXiv:gr-qc/0003025.
\bibitem{WormScalarW3}
A.~Anabalon and A.~Cisterna,
``Asymptotically (anti) de Sitter Black Holes and Wormholes with a Self Interacting Scalar Field in Four Dimensions,''
Phys. Rev. D \textbf{85}, 084035 (2012),
arXiv:1201.2008 [hep-th].
\bibitem{WormScalarW4}
S.~V.~Bolokhov, K.~A.~Bronnikov and M.~V.~Skvortsova,
``Magnetic black universes and wormholes with a phantom scalar,''
Class. Quant. Grav. \textbf{29}, 245006 (2012),
arXiv:1208.4619 [gr-qc].
\bibitem{WormScalarW5}
G.~Franciolini, L.~Hui, R.~Penco, L.~Santoni and E.~Trincherini,
``Stable wormholes in scalar-tensor theories,''
JHEP \textbf{01}, 221 (2019)
arXiv:1811.05481 [hep-th].
\bibitem{WormScalarW6}
G.~Antoniou, A.~Bakopoulos, P.~Kanti, B.~Kleihaus and J.~Kunz,
``Novel Einstein--scalar--Gauss--Bonnet wormholes without exotic matter,''
Phys. Rev. D \textbf{101}, no.2, 024033 (2020),
arXiv:1904.13091 [hep-th].
\bibitem{WormScalarW7}
J.~L.~Bl\'azquez-Salcedo, X.~Y.~Chew, J.~Kunz and D.~H.~Yeom,
``Ellis wormholes in anti-de Sitter space,''
Eur. Phys. J. C \textbf{81}, no.9, 858 (2021)
arXiv:2012.06213 [gr-qc].
\bibitem{WormScalarW8}
W.~Harvey and K.~Jensen,
``Eternal traversable wormholes in three dimensions,''
JHEP \textbf{10}, 178 (2023),
arXiv:2302.14049 [hep-th].
%
%
\bibitem{EREPRw1}
J.~Maldacena and L.~Susskind,
``Cool horizons for entangled black holes,''
Fortsch. Phys. \textbf{61}, 781-811 (2013),
arXiv:1306.0533 [hep-th].
%
\bibitem{EREPRw2}
K.~Jensen and A.~Karch,
``Holographic Dual of an Einstein-Podolsky-Rosen Pair has a Wormhole,''
Phys. Rev. Lett. \textbf{111}, no.21, 211602 (2013),
arXiv:1307.1132.
%
\bibitem{EREPRw3}
L.~Susskind,
``ER=EPR, GHZ, and the consistency of quantum measurements,''
Fortsch. Phys. \textbf{64}, 72-83 (2016),
arXiv:1412.8483 [hep-th].
%
\bibitem{EREPRw4}
J.~Maldacena, D.~Stanford and Z.~Yang,
``Diving into traversable wormholes,''
Fortsch. Phys. \textbf{65}, no.5, 1700034 (2017),
arXiv:1704.05333 [hep-th].
\bibitem{EREPRw5}
J.~Maldacena, A.~Milekhin and F.~Popov,
``Traversable wormholes in four dimensions,''
arXiv:1807.04726 [hep-th].
\bibitem{4dWHw1}
J.~Maldacena and X.~L.~Qi,
``Eternal traversable wormhole,''
arXiv:1804.00491.
\bibitem{4dWHw2}
G.~T.~Horowitz, D.~Marolf, J.~E.~Santos and D.~Wang,
``Creating a Traversable Wormhole,''
Class. Quant. Grav. \textbf{36}, no.20, 205011 (2019),
arXiv:1904.02187.
\bibitem{4dWHQw1}
S.~W.~Kim and H.~Lee,
``Exact solutions of a charged wormhole,''
Phys. Rev. D \textbf{63}, 064014 (2001),
arXiv:gr-qc/0102077 [gr-qc].
\bibitem{4dWHQw2}
K.~A.~Bronnikov, L.~N.~Lipatova, I.~D.~Novikov and A.~A.~Shatskiy,
``Example of a stable wormhole in general relativity,''
Grav. Cosmol. \textbf{19}, 269-274 (2013),
arXiv:1312.6929 [gr-qc].
\bibitem{4dWHQw3}
%
A.~Cisterna, K.~M\"uller, K.~Pallikaris and A.~Vigan\`o,
``Exact rotating wormholes via Ehlers transformations,''
Phys. Rev. D \textbf{108}, no.2, 024066 (2023),
arXiv:2306.14541.
\bibitem{4dWHQw4}
B.~Meiring, I.~Shyovitz, S.~Waeber and A.~Yarom,
``Multiply charged magnetic black branes,''
JHEP \textbf{06}, 196 (2024)
arXiv:2312.02802 [hep-th].
\bibitem{4dWHQw5}
S.~Antonini and L.~G.~C.~Bariuan,
``Magnetic Braneworlds: Cosmology and Wormholes,''
arXiv:2405.18465 [hep-th].
%
\bibitem{Example4dW1}
A.~Anabal\'on and J.~Oliva,
``Four-dimensional Traversable Wormholes and Bouncing Cosmologies in Vacuum,''
JHEP \textbf{04}, 106 (2019),
arXiv:1811.03497 [hep-th].
%
\bibitem{Example4dW2}
A.~Anabal\'on, B.~de Wit and J.~Oliva,
``Supersymmetric traversable wormholes,''
JHEP \textbf{09}, 109 (2020),
arXiv:2001.00606 [hep-th].
\bibitem{Example4dW3}
%
T.~Tangphati, B.~Chaihao, D.~Samart, P.~Channuie and D.~Momeni,
``Rotating traversable wormhole geometries in the presence of three-form fields,''
Nucl. Phys. B \textbf{999}, 116446 (2024),
arXiv:2307.13968 [gr-qc].
%
\bibitem{Example4dW4}
P.~Panyasiripan, F.~Felegary and P.~Channuie,
``Charged Wormhole Solutions in 4D Einstein-Gauss-Bonnet Gravity,''
arXiv:2406.17362 [gr-qc].
%
%
\bibitem{WHdimW1}
J.~Y.~Kim, H.~W.~Lee and Y.~S.~Myung,
``Classical instanton and wormhole solution of type IIB string theory,''
Phys. Lett. B \textbf{400}, 32-36 (1997),
arXiv:hep-th/9612249.
\bibitem{WHdimW2}
G.~Dotti, J.~Oliva and R.~Troncoso,
``Static wormhole solution for higher-dimensional gravity in vacuum,''
Phys. Rev. D \textbf{75}, 024002 (2007),
arXiv:hep-th/0607062.
\bibitem{WHdimW3}
%
M.~H.~Dehghani and S.~H.~Hendi,
``Wormhole Solutions in Gauss-Bonnet-Born-Infeld Gravity,''
Gen. Rel. Grav. \textbf{41}, 1853-1863 (2009),
arXiv:0903.4259 [hep-th].
%
\bibitem{WHdimW4}
F.~Rahaman, S.~Islam, P.~K.~F.~Kuhfittig and S.~Ray,
``Searching for higher dimensional wormhole with noncommutative geometry,''
Phys. Rev. D \textbf{86}, 106010 (2012),
arXiv:1209.2917.
\bibitem{WHdimW5}
%
F.~R.~Klinkhamer,
``Higher-dimensional extension of a vacuum-defect wormhole,''
arXiv:2307.12876 [gr-qc].
%
%
\bibitem{GSR4w1}
B.~de Wit and H.~Nicolai,
``N=8 Supergravity with Local SO(8) x SU(8) Invariance,''
Phys. Lett. B \textbf{108}, 285 (1982).
\bibitem{GSR4w2}
B.~de Wit and H.~Nicolai,
``N=8 Supergravity,''
Nucl. Phys. B \textbf{208}, 323 (1982).

\bibitem{GSR5w1}
M.~Gunaydin, L.~J.~Romans and N.~P.~Warner,
``Gauged N=8 Supergravity in Five-Dimensions,''
Phys. Lett. B \textbf{154}, 268-274 (1985).
\bibitem{GSR5w2}
L.~J.~Romans,
``Gauged $N=4$ Supergravities in Five-dimensions and Their Magnetovac Backgrounds,''
Nucl. Phys. B \textbf{267}, 433-447 (1986).

\bibitem{GSR6}
L.~J.~Romans,
``The F(4) Gauged Supergravity in Six-dimensions,''
Nucl. Phys. B \textbf{269}, 691 (1986)

\bibitem{GSR7w1}
P.~K.~Townsend and P.~van Nieuwenhuizen,
``Gauged seven-dimensional supergravity,''
Phys. Lett. B \textbf{125}, 41-46 (1983).
\bibitem{GSR7w2}
M.~Pernici, K.~Pilch and P.~van Nieuwenhuizen,
``Gauged Maximally Extended Supergravity in Seven-dimensions,''
Phys. Lett. B \textbf{143}, 103-107 (1984).
%
\bibitem{GSR4slnW1}
M.~J.~Duff and J.~T.~Liu,
``Anti-de Sitter black holes in gauged N = 8 supergravity,''
Nucl. Phys. B \textbf{554}, 237-253 (1999),
arXiv:hep-th/9901149.
%
\bibitem{GSR4slnW2}
%
W.~A.~Sabra,
``Anti-de Sitter BPS black holes in N=2 gauged supergravity,''
Phys. Lett. B \textbf{458}, 36-42 (1999),
arXiv:hep-th/9903143.
%
\bibitem{GSR4snRt}
D.~D.~K.~Chow,
``Single-charge rotating black holes in four-dimensional gauged supergravity,''
Class. Quant. Grav. \textbf{28}, 032001 (2011),
arXiv:1011.2202 [hep-th].
%
%
%
\bibitem{GSR5sln}
K.~Behrndt, M.~Cvetic and W.~A.~Sabra,
``Nonextreme black holes of five-dimensional N=2 AdS supergravity,''
Nucl. Phys. B \textbf{553}, 317-332 (1999),
arXiv:hep-th/9810227.
\bibitem{GSR5snRotW1}
D.~Klemm and W.~A.~Sabra,
``Charged rotating black holes in 5-D Einstein-Maxwell (A)dS gravity,''
Phys. Lett. B \textbf{503}, 147-153 (2001),
arXiv:hep-th/0010200.
\bibitem{GSR5snRotW2}
M.~Cvetic, H.~Lu and C.~N.~Pope,
``Charged rotating black holes in five dimensional U(1)**3 gauged N=2 supergravity,''
Phys. Rev. D \textbf{70}, 081502 (2004),
arXiv:hep-th/0407058.
\bibitem{GSR5snRotW3}
Z.~W.~Chong, M.~Cvetic, H.~Lu and C.~N.~Pope,
``Five-dimensional gauged supergravity black holes with independent rotation parameters,''
Phys. Rev. D \textbf{72}, 041901 (2005),
arXiv:hep-th/0505112.
\bibitem{GSR5snRotW4}
Z.~W.~Chong, M.~Cvetic, H.~Lu and C.~N.~Pope,
``General non-extremal rotating black holes in minimal five-dimensional gauged supergravity,''
Phys. Rev. Lett. \textbf{95}, 161301 (2005),
arXiv:hep-th/0506029.
\bibitem{GSR5snRotW5}
Z.~W.~Chong, M.~Cvetic, H.~Lu and C.~N.~Pope,
``Non-extremal rotating black holes in five-dimensional gauged supergravity,''
Phys. Lett. B \textbf{644}, 192-197 (2007),
arXiv:hep-th/0606213;
%
%
%
\bibitem{GSR6sln}
M.~Cvetic, H.~Lu and C.~N.~Pope,
``Gauged six-dimensional supergravity from massive type IIA,''
Phys. Rev. Lett. \textbf{83}, 5226-5229 (1999),
arXiv:hep-th/9906221.
\bibitem{GSR6snRot}
D.~D.~K.~Chow,
``Charged rotating black holes in six-dimensional gauged supergravity,''
Class. Quant. Grav. \textbf{27}, 065004 (2010),
arXiv:0808.2728 [hep-th].
%
\bibitem{AdCFbhW1}
S.~W.~Hawking, C.~J.~Hunter and M.~Taylor,
``Rotation and the AdS / CFT correspondence,''
Phys. Rev. D \textbf{59}, 064005 (1999),
arXiv:hep-th/9811056.
\bibitem{AdCFbhW2}
S.~W.~Hawking and H.~S.~Reall,
``Charged and rotating AdS black holes and their CFT duals,''
Phys. Rev. D \textbf{61}, 024014 (2000),
arXiv:hep-th/9908109.
\bibitem{CvetGub}
M.~Cvetic and S.~S.~Gubser,
``Phases of R charged black holes, spinning branes and strongly coupled gauge theories,''
JHEP \textbf{04}, 024 (1999)
arXiv:hep-th/9902195.
%
\bibitem{GSR7snRot}
Z.~W.~Chong, M.~Cvetic, H.~Lu and C.~N.~Pope,
``Non-extremal charged rotating black holes in seven-dimensional gauged supergravity,''
Phys. Lett. B \textbf{626}, 215-222 (2005),
arXiv:hep-th/0412094.
%



\bibitem{10auth}
M.~Cvetic, M.~J.~Duff, P.~Hoxha, J.~T.~Liu, H.~Lu, J.~X.~Lu, R.~Martinez-Acosta, C.~N.~Pope, H.~Sati and T.~A.~Tran,
``Embedding AdS black holes in ten-dimensions and eleven-dimensions,''
Nucl. Phys. B \textbf{558}, 96-126 (1999),
arXiv:hep-th/9903214.
%
\bibitem{Wu5d}
S.~Q.~Wu,
``General Nonextremal Rotating Charged AdS Black Holes in Five-dimensional $U(1)^3$ Gauged Supergravity: A Simple Construction Method,''
Phys. Lett. B \textbf{707}, 286-291 (2012)
arXiv:1108.4159 [hep-th].
\bibitem{Wu7d}
S.~Q.~Wu,
``Two-charged non-extremal rotating black holes in seven-dimensional gauged supergravity: The Single-rotation case,''
Phys. Lett. B \textbf{705}, 383-387 (2011).
arXiv:1108.4158 [hep-th].
%
%
\bibitem{LuGuesPot}
H.~Lu,
``Charged dilatonic ads black holes and magnetic AdS$_{D-2} \times R^{2}$ vacua,''
JHEP \textbf{09}, 112 (2013)
arXiv:1306.2386 [hep-th].
%
\bibitem{Schw}
J.~Maharana and J.~H.~Schwarz,
``Noncompact symmetries in string theory,''
Nucl. Phys. B \textbf{390}, 3-32 (1993),
arXiv:hep-th/9207016;\\
%
J.~H.~Schwarz,
``An SL(2,Z) multiplet of type IIB superstrings,''
Phys. Lett. B \textbf{360}, 13-18 (1995)
[erratum: Phys. Lett. B \textbf{364}, 252 (1995)],
arXiv:hep-th/9508143.
%
\bibitem{WuAllDim}
S.~Q.~Wu,
``General rotating charged Kaluza-Klein AdS black holes in higher dimensions,''
Phys. Rev. D \textbf{83}, 121502 (2011)
arXiv:1108.4157 [hep-th].
%
\bibitem{PolchStrW1}
J.~H.~Schwarz,
``Covariant Field Equations of Chiral N=2 D=10 Supergravity,''
Nucl. Phys. B \textbf{226}, 269 (1983).
\bibitem{PolchStrW2}
J.~H.~Schwarz and P.~C.~West,
``Symmetries and Transformations of Chiral N=2 D=10 Supergravity,''
Phys. Lett. B \textbf{126}, 301-304 (1983).
\bibitem{PolchStrW3}
P.~S.~Howe and P.~C.~West,
``The Complete N=2, D=10 Supergravity,''
Nucl. Phys. B \textbf{238}, 181-220 (1984).
\bibitem{PolchStrW4}
J.~Polchinski and M.~J.~Strassler,
``The String dual of a confining four-dimensional gauge theory,''
arXiv:hep-th/0003136.
%
\bibitem{KlebTseytl}
I.~R.~Klebanov and A.~A.~Tseytlin,
``Intersecting M-branes as four-dimensional black holes,''
Nucl. Phys. B \textbf{475}, 179-192 (1996),
arXiv:hep-th/9604166.
%
\bibitem{LiuSabra}
J.~T.~Liu and W.~A.~Sabra,
``Charged configurations in (A)dS spaces,''
Nucl. Phys. B \textbf{679}, 329-344 (2004),
arXiv:hep-th/0307300.
%
%
\bibitem{FrRub}
P.~G.~O.~Freund and M.~A.~Rubin,
``Dynamics of Dimensional Reduction,''
Phys. Lett. B \textbf{97}, 233-235 (1980)
%
\bibitem{AsympLocAdSw1}
K.~Skenderis,
``Lecture notes on holographic renormalization,''
Class. Quant. Grav. \textbf{19}, 5849-5876 (2002),
arXiv:hep-th/0209067.
\bibitem{AsympLocAdSw2}
D.~Marolf, W.~Kelly and S.~Fischetti,
``Conserved Charges in Asymptotically (Locally) AdS Spacetimes,''
arXiv:1211.6347 [gr-qc].
%
%
 \bibitem{D2D6w1}
 G.~Papadopoulos, P.~K.~Townsend and P.~V.~Landshoff,
``Intersecting M-branes,''
Phys. Lett. B \textbf{380}, 273-279 (1996),
arXiv:hep-th/9603087.
\bibitem{D2D6w2}
A.~A.~Tseytlin,
``Harmonic superpositions of M-branes,''
Nucl. Phys. B \textbf{475}, 149-163 (1996),
arXiv:hep-th/9604035.
\bibitem{D2D6w3}
A.~A.~Tseytlin,
``'No force' condition and BPS combinations of p-branes in eleven-dimensions and ten-dimensions,''
Nucl. Phys. B \textbf{487}, 141-154 (1997),
arXiv:hep-th/9609212.
\bibitem{D2D6w4}
R.~Argurio, F.~Englert and L.~Houart,
``Intersection rules for p-branes,''
Phys. Lett. B \textbf{398}, 61-68 (1997),
arXiv:hep-th/9701042.
%
\bibitem{D2D6w5}
N.~Ohta,
``Intersection rules for nonextreme p-branes,''
Phys. Lett. B \textbf{403}, 218-224 (1997),
arXiv:hep-th/9702164.
%
\bibitem{2301Pre}
M.~Rakhmanov,
``Dilaton black holes with electric charge,''
Phys. Rev. D \textbf{50}, 5155-5163 (1994),
arXiv:hep-th/9310174 [hep-th].
%
\bibitem{2301}
A.~P.~Porfyriadis and G.~N.~Remmen,
``Charged dilatonic spacetimes in string theory,''
JHEP \textbf{03}, 125 (2023),
arXiv:2301.08256 [hep-th].
%
\bibitem{Abel}
N.~H.~Abel. ``Precis d'une theorie des fonctions elliptiques,'' J. Reine Angew. Math. {\bf 4}, 309-348 (1829).
%
\bibitem{LiuSabraPre}
D.~Klemm and W.~A.~Sabra,
``Supersymmetry of black strings in D = 5 gauged supergravities,''
Phys. Rev. D \textbf{62}, 024003 (2000),
arXiv:hep-th/0001131.
%
%
%
%








\end{thebibliography}
\end{document}